\documentclass[10pt]{iopart}
\usepackage{xcolor}
\usepackage{graphicx}
\usepackage[normalem]{ulem} 
\usepackage{amssymb}
\usepackage{tikz}
\usepackage{comment}
\newcommand{\wjr}[1]{\textcolor{cyan}{#1}}

\newcommand{\sear}[1]{\textcolor{magenta}{#1}}
\newcommand{\badri}[1]{\textcolor{brown}{#1}}
\usepackage{hyperref}

\begin{document}

\title{Lumens as active balloons: a biological physics review}

\author{S Echeverr\'ia-Alar$^1$, B Narayanan Narasimhan$^2$, 
S I Fraley$^2$ and W-J Rappel$^1$\footnote{To whom correspondence should be addressed.}}

\address{$^1$ Department of Physics, University of California, San Diego, California 92093, USA}
\address{$^2$ Department of Bioengineering, University of California, San Diego, California 92093, USA}
\ead{rappel@physics.ucsd.edu}

\vspace{10pt}
\begin{indented}
\item[]\today
\end{indented}

\begin{abstract}
Lumens are cavities enclosed by polarized cells that are essential for organ function, from nutrient transport in the gut to gas exchange in the lungs.  Defects in lumen formation are associated with severe diseases, including polycystic kidney disease and respiratory malformations. The emergence, growth, and maintenance of lumens involve a rich set of phenomena that can be framed within out-of-equilibrium physics and biological active matter, including osmotically driven hydraulic flows, coarsening-like dynamics, morphological instabilities, and mechanochemical feedbacks linking luminal pressure to tissue response. Yet experimental and theoretical efforts to study these phenomena have largely developed within specific biological systems, complicating the identification of shared physical principles across them. In this review, we bring these efforts together and present lumenogenesis within a biological physics framework in which lumens are viewed as active balloons: pressurized cavities that are inflated, sculpted, 
and maintained through tightly coupled active processes. We first introduce the main biological constituents of lumenogenesis, from molecular pumps and polarity complexes to epithelial architecture, luminal fluid, and extracellular matrix. We then review the physics of lumenogenesis, including the routes by which cavities nucleate, the hydraulic machinery that inflates them, the mechanical balance that accommodates  expansion, and the mechanochemical feedbacks between the lumen and its epithelial container. 
Next, we discuss theoretical and computational frameworks developed to formalize these processes, including network models for pressure-driven coarsening, coarse-grained theories of lumen-tissue interactions, and agent-based models that resolve lumen dynamics at the cellular level. 
We also examine how engineered extracellular matrices provide experimental control over adhesion, mechanics, degradability, and confinement, thereby bridging tissue engineering and quantitative modeling.  
Finally, we highlight open questions---from whether single-lumen selection can be understood within general theories of active coarsening to how lumen positioning, shape, and stability are controlled across systems---that make lumenogenesis a promising problem for physics-based approaches to morphogenesis.
\end{abstract}

%
\vspace{2pc}
\ioptwocol

\tableofcontents
\thispagestyle{empty}
\markboth{}{}

\section{Introduction}\label{intro}

The origin of form in living systems---and the role that form plays in governing biological function---has long fascinated scientists. As early as Ancient Greece, Aristotle proposed the concept of vitalism, positing that a guiding force, or entelechy, directed the development of organisms. Though philosophical in nature, this idea later influenced early biological theories. In the early 1900s, Hans Driesch showed that separated cells from early sea urchin embryos could each develop into a complete organism \cite{driesch1900isolirten}. To explain this developmental robustness, he invoked the idea of entelechy as well. Nearly two decades later, D’Arcy W. Thompson shifted the discourse by proposing that biological forms could be explained through physical principles alone \cite{thomson1917growth}. The transition to a more mechanistic view was further advanced by Alan Turing, who introduced a mathematical theory of morphogen patterning via diffusion-driven instabilities \cite{turing1990chemical}. Later, advances in molecular biology revealed the machinery that translate genetic information into spatial organization \cite{wolpert1969positional,lewis1978gene}. From a statistical physics perspective, Ilya Prigogine and the Brussels school developed the concept of dissipative structures trying to explain how biological order arises from disorder \cite{prigogine1967symmetry,prigogine1968symmetry,nicolis1971fluctuations}. They showed that systems driven far from equilibrium can undergo spontaneous symmetry breaking and self-organize.

Today, morphogenesis remains an active area of research and continues to inspire multidisciplinary research \cite{grimes2017left,zhang2018symmetry,erzberger2020mechanochemical,bordeu2023inflationary,plum2025dynamical,bischoff2025patterning}. In particular, physicists and biologists are increasingly collaborating to uncover the physical, biochemical, and genetic mechanisms that govern form generation \cite{gilmour2017morphogen,morelli2012computational}. From a physics perspective, morphogenesis presents essential challenges and raises fundamental questions. How do collectives of cells robustly generate reproducible tissue-level structures? What physical principles underlie the emergence of shapes when genetic programs, mechanical forces, and biochemical gradients act simultaneously across various length and time scales? These questions lie at the interface of active matter physics \cite{marchetti2013hydrodynamics}, non-equilibrium thermodynamics \cite{prost2015active,prigogine2017non}, and mechano-chemical pattern formation \cite{turing1990chemical,howard2011turing}. Despite decades of progress in understanding dissipative structures in macroscopic systems and the mechanics of passive materials, we still lack a predictive framework for how cells build and maintain dynamic three-dimensional morphologies.

\begin{figure*}
\centering
\begin{tikzpicture}
  \node[inner sep=0] (img)
    {\includegraphics[width=0.8\linewidth]{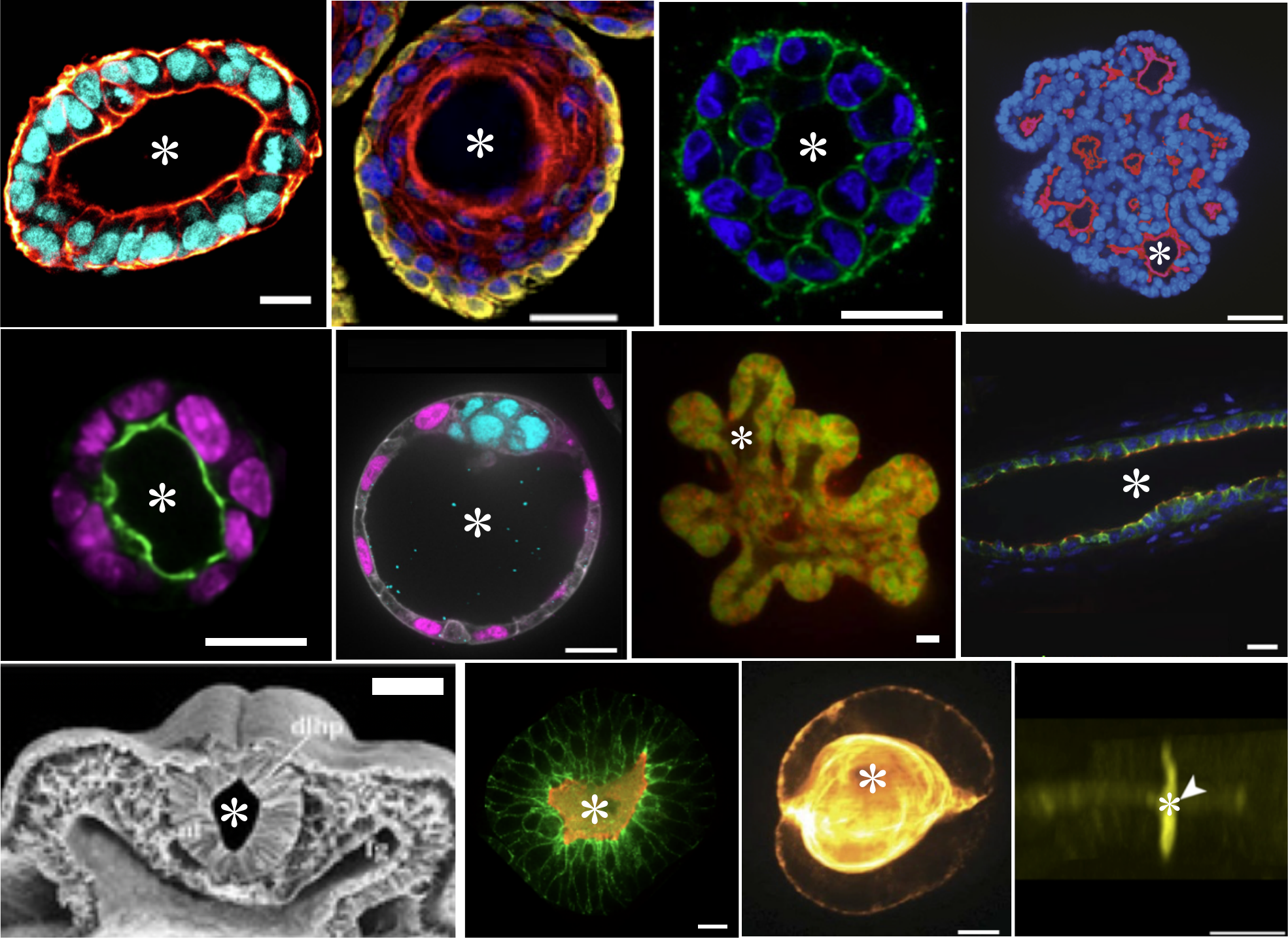}};
  \node at (img.north west) [xshift=3mm, yshift=-3mm, text=white]{\large\bfseries A};
  \node at (img.north west) [xshift=38mm, yshift=-3mm, text=white]{\large\bfseries B};
    \node at (img.north west) [xshift=73mm, yshift=-3mm, text=white]{\large\bfseries C};
      \node at (img.north west) [xshift=105.5mm, yshift=-3mm, text=white]{\large\bfseries D};
       \node at (img.north west) [xshift=3mm, yshift=-38mm, text=white]{\large\bfseries E};
       \node at (img.north west) [xshift=38mm, yshift=-38mm, text=white]{\large\bfseries F};
       \node at (img.north west) [xshift=71mm, yshift=-38mm, text=white]{\large\bfseries G};
        \node at (img.north west) [xshift=105.5mm, yshift=-38mm, text=white]{\large\bfseries H};
        \node at (img.north west) [xshift=2.5mm, yshift=-74mm, text=white]{\large\bfseries I};
          \node at (img.north west) [xshift=52mm, yshift=-74mm, text=white]{\large\bfseries J};
            \node at (img.north west) [xshift=82mm, yshift=-74mm, text=white]{\large\bfseries K};
              \node at (img.north west) [xshift=111mm, yshift=-74mm, text=white]{\large\bfseries L};
\end{tikzpicture}
    \caption{Lumens are ubiquitous structures in biological contexts. (A) Lumen in human induced pluripotent stem cell epiblasts (scale bar, $20\ \mu$m). Adapted from \cite{indana2024lumen}. (B) Lumen in murine epidermal organoids (scale bar, $50\ \mu$m). Adapted from \cite{boonekamp2019long}. (C) Lumen in the acinar structure of MDA-MB-231 cells (scale bar, $10\ \mu$m). Adapted from \cite{ranamukhaarachchi2025global}. (D) Multiple lumens in a pancreatic organoid (scale bar, $40\ \mu$m). Adapted from \cite{tanida2025predicting}. (E) Lumen in pancreatic spheroid (scale bar, $10\ \mu$m). Adapted from \cite{lu2025generic}. (F) Lumen in the mouse blastocyst (scale bar, $20\ \mu$m). Adapted from \cite{schliffka2021multiscale}. (G-H) Lumens in the mammary ducts of a mouse model (scale bar, $20\ \mu$m). Adapted from \cite{ewald2008collective}. (I) Lumen after neural folding in the chick embryo (scale bar inferred from  \cite{hamburger1951series}, $100\mu$m). Adapted from \cite{colas2001towards}. (J) Lumen in epiblast (scale bar, $10\ \mu$m). Adapted from \cite{lu2025generic}. (K) Lumen within two hepatocytes (scale bar, $2\ \mu$m). Adapted from \cite{dasgupta2018physics}. (L) Intracellular lumen in the excretory cell of C. elegans (scale bar, $5\ \mu$m). Adapted from \cite{meyer2026integrin}. The asterisk (*) indicates the location of the cavities and, for simplicity, in (D) and (F) only one lumen is marked. Color codes for the fluorescent images are detailed in the respective references.}
    \label{fig1}
\end{figure*}
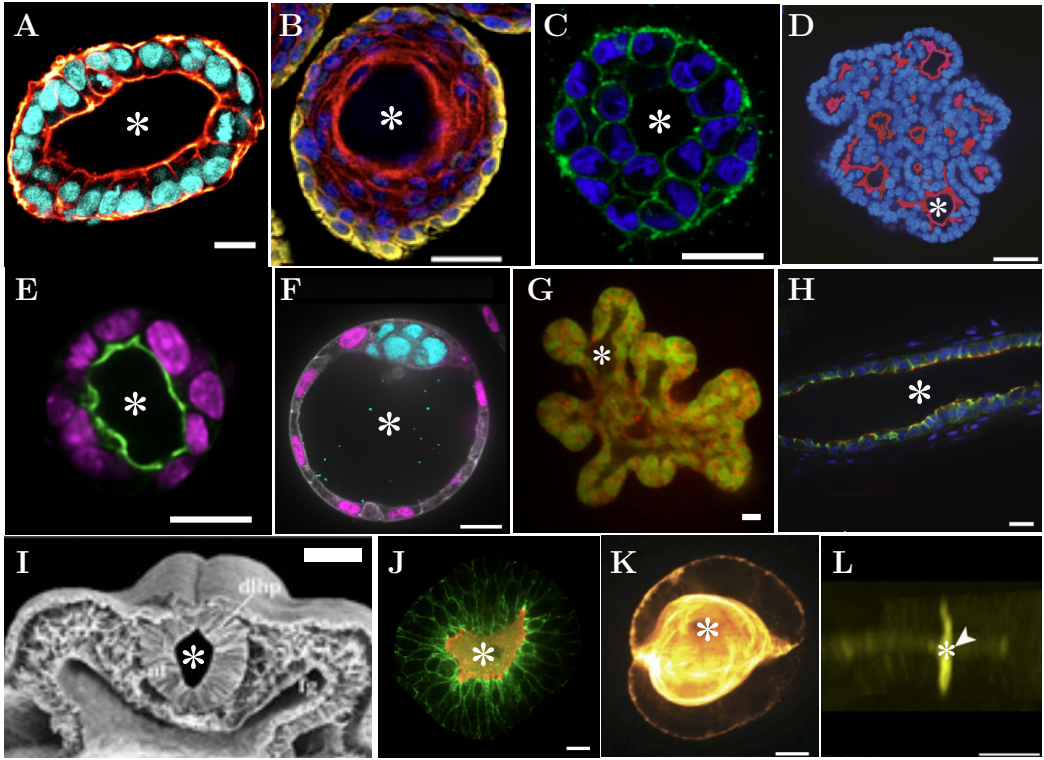

Among the many components required to produce functional organs, a central and often overlooked structure is the lumen: a fluid- or air-filled cavity enclosed by a polarized epithelial layer connected via tight junctions \cite{NAVIS2016139,mukenhirn2024tight}. From a physics perspective, the lumen can be understood as an \textit{active balloon}: a pressurized cavity surrounded by a material that consumes energy and realizes work to maintain and regulate its shape. The hydrostatic pressure of the luminal fluid provides mechanical support to the lumen-container  \cite{lemke2021dynamic,yu2024mechanochemical}. Unlike a passive elastic shell, this container is made of epithelial cells that allow ion fluxes \cite{dasgupta2018physics}, regulate spatiotemporally contractile forces \cite{salbreux2012actin}, and remodel contacts with their neighbors \cite{takeichi1991cadherin}. At the same time, this active balloon is receiving mechanochemical cues from the interior fluid and the exterior matrix. This \textit{active} behavior places lumenogenesis within the realm of out-of-equilibrium physics. Lumens serve essential physiological roles in mature organisms, mediating the flow of gases, fluids, and cells \cite{lubarsky2003tube}. During early development, however, their importance is even more fundamental: lumen formation guides tissue folding during organogenesis \cite{martin2008regulation}, facilitates spatially organized signaling \cite{durdu2014luminal,zhang2019mouse}, and influences cell fate decisions \cite{kim2021deciphering}. Fig.~\ref{fig1} shows different lumen types across biological systems.

Lumenogenesis unfolds through a wide range of spatial and temporal scales. At the molecular level (nanometers), ion pumps, ion channels and water channels establish osmotic gradients \cite{dasgupta2018physics}, generating fluxes that last from minutes to hours \cite{torres2021tissue}. At the cellular level (micrometers), cells coordinate an apicobasal polarity over the course of hours \cite{indana2024lumen}. Lumens can nucleate within hours to a day after cell division in many \textit{in-vitro} systems \cite{lu2025generic} and  initially span a few microns in diameter before expanding to tens of microns in the span of days \cite{martin2008regulation,wang1990steps}.

Defects in lumenogenesis can severely compromise organ function and are implicated in life-threatening conditions. For instance, autosomal dominant polycycstic kidney disease (ADPKD), one of the most common genetic kidney disorders, affects approximately 1 in $400$--$1000$ individuals worldwide \cite{harris2009polycystic}. In this disease, aberrant lumen sculpting leads to the progressive development of fluid-filled cysts---three-dimensional structures comprising a central lumen surrounded by a layer of polarized epithelial cells---throughout the kidney \cite{ma2021cilia}. These cysts expand over time, replacing normal tissue and eventually resulting in renal failure. The disease is caused by mutations in genes encoding polycystin proteins, which are thought to regulate mechanosensation in response to fluid flow within the lumen. Similar lumens abnormalities can restrict blood flow in the aorta \cite{cartlidge2021contrast}, cause neonatal respiratory distress due to lung malformation \cite{appuhn2021capillary}, obstruct nutrient passage in the intestines \cite{liu2022gut}, or disrupt sensory function through malformation of the inner ear \cite{mosaliganti2019size}.


As with many aspects of morphogenesis, a deeper quantitative understanding of lumenogenesis will be critical for both advancing basic science and achieving reliable control over engineered tissues \cite{datta2011molecular,taniguchi2015lumen}. In this review, we will present the current state-of-the-art of lumen formation from an
experimental, theoretical, and computational  point of view. Our goal is to highlight the central physical principles underlying this process and to encourage broader engagement from the biological physics community. To provide both a roadmap for future investigation and an accessible entry point for physicists entering the field of lumenogenesis, we frame the review around four key questions: How does a single lumen emerge from many through coarsening-like dynamics? How is the active balloon inflated, sculpted, and maintained out of equilibrium? What is the role of the extracellular matrix in this process? How does the lumen feed back on the surrounding epithelial tissue?

Our review is organized as follows: we will first describe the principal biological units associated with the creation of lumens from small to large scales, highlighting lumenogenesis as a multiscale phenomena (Section II). Then, we will review the physics underlying lumenogenesis, addressing the emergence and maintenance of the cavity, framing the lumen as an active balloon whose dynamic behavior depends of an interplay between hydraulic machinery, cell membrane mechanics, out-of-equilibrium cell processes, and extracellular matrix (ECM) properties. Along the way, we will underscore key experimental observations and techniques that have helped build the physical picture of lumenogenesis (Section III). Next, we  review the theoretical rationale of coarse-grained models used to describe the active stabilization of lumens and the role of cavities at the tissue level, together with the  computational techniques currently used to analyze these models (Section IV). After that, we will discuss recent advances in engineered ECMs, bridging tissue engineering with physical inquiry; a critical crosstalk to constrain mathematical models in the context of lumenogenesis (Section V). Finally, we will conclude with an outlook of the field, with a particular focus on
open problems that can be addressed through physical modeling and innovative
experimentation (Section VI). 

\section{Orchestrating lumenogenesis: Biological constituents across scales}

Lumenogenesis is a multiscale process that cannot be understood from any single biological level in isolation. The emergence of a cavity requires the coordinated action of molecular machinery that establishes polarity, regulates permeability, drives transport, generates forces, and couples cells to their surroundings. These activities unfold across a wide range of length scales, from nanometer-scale protein complexes to micrometer-scale cellular organization and, ultimately, to tissue-scale structures. In this section, we briefly introduce the principal biological constituents that participate in lumen formation and maintenance, with the goal of providing a coherent framework for the physical picture developed in later sections. 
For readers seeking a more detailed biological treatment, we refer to dedicated reviews \cite{datta2011molecular,bryant2008cells,sigurbjornsdottir2014molecular}

We begin at the subcellular level by reviewing the molecular machinery that makes lumenogenesis possible. This includes polarity complexes, junctional proteins, ion channels, pumps, and adhesion complexes that define membrane identity and regulate exchange of ions and fluid across the epithelial layer (Section~\ref{s21}). We then consider vesicular transport, which directs key proteins and membrane components to their proper destinations, and the cytoskeleton, which acts both as mechanical support and as a source of active force generation (Sections~\ref{s22} and~\ref{s23}). Together, these subsections describe how the apical domain is specified, supplied, and mechanically supported before a stable lumen can emerge.

We then move to larger scales, where these molecular processes are integrated into cellular and tissue architecture. This includes the organization of epithelial cells into apical, lateral, and basal domains, together with the adhesive and mechanical coupling they establish with their neighbors (Section~\ref{s24}). We also discuss the luminal fluid itself, not as a passive filler, but as a dynamic compartment containing ions, water, and signaling molecules that contribute to both mechanics and communication, as well as the structural scaffold provided by the ECM, which supplies biochemical cues and mechanical support (Sections~\ref{hub} and~\ref{sECM}). Taken together, these subsections highlight that lumenogenesis arises from strongly coupled processes across scales, and they set the biological stage for the physical discussion of how lumens are created, shaped, and maintained.

\begin{figure*}
    \centering
    \includegraphics[width=0.8\linewidth]{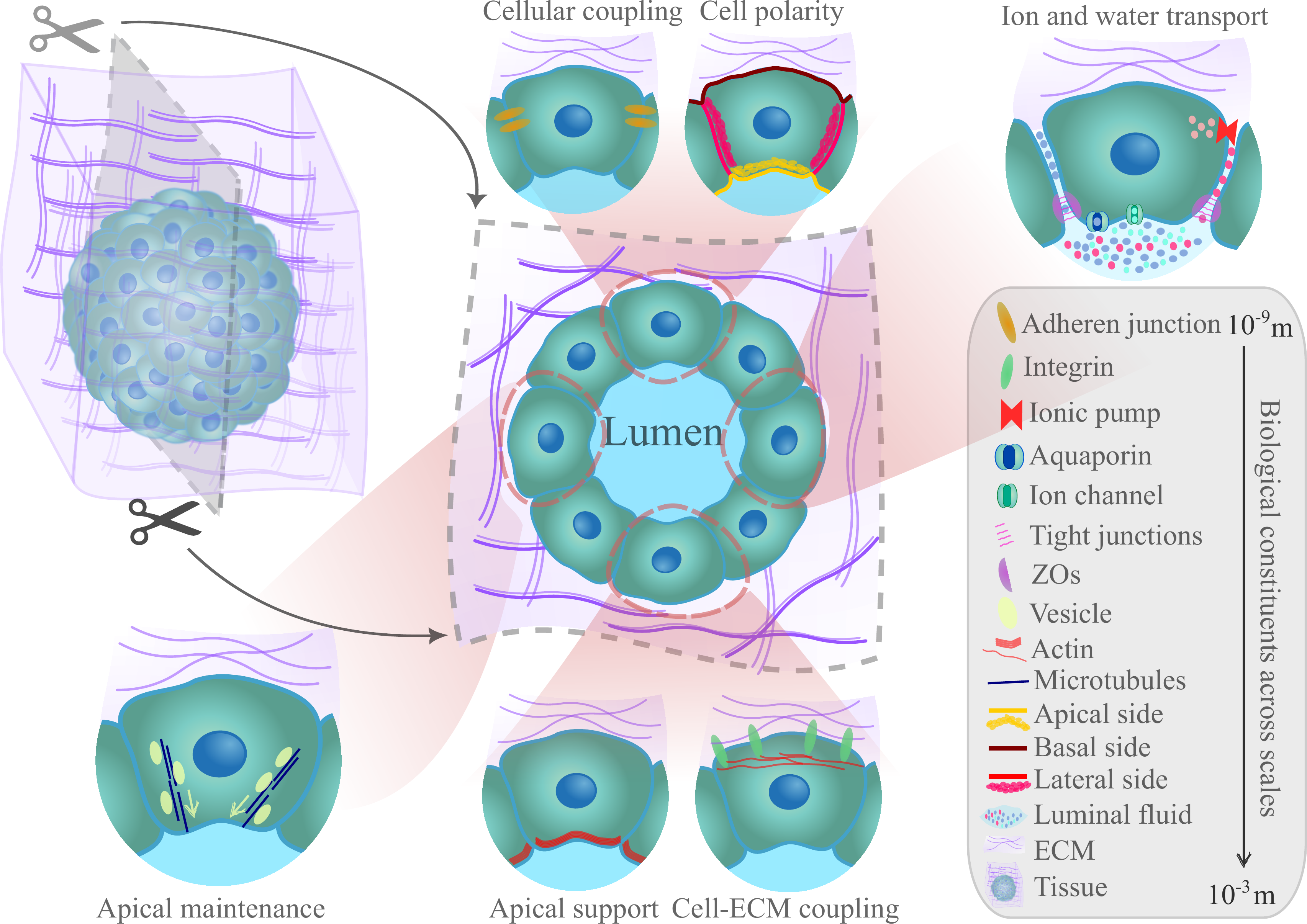}
    \caption{Schematic representation of the biological constituents sculpting a lumen, inspired by cavity structures observed in MDCK cyst embedded in a collagen matrix. The circle symbols in the luminal fluid illustrate key ions (Na$^{+}$, hot pink; Cl$^{-}$, mint) and molecules (H$_{2}$O, lavender). The ellipsoidal symbols in the cell polarity description represent key apical (Crumbs) and lateral proteins (Scribble).}
    \label{fig2}
\end{figure*}

\subsection{Molecular machinery}\label{s21}

At the smallest spatial scale, lumenogenesis depends on a set of 
protein complexes that establish polarity, regulate permeability, and organize transport across the epithelial layer. These proteins act together to define which membrane domain will face the lumen, control transport across the intercellular barrier, and provide the molecular basis for  mechanical and hydraulic processes. Thus, the molecular machinery of lumenogenesis does not merely act on an already formed cavity, but helps specify where and how the cavity can emerge.

A first class of molecules is formed by tight-junction proteins (Fig.~\ref{fig2}), including claudins and occludins, which act as selectively permeable sealants between neighboring epithelial cells \cite{furuse1998claudin,tsukita2001multifunctional,anderson2009physiology,shen2011tight,krug2014tight}. By regulating paracellular permeability, these proteins allow lumens to retain fluid and build hydrostatic pressure \cite{mukenhirn2024tight}. Tight junctions are further organized by scaffold proteins such as Zonula Occluden proteins (ZO-1 and ZO-2). Importantly, these structures are not static. Beutel \textit{et al}.~\cite{beutel2019phase} showed that ZO-1 and ZO-2 can form phase-separated assemblies at tight junctions, and that both proteins display fast and slow diffusive components in the cytoplasm, consistent with a mixture of monomeric and larger associated states. 
Complementary Fluorescence Recovery After Photobleaching (FRAP) \cite{axelrod1976mobility} measurements revealed substantial fluorescence recovery on the timescale of minutes, indicating active molecular turnover. Together, these observations support the view that tight junctions are dynamically regulated sealants rather than rigid barriers, a property that is central for osmotic sealing (Section~\ref{pumping}) and the regulation of cortical tension in Madin-Darby Canine Kidney (MDCK) cysts (Section~\ref{sculpting}).

The polarity machinery is organized around the Par, Crumbs, and Scribble complexes and establishes the apicobasal axis of epithelial cells 
\cite{martin2008regulation,rodriguez2005organization}. This polarity is reinforced by signals from the ECM. In MDCK cysts, for example, integrin-mediated adhesion to the ECM helps orient the apical domain toward the future lumen 
\cite{myllymaki2011two,development_campanale_2017}. In particular, inhibition of integrin $\beta_1$ disrupts polarity reorientation, leaving podocalyxin at the ECM-facing surface and preventing its relocalization, together with other apical factors such as ezrin and NHERF1, to the center of the cyst \cite{bryant2014molecular}. Downstream, this reorganization proceeds through the FAK/p190RhoGAP pathway, which suppresses RhoA-ROCK-ezrin activity and facilitates release of the podocalyxin complex from the outer membrane toward the apical membrane initiation site \cite{bryant2010molecular}. Thus, polarity proteins and ECM-sensing receptors work together to define the membrane domain from which the lumen will emerge.

These molecular systems also prepare the epithelial layer for the transport processes required for lumen growth. Once apicobasal polarity is established, ion pumps, ion channels, and aquaporins can be asymmetrically positioned across the membrane, enabling directional ion and water fluxes \cite{navis2015developing,torres2021tissue}. In particular, the spatial localization of Na$^+$/K$^+$-ATPase pumps and aquaporins is essential for generating the osmotic gradients that drive lumen expansion and regulate lumen size 
\cite{datta2011molecular,bryant2010molecular} (see Section~\ref{pumping}). More broadly, the molecular machinery discussed here provides the foundation on which later stages of lumenogenesis depend: tight junctions regulate sealing, polarity complexes define membrane identity, ECM-linked receptors orient epithelial polarity, and transport proteins convert chemical energy into osmotic work. Together, these components establish the molecular conditions under which a stable lumen can be created and maintained.

\subsection{Vesicular transport}\label{s22}
One of the processes actively shaping the apical surface lining the lumen is vesicular trafficking mediated by Rab proteins \cite{bryant2010molecular} (Fig.~\ref{fig2}). Membrane-bound vesicles, with diameters within the hundreds of nanometers, bud from intracellular compartments and are directed toward the apical membrane \cite{bryant2010molecular,martin2008cell}. These vesicles carry the essential cargo for lumenogenesis: membrane lipids to expand the apical surface, the ion pumps and water channels mentioned above, and proteins to maintain the apicobasal polarity. Interestingly, vesicles can also trigger lumen formation from within the cell by coalescing in the cytoplasm into a pre-apical compartment, which then expands into a cavity \cite{sigurbjornsdottir2014molecular}.

\subsection{Cytoskeleton}\label{s23}
A main driver of spatial structure in biological systems is the cytoskeleton, consisting of actin filaments, microtubules, and intermediate filaments. Not suprisingly, it also plays an important role in lumenogenesis (Fig.~\ref{fig2}). 
During vesicle trafficking, an actin meshwork localized at the nascent apical domain---reported in MDCK cysts and tracheal cells \cite{meder2005gp135,gervais2010vivo}---guides the delivery of Rab proteins to the lumen initiation site \cite{sigurbjornsdottir2014molecular}. In excretory cells---responsible of removing metabolic waste in organisms including C. elegans---after lumen initiation, detachment of the actin network from the apical surface results in cavity enlargement \cite{mckeown1998sma,buechner1999cystic}.
On the basal side, polymerized actin binds to the cytoplasmic tails of integrin receptors establishing cell-ECM communication \cite{mechanisms_legate_2009}, and  supports  the epithelial layer that encloses the lumen \cite{drosophila_levi_2006}.
Microtubules have also been implicated in vesicle trafficking in MDCK cells \cite{kif3a_boehlke_2013} (Fig.~\ref{fig2}), and in subcellular lumen growth of tracheal terminal cells \cite{gervais2010vivo}.

Later in this review, the cytoskeleton is addressed beyond its trafficking role. We discuss how actin polymerization can act as a lumen growth mechanism in epiblasts (Section~\ref{pumping}), and how actomyosin cortical tension balances luminal pressure (Section~\ref{sculpting}).

\subsection{Cellular architecture}\label{s24}
At the micrometer scale, a sheet of epithelial cells constitutes the material that encloses the lumen (Fig.~\ref{fig2}). Each cell must produce and maintain apicobasal polarity, dividing the cell membrane into apical, basal and lateral compartments \cite{rodriguez2005organization}. The apical surface, besides being enriched with specific lipids and proteins, is often covered with microvilli or cillia that contribute to the mechanosensation of fluid flow \cite{nauli2003polycystins}. The lateral surfaces form adhesive contacts with adjacent cells through cadherin-mediated adherens junctions and tight junctions (Fig.~\ref{fig2}). Both adherens and tight junctions regulate the apical actin cortex through force redistribution \cite{zonula_fanning_2012,control_bazellires_2015,remodeling_choi_2016}, and establish a cohesive epithelial sheet \cite{takeichi1991cadherin,harris2014formation}. Importantly, during lumenogenesis, this sheet operates far from equilibrium, with coordinated actomyosin contractions in the cortex of each cell \cite{salbreux2012actin}. These contractions generate membrane tension that can trigger cell rearrangements, influence lumen geometry, and balance pressure from the luminal fluid \cite{casares2015hydraulic}. Last, but not least, the basal surface anchors the epithelium to the ECM via integrin adhesions \cite{yu2005beta1,buckley2022apical}.

The apical, basal, and lateral organization of epithelial cells, together with the capacity of the epithelial layer to contract and rearrange, are key ingredients driving several of the theoretical and computational models reviewed in Section~\ref{theocomp}.

\subsection{Luminal fluid}\label{hub}
The common picture of the lumen is not a simply hollow cavity, but a dynamical fluid-filled compartment \cite{dasgupta2018physics}. The fluid accumulates via ion and water transport (Fig.~\ref{fig2}): cells pump ions into the lumen (particularly Na$^{+}$ and Cl$^{-}$), introducing an osmotic gradient that draws water through both transcellular (aquaporins) and paracellular (tight junctions) pathways. The luminal fluid is not simply water, it contains secreted macromolecules such as growth factors, morphogens, and ECM components \cite{cohen2020multi}. These biochemical ingredients can regulate the fluid viscosity and serve as diffusible biochemical signals \cite{durdu2014luminal}. Measurements of luminal fluid viscosity in MDCK-II cysts have reported values $5\times$ higher than that of water \cite{mukenhirn2024tight}. The rich composition of the luminal fluid enables the lumen to act as both an active mechanical actuator and a biochemical signaling hub \cite{durdu2014luminal}. This creates long-range coupling between distant epithelial cells, creating tissue-scale coordination without the need of direct cell-cell contacts. 
The dual role of the luminal fluid is reviewed in Sections~\ref{mechanoprocess} and~\ref{feedbackModels}, where we discuss lumen-centered mechanochemical feedbacks.

\subsection{Structural scaffold}\label{sECM}
At the largest scale (from hundreds of microns to millimeters), the ECM---a meshwork of proteins, such as collagen (Fig.~\ref{fig2}), laminin and fibronectin, along with other macromolecules---provides mechanical support and biochemical cues to the epithelial layer containing the lumen \cite{frantz2010extracellular,rodriguez2012cell}. In \textit{in vivo} contexts, the ECM attaches epithelial cells via integrin receptors (Fig.~\ref{fig2}), allowing the maintenance of tissue architecture \cite{yurchenco2011basement}. The mechanical properties of the ECM affect lumen morphogenesis by shaping a dynamic boundary that acts against tissue expansion, and by modulating cellular cortical tension \cite{hofer2021engineering}. In \textit{in vitro} contexts, cells are usually embedded in matrigel \cite{kleinman2005matrigel}, collagen \cite{jabaji2014type}, or designed hydrogels that emulate the surrounding ECM, enabling experimental control of biochemical and mechanical cues \cite{gjorevski2016designer}.

In Section~\ref{sculpting}, we outline how the ECM constrains lumen inflation in MDCK cysts and helps 
sculpt Drosophila tracheal cavities. In Section~\ref{matrices}, we describe recent experimental advances in engineered matrices for controlling lumenogenesis.

\section{Creating, shaping, and  maintaining the active balloon}\label{sec4}

Having reviewed the biological constituents of lumenogenesis across scales, we now turn to the physical processes by which these ingredients generate, expand, and stabilize a cavity. The central perspective of this section is that a lumen is not a passive cavity, but an active balloon: a compartment whose emergence and maintenance require the coordinated action of polarity cues, intracellular trafficking, osmotic transport, sealing regulation, tissue mechanics, and extracellular constraints \cite{chan2020integration}. Accordingly, lumenogenesis is not governed by a single mechanism, but by a sequence of tightly coupled events in which luminal space must first be created, then inflated, and finally maintained against mechanical and biochemical perturbations \cite{koser2016mechanosensing,rouaud2020scaffolding,lee2014integrins}. This physical picture emphasizes that lumen formation and stabilization arise from reciprocal interactions between molecular machinery, cell-level forces, and tissue-scale constraints.

We begin by discussing how luminal space is initiated. This includes the distinct morphogenetic routes by which cells generate a cavity, such as cord hollowing, cell hollowing, and cavitation, as well as situations in which multiple nascent lumens appear and later coalesce into a single cavity (Section~\ref{routes}). We then examine the inflationary engine that drives lumen growth, focusing on ion pumping, osmotic pressure differences, water permeation, and leakage
(Section~\ref{pumping}). A key message is that lumen expansion is fundamentally an out-of-equilibrium process: active ionic pumping sustains chemical gradients, passive water flow converts them into mechanical work, and the degree of sealing of the epithelial layer determines whether lumens grow steadily, remain small, or display more complex dynamics.

The second half of the section addresses how the expanding cavity is sculpted and stabilized by its material surroundings. We first review the mechanical response of the epithelial layer and, when relevant, the ECM, emphasizing why simple Laplace-law intuition is often insufficient to describe real lumens (Section~\ref{sculpting}). We then broaden the discussion to mechanochemical feedbacks, where the lumen acts not only as a pressurized compartment but also as a regulator of the surrounding tissue, influencing contractility, junction maturation, gene expression, and cell fate
(Section~\ref{mechanoprocess}). Taken together, this section establish the central idea that lumenogenesis is a reciprocal process: cells build the cavity, but the cavity, once formed, feeds back on the cells that contain it.

\subsection{Morphogenetic routes to lumen nucleation}\label{routes}
\begin{figure*}
    \centering
    \includegraphics[width=0.8\linewidth]{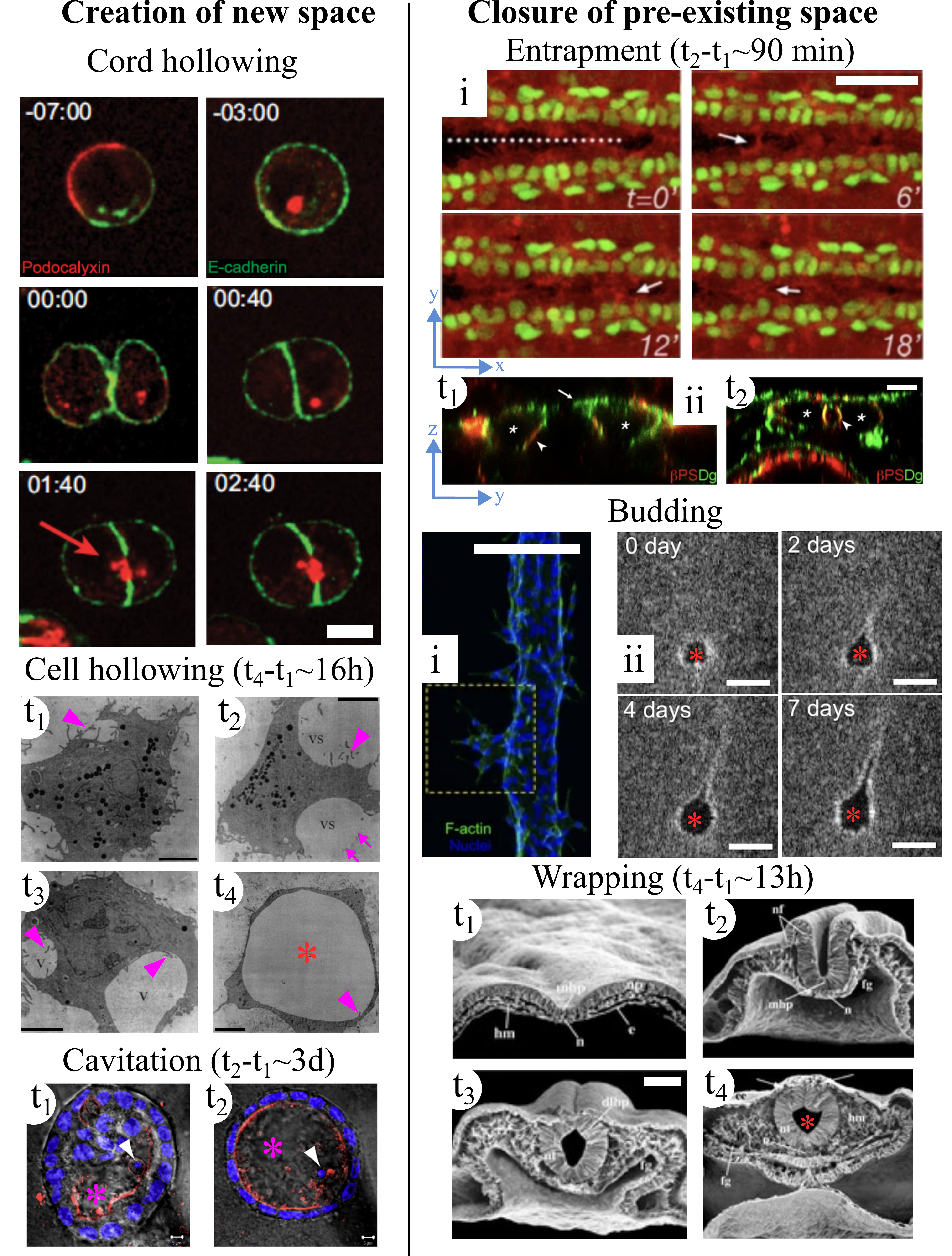}
\caption{Examples of lumen creation in different systems. Left column (creation of new space). Top panel: Cord hollowing after the first cell division (occurring at $00:00$) of MDCK cells within matrigel. The accumulation of apical proteins at the center of the cell-cell contact (AMIS; red arrow) illustrates the lumen birth (Scale bar, $10\ \mu$m). Adapted from \cite{lu2025generic}. Middle panel: Cell hollowing inside an endothelial cell within a collagen matrix (scale bar, $5\ \mu$m). The pink arrowheads point to small finger-like membrane protrusions extending to create the vacuole space (vs; pink arrows). After vacuoles are internalized (v), they coalesce creating the lumen (red asterisk). Adapted from \cite{davis2003integrin}. Bottom panel: Apoptotic cavitation in an aggregate of MDCK cells within collagen (scale bar, $5\ \mu$m). A cavity (pink asterisk) develops on a span of $3$ days central clearance (arrowheads indicate apoptotic cells), and the surrounding cellular layer establishes apicobasal polarity. Adapted from \cite{martin2008cell}. Right column (closure of pre-existing space). Top pannel: Entrapment process during Drosophila development. Two cardioblast cords create a cavity by generating adheren junctions only at their dorsal-most and ventral-most sides. (i) Dorsal view (dotted line: dorsal midline; arrows: contact processes between cells; scale bar, $25\ \mu$m). (ii) Cross view (time scale inferred from \cite{medioni2008genetic}; scale bar, $5\ \mu$m) of the endorsement process, illustrating dorsal adhesion (white arrow) between two cells (asterisks mark the nuclei) and the subsequent lumen establishment (arrowhead). Adapted from \cite{vanderploeg2012integrins}, where the reader is referred for the color codes of the fluorescent images. Middle panel: Budding during the process of angiogenesis; creation of new blood vessels from pre-existing ones. (i) A localized region of an endonthelial tube, enclosing a lumen, buds outward generating a larger and continuous luminal structure (yellow square; scale bar, $200\ \mu$m). (ii) A cross section of the luminal tube (red asterisk) showing the temporal growth of the bud. Adapted from \cite{takahashi2017visualizing}. Bottom panel: Wrapping during neural tube formation in the chick embryo (scale bar inferred from  \cite{hamburger1951series}, $100\mu$m). A flat epithelial layer undergoes spatially patterned bending, bringing opposite edges together, which then fuse to form a closed tube (red asterisk). Adapted from \cite{colas2001towards}.}
    \label{fig3}
\end{figure*}

Lumens are found within the interiors of tubular organs, spherical cysts, glandular acini, and even single cells. Classical embryological studies identified morphologically distinct routes to lumenogenesis \cite{pollack1998morphogenetic,colas2001towards,affolter2003tube,meyer2004spatiotemporal,martin2008cell,tung2012tips,lammert2012vascular,baumholtz2017claudins,nelson2017microfluidic,jewett2018insane}. In the specific context of tube formation, Lubarsky and Krasnow grouped these routes into five categories: cavitation, cord hollowing, cell hollowing, wrapping, and budding \cite{lubarsky2003tube} (Fig.~\ref{fig3}). These categories can be reduced into two underlying physical processes: the creation of new space---\textit{de novo} lumenogenesis (hollowing and cavitation mechanisms)---and the closure of pre-existing space (wrapping and budding). Other authors suggest a sixth category; entrapment or closure \cite{medioni2008genetic,santiago2008repulsion,vanderploeg2012integrins,liang2022cadherin,romero2025dynamics}, which corresponds to a different mode of closing pre-existing space mediated by adhesive mechanisms (Fig.~\ref{fig3}). Here, we focus on \textit{de novo} lumenogenesis, as this process extends beyond tubular organs \cite{sigurbjornsdottir2014molecular,blasky2015polarized,lu2025generic}.

In cord hollowing, an aggregate of two or more cells creates a lumen at their mutual interfaces. Before the emergence of the lumen, cells must undergo an apicobasal symmetry breaking, defining an apical domain: the membrane surface that will face towards the cavity. The Par-aPKC polarity complex, working in mutual antagonism with basolateral determinants such as Lgl and Scribble, provides key biochemical cues that define molecularly distinct membrane domains \cite{st2010cell,rodriguez2014organization}. The particular region of the apical domain where lumen birth takes place is called the apical membrane initiation site (AMIS) \cite{ferrari2008rock,bryant2010molecular}. In MDCK cells, for example, the AMIS appears at the first cell-cell interface after cell division (Fig.~\ref{fig3}), marked by recruitment of tight junction proteins and the exocyst complex. Then, a one-way intracellular transport traffics apical proteins, including podocalyxin and Crumbs, to the site specified by the cell-cell contact. Finally, the cytoskeleton sculpts the newly delivered material into the correct luminal shape \cite{sigurbjornsdottir2014molecular}. Cord hollowing also operates in other systems including the C. elegans gut \cite{leung1999organogenesis}, the Drosophila heart \cite{rugendorff1994embryonic}, and epiblasts \cite{lu2025generic}. In the latter, the hollowing process is triggered only after a minimum of $\sim10$ cells, arranged in a rosette formation, is reached.

In cell hollowing, single cells produce intracellular vacuoles that coalesce, form an intracellular lumen \cite{davis2003integrin,kuvcera2009ancestral} (Fig.~\ref{fig3}), and eventually fuse with the plasma membrane to create a transcellular lumen upon contact with neighboring cells \cite{sigurbjornsdottir2014molecular}. For example, terminal cells of the Drosophila tracheal system extend long cytoplasmic projections and then organize a membrane-bound lumen within each projection through hollowing \cite{samakovlis1996development,shaping_pradhan_2022}. Endothelial cells forming fine capillaries employ a similar strategy \cite{folkman1980angiogenesis,davis1996alpha2beta1}.  

In cavitation, after many rounds of cell division, the interior cells within a multicellular aggregate are removed via caspase-dependent (Fig.~\ref{fig3}) and caspase-independent mechanisms, leaving a hollow shape surrounded by epithelial cell that eventually establish an apicobasal polarity. When the removal is triggered via apoptosis (caspase-dependent) \cite{martin2008cell,qi2012bnip3}, cavitation underlies the formation of  proamniotic cavities in mouse embryos \cite{coucouvanis1995signals} and mammary acini \cite{debnath2002role}. However, central cell death can also be caspase-independent, as observed in mice mammary gland lumenogenesis upon disruption of the proapoptotic factor BIM, where central cells display necrotic features \cite{mailleux2007bim}. More recently, studies in eccrine gland organoids have shown that an autophagy-mediated clearance of interior cells can likewise trigger lumen formation \cite{du2022autophagy}.

Regardless of the specific mechanism of \textit{de novo} lumen formation, creating luminal space requires physical separation of cellular material. In the case of multicellular aggregates, three possibilities have been proposed to shed light on the forces responsible of separating the apical membranes. First, the surface of these membranes may be non-adhesive: the apical surface is depleted of cadherins and enriched in negatively charged macromolecules that generate electrostatic repulsion \cite{Strilic2010Electrostatic}, stabilizing lumen formation. Second, the polymerization of actin in the apical side of epithelial cells can mechanically open the nascent cavity \cite{indana2024lumen}. Lastly, is the concept of hydraulic fracturing presented by Dumortier and colleagues \cite{dumortier2019hydraulic}. They found that in the blastocyst, microlumens form when pressurized fluid mechanically fractures cell-cell adhesions. The intercellular fluid pressure, measured at approximately $300$ Pa, is sufficient to displace E-cadherin from cell-cell contacts, promoting the cavity formation.

\begin{figure*}
    \centering
    \includegraphics[width=1\linewidth]{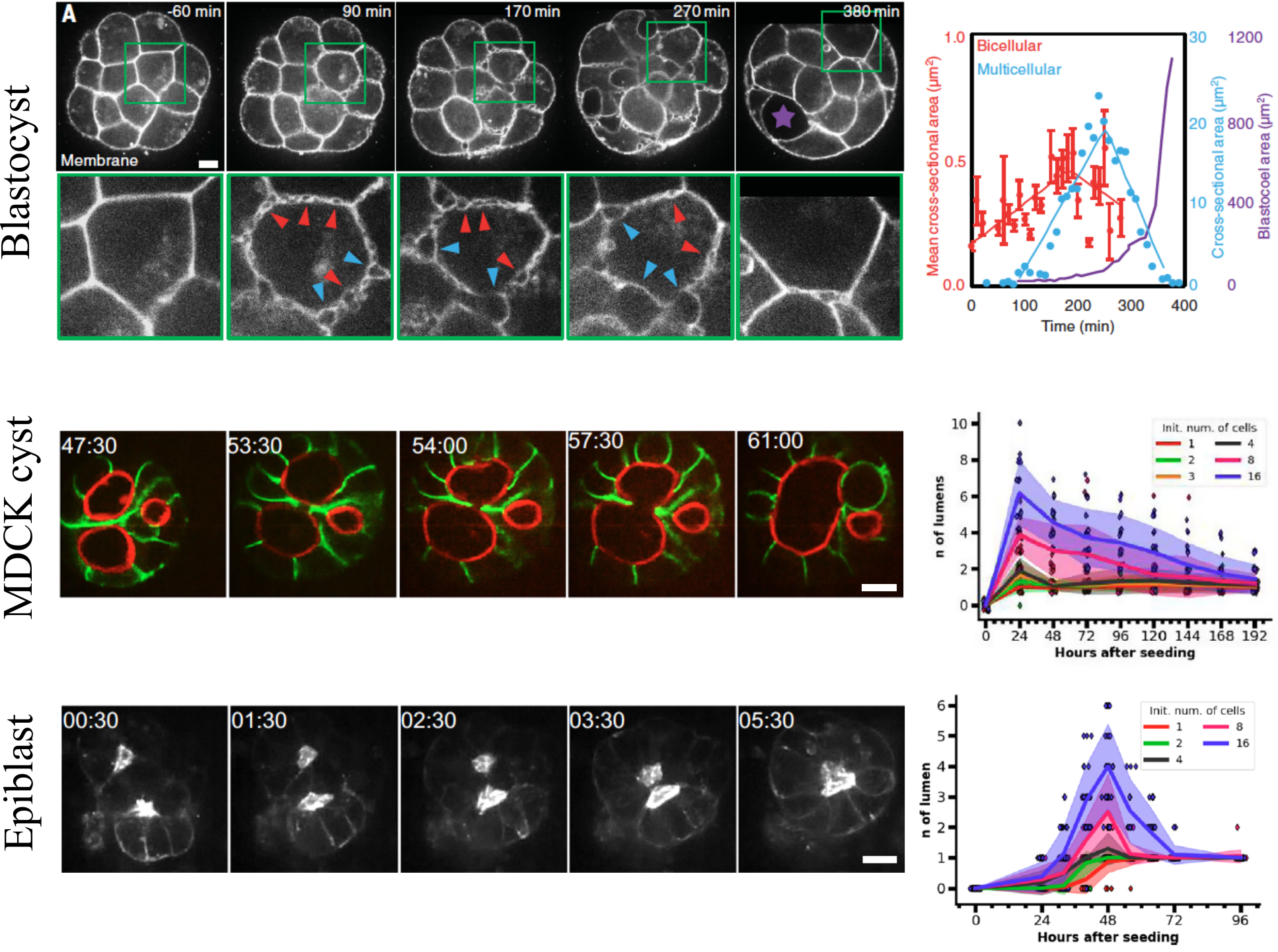}
    \caption{Coarsening-like dynamics select a single lumen in different biological systems. Top panel: Blastocoel formation in the blastocyst (scale bar, $10\ \mu$m). The insets are $3$x magnifications of the green squares. As shown in the right-most panel, swelling of the microlumens is first observed at bicellular contacts (red arrows), and $100$ minutes later, lumens at multicellular junctions (blue arrows) start to grow. In both cases, the coarsening dynamics begins $\sim200$ minutes after the onset of swelling, with lumens discharging their content into the blastocoel (indicated by the purple symbol). Adapted from \cite{dumortier2019hydraulic}. Middle panel: Lumen coalescence in MDCK cysts (E-cadherin in green and Podocalyxin in red. Scale bar, $10\ \mu$m).  Bottom panel: Lumen coalescence in epiblasts (the polymerized actin is imaged. Scale bar, $10\ \mu$m). For both the MDCK cysts and epiblast the temporal dynamics of the process depends on the initial number of cells (right-most panels). Adapted from \cite{lu2025generic}.}
    \label{fig4}
\end{figure*}

Notably, Dumortier \textit{et al.}~\cite{dumortier2019hydraulic} also showed that the fracture-based mechanism does not nucleate a single cavity. Instead, multiple small lumens form throughout the tissue and subsequently resolve into a single dominant cavity, a feature conserved among many systems \cite{bebelman2024hepatocyte,lu2025generic} (Fig.~\ref{fig4}). 
In the  blastocyst, blastocoel---the fluid filled cavity---formation, begins with the near-simultaneous emergence of hundreds of microlumens throughout the embryo \cite{dumortier2019hydraulic,le2021hydro}, either at bicellular or multicellular (more than two cells) contacts between trophectoderm (TE) cells and inner mass cells (ICM). After growing for $200$--$300$ minutes, pressure-driven fluid redistribution takes over, leading to a coarsening-like process in which most lumens shrink, while one continues to expand and ultimately becomes the blastocoel (Fig.~\ref{fig4}). The authors described the process as analogous to Ostwald ripening \cite{ostwald1901blocking}: the coarsening phenomenon that drives the foam and emulsion dynamics via pressure differences \cite{cahn1958free,lifshitz1961kinetics,wagner1961theory}. In the blastocyst, the lumens are connected across the intercellular space, allowing pressure-driven fluid flow between them. Importantly, this process is not purely passive: the embryo exhibits active control through differential cell contractility, biasing fluid toward a specific location (Fig.~\ref{fig4}). Similar phenomenology, coalescence of multiple lumens, has been observed during formation of the Kupffer's vesicle \cite{origin_oteza_2008}, and in the zebrafish gut lumenogenesis \cite{bagnat2007genetic,alvers2014single}. More recently, coarsening-like dynamics were shown to drive lumen formation across distinct organoid systems, including MDCK cysts, epiblasts and pancreatic spheroids \cite{lu2025generic}. Specifically, using microwell cavities that allow multiple cells to be seeded simultaneously, the authors induced conditions in which multiple microlumens could nucleate. Independently of the organoid system, when the initial condition consisted in $\gtrsim 4$ cells, a coarsening-like process followed microlumen nucleation. This coalescence was slower in MDCK cysts compared to epiblasts: $168$ h vs $24$ h when the initial number of cells was $16$ (Fig.~\ref{fig4}). This distinct time scale could be a consequence of the different physical mechanisms underlying lumen coalescence. In MDCK cysts, it is driven by hydrostatic pressure (similar to blastocoel formation), whereas in epiblasts, cell migration---tied to shape deformation of the whole cell aggregate---drives the fusion of small lumens into a large one. 

\subsection{The inflationary engine: ion pumping and water influx}\label{pumping}
Once luminal spaces have been created, they must grow to physiological sizes, expanding from micrometer to tens or even hundreds of micrometers, while maintaining their structural integrity. The standard picture of lumen creation and expansion is that of a chemically regulated hydraulic process. Cells convert chemical free energy into osmotic work by transporting ions, enabling passive water flow into the lumen \cite{kedem1963permeability,de2013non}. This influx increases luminal pressure, which performs mechanical work in the surrounding epithelial container, leading to lumen growth.
This out-of-equilibrium process is a pervasive feature of epithelial morphogenesis \cite{navis2015developing}. Here, we discuss the relevant biophysical processes driving the osmotic flows, building on concepts synthetized by Torres-S\'anchez \textit{et al}.~\cite{torres2021tissue}.

\begin{figure}
    \centering
    \includegraphics[width=0.9\linewidth]{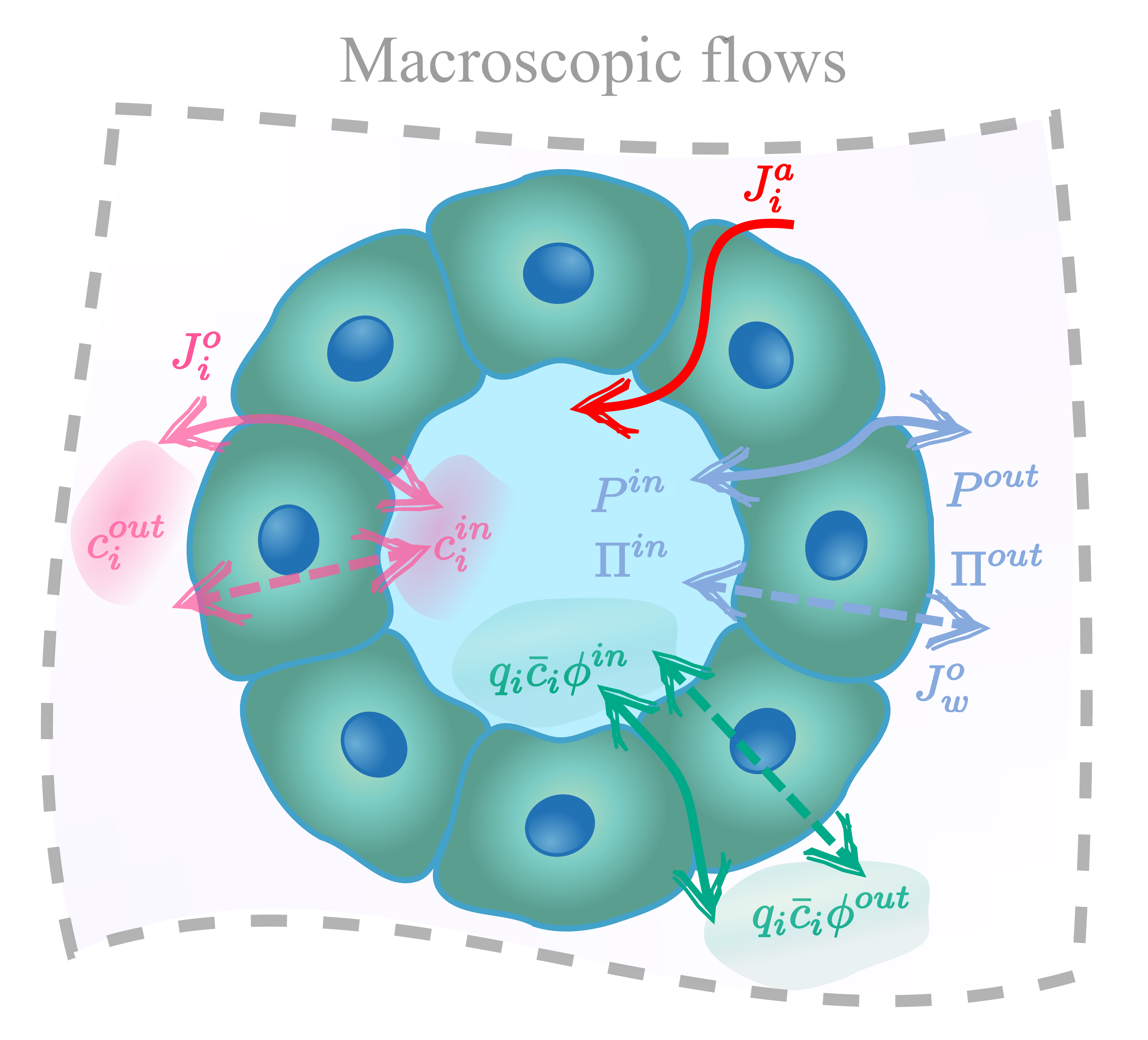}
    \caption{Schematic of the ion and water fluxes across a cross section of a spherical acinar structure. This is a coarse-grained (macroscopic) picture of the ion and water transport mechanisms illustrated in Fig.~\ref{fig2}. Solid lines represent intercellular fluxes, while dashed lines denote transcellular fluxes. Here, the hydrostatic pressure difference is given by $\Delta P = P^{in}-P^{out}$, while the osmotic pressure difference is $\Delta\Pi = \Pi^{in}-\Pi^{out}$. All passive, gradient-dependent fluxes are bidirectional.}
    \label{fig5}
\end{figure}

At the macroscopic level, the osmotic pressure difference between the luminal cavity and the extracellular space produced by $N$ ion types can be approximated by the Van~'t Hoff relation \cite{bagnat2022morphogenetic}, $\Delta\Pi=RT\sum_{i=1}^{N}\Delta c_i$, where $R$ is the ideal gas constant, and $\Delta c_i=c_{i}^{in}-c_{i}^{out}$, with $c_{i}^{in}$ ($c_{i}^{out}$) the ion concentration in the lumen (extracellular space). At physiological temperatures ($T\approx37^{\circ}$C), $RT\simeq2.6\times10^3\ \mathrm{J\ mol^{-1}}$, so a concentration difference of order $10^{-1}$ mM can yield a pressure difference of $\Delta\Pi\sim10^2$ Pa, a magnitude comparable to luminal pressures reported in both developing systems and engineered cysts \cite{dumortier2019hydraulic, mosaliganti2019size}. The imbalance in osmotic pressure triggers a passive water flux through the permeable epithelial layer into the lumen, resulting in its inflation, which in turn builds up hydrostatic pressure $\Delta P$ in the cavity with respect to the outside (Fig.~\ref{fig5}). In the dilute limit $c_{i}/c_{w}\ll1$ ($c_{w}$; concentration of water molecules), an intuitive description of the passive water flux density is $J_{w}^{o}\approx\mathcal{L}_w(\Delta\Pi-\Delta P)$, where $\mathcal{L}_w$ is an effective hydraulic permeability determined by the density of aquaporins on the cell membrane \cite{kedem1963permeability,dasgupta2018physics}. However, the out-of-equilibrium nature of this biological system requires additional correction terms to account for the irreversibility of the transport process:
\begin{equation}
J_{w}=\mathcal{L}_w(\Delta\Pi-\Delta P)-\sum_{i}^{N}\mathcal{\lambda}_{w,i}\!\left(\Delta c_{i}+\frac{q_{i}\bar{c}_i}{RT}\Delta\phi\right)+J_{w}^{a}.
\label{eqJW}
\end{equation}
The second term in Eq.~(\ref{eqJW}) characterizes the water flows driven by ion chemical potential gradients across the epithelial layer, where $\lambda_{w,i}$ are the cross-coupling (or Onsager) coefficients, $\bar{c}_{i}=(c_{i}^{in}+c_{i}^{out})/2$, $q_{i}$ are the ion charges, and $\Delta\phi=\phi^{in}-\phi^{out}$ is the difference in electrostatic potential. This difference can produce ionic fluxes, and it depends on the tightness of the epithelial layer, with $\Delta\phi\sim\pm50$ mV for tight and $\Delta\phi\sim\pm2$ mV for leaky epithelia \cite{SACKIN2013177}. The last flux term in Eq.~(\ref{eqJW}), $J_{w}^{a}$, accounts for possible active mechanisms used by epithelial cells to control water transport, which could be related to apicobasal cortical flows \cite{torres2021tissue}. Ionic fluxes $J_i$ can be modeled with a similar flux equation presented in Eq.~\ref{eqJW}, with a ionic permeability $\mathcal{I}_p$ controlling passive ionic fluxes  $J_{i}^{o}$, Onsager coefficients $\lambda_{i,w}$ coupling water and ion transport, and a highly relevant active flux term $J_{i}^{a}$ (see below and further details in the review of Torres-S\'anchez \textit{et al}.~\cite{torres2021tissue}).

The effective hydraulic permeability $\mathcal{L}_{p}$ takes into account transcellular routes, such as aquaporins and the permeable membrane, and intercellular pathways as controlled leaks through tight junctions \cite{baumholtz2017claudins,carleton2022human} (Fig.~\ref{fig5}). Experimental measurements indicate that water permeabilities in MDCK cells are within the range of $10^{-6}$--$10^{-7}$ ($\mu$m s$^{-1}$ Pa$^{-1}$) \cite{timbs1996hydraulic}, while specialized epithelial layers can be tens to hundreds of times more permeable \cite{fischbarg2010fluid}. This implies that similar osmotic forces could produce strikingly different lumen growth rates depending on the biological context. In the zebrafish otic vesicle---a fluid-filled cyst---water fluxes have been measured in the range $1$--$8$ $\mu$mh$^{-1}$, which, combined with permeability estimates and neglecting the osmo-ionic coupling in Eq.~\ref{eqJW}, corresponds to a driving force of $(\Delta\Pi-\Delta P)\sim3$--$200$ kPa \cite{mosaliganti2019size}. Hydrostatic pressure measurements inside inflating lumens reveal much smaller numbers: $100$--$300$ Pa in blastocysts, $100$ Pa in epithelial domes, and around $40$ Pa in some MDCK cysts \cite{mosaliganti2019size,dumortier2019hydraulic,narayanan2020osmotic}, suggesting that a robust lumen growth mechanism occurs in the regime $\Delta\Pi\gg\Delta P$ \cite{torres2021tissue}. 

In the blastocyst case, $\Delta P$ was quantified via micropipette aspiration, in which a sub-millimeter glass pipette applies a controlled negative pressure to induce deformation of the lumen \cite{dumortier2019hydraulic}. This is not the only way of measuring luminal pressures \cite{chan2020integration}. For example, traction force microscopy can be used to indirectly estimate $\Delta P$ \cite{style2014traction,chan2020integration}, through fluorescent beads incorporated in an elastic deformable substrate that balances lumen expansion, provided that this growth happens on timescales over which substrate deformations can be reliably measured \cite{latorre2018active}.

The ionic permeability can be estimated from epithelial electrical resistance measurements, because the ionic current of an ion $i$, $q_{i}J_{i}$, can be related to a resistance through the epithelial layer \cite{torres2021tissue}. Epithelial resistances span a wide range of values, from approximately $10\ \Omega\ cm^{2}$ for leaky epithelia to $10^{5}\ \Omega\ cm^{2}$ for tight ones \cite{SACKIN2013177}. This translates to $\mathcal{I}_{p}$ values ranging from $3\times10^{-3}$--$3$ $\mu$ms$^{-1}$, considering an ionic concentration of $100$ mM \cite{butt2023physics} and the electron charge ($q_{i}=e$), consistent with direct measurements of Na$^{+}$ permeability in MDCK monolayers ($\mathcal{I}_{p}\sim10^{-1}$ $\mu$m s$^{-1}$) \cite{fernandez1985ion}.

The permeability of the ions impose the timescale to reach a steady state in lumen size: if an ionic concentration difference is imposed across the epithelial layer, a passive ionic flux dissipates this gradient on the order of $\sim R_{l}/\mathcal{I}_{p}$, where $R_{l}$ is the lumen radius. For $R_{l}\sim10^{2}\ \mu$m, the time to reach steady state is thus between a few seconds and several hours, depending on $\mathcal{I}_{p}$ \cite{torres2021tissue}. In the mouse blastocyst, the timescale for blastocoel expansion is on the order of hours \cite{dumortier2019hydraulic,chan2019hydraulic}, while MDCK cysts reach their mature size after $1$--$3$ days \cite{vasquez2021physical}. Therefore, to sustain osmotic gradients and grow a lumen to their physiological size, the balloon requires  active pumping of ions to balance the ionic dissipation. 

Active ionic fluxes $J_{i}^{a}$ lie at the heart of the osmotic engine. Several molecular processes contribute to these fluxes in the context of lumen size control \cite{navis2015developing,frizzell2012physiology}, including the Na$^{+}$/K$^{+}$-ATPase pump (Fig.~\ref{fig5}). By hydrolyzing ATP, this pump exports three Na$^{+}$ ions while importing two K$^{+}$ ions per cycle, thereby maintaining the transmembrane ionic gradients that drive osmotic water influx \cite{reuss2008mechanisms}. It operates at rates of approximately $10^{2}$ s$^{-1}$, and, given typical membrane densities of $10^{3}$--$10^{4}$ pumps per $\mu$m$^{2}$, generates an active flux density of order $J_{i}^{a}\sim10^{5}$--$10^{6}$ $\mu$m$^{2}$s$^{-1}$, consistent with pumping rates measured in MDCK monolayers \cite{ewart1995hormonal,cereijido1981occluding,simmons1981ion}. The central role of this pump is underscored by ouabain-inhibition experiments, which block lumenogenesis across diverse systems \cite{bagnat2007genetic}. Additionally, the Na$^{+}$ gradient established by the pump drives the accumulation of additional ions, including Cl$^{-}$, which further contributes to setting the target size of the active balloon \cite{bagnat2007genetic,li2004relationship}.

The osmotic machinery, however, extends beyond Na$^{+}$/K$^{+}$-ATPase alone. Additional transporters, including the Na-K-Cl cotransporter NKCC1 and the Na$^{+}$/H$^{+}$ exchanger NHE1, reinforce the ionic gradients required for lumen growth \cite{wu2025regulation}. Consistent with this role, inhibition of NKCC1 with bumetanide or of NHE1 with ethyl isopropyl amiloride reduces lumen expansion, as does perturbation of aquaporin-mediated water transport by copper sulfate \cite{wu2025regulation}. Osmotic gradients can also be modulated more directly. For example, forskolin activates cystic fibrosis transmembrane conductance regulator channels, promoting secretion of Cl$^{-}$ and HCO$_3^{-}$ into the lumen \cite{dekkers2013functional}. More recently, Shim \textit{et al}.~\cite{shim2024bioelectric} showed that electrical stimulation ($5$--$10$ V\,cm$^{-1}$) of MDCK cysts induces rapid lumen inflation through ion accumulation and the ensuing water influx. Together, these results emphasize that lumen growth is controlled by a broader transport network in which ionic pumping, secondary transport, and water permeability act in concert.

\begin{figure*}
 \centering
\begin{tikzpicture}
  \node[inner sep=0] (img)
    {\includegraphics[width=0.8\linewidth]{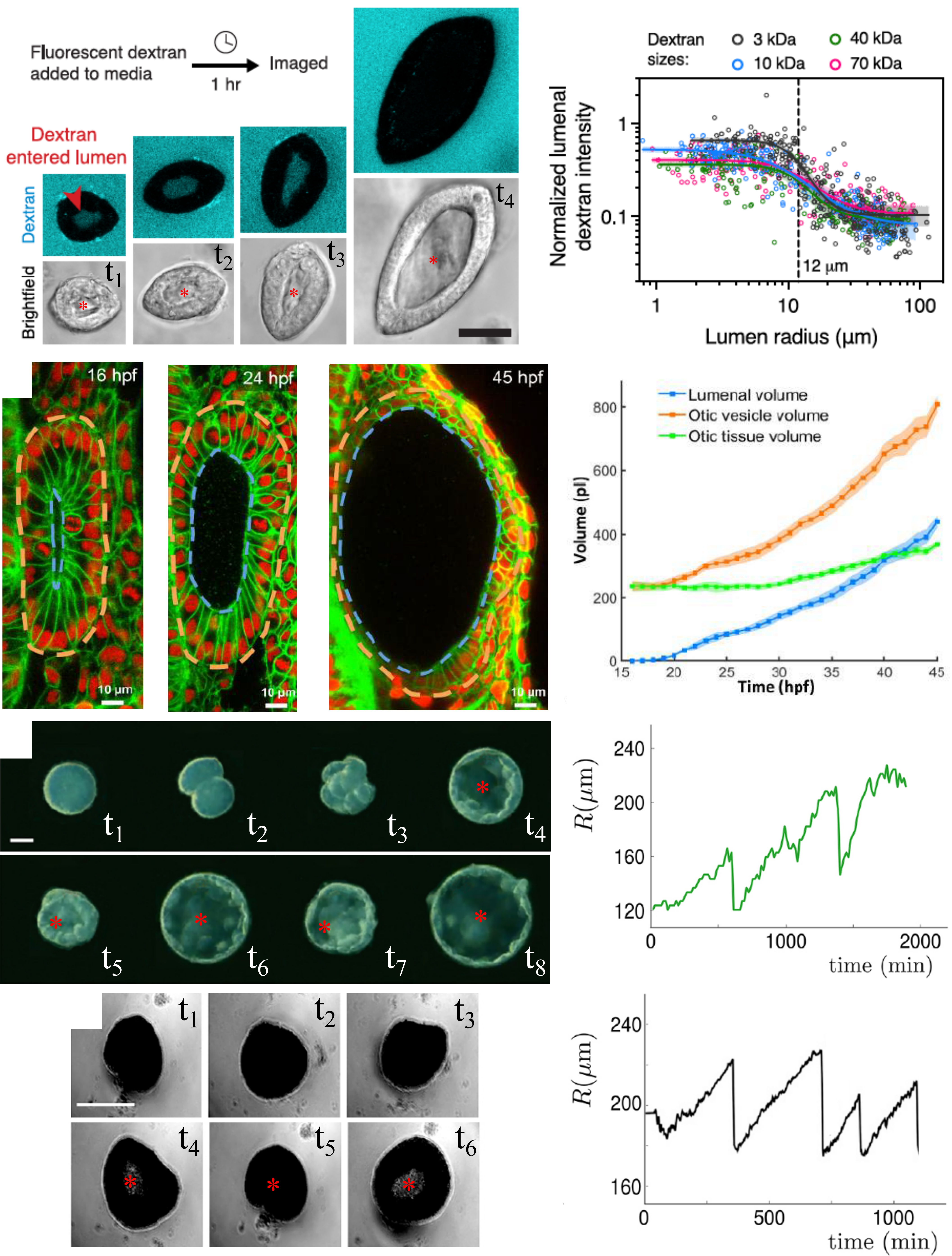}};
  \node at (img.north west) [xshift=2mm, yshift=-3mm]{\large\bfseries A};
    \node at (img.north west) [xshift=85mm, yshift=-3mm]{\large\bfseries B};
    \node at (img.north west) [xshift=2mm, yshift=-54mm]{\large\bfseries C};
     \node at (img.north west) [xshift=85mm, yshift=-54mm]{\large\bfseries D};
      \node at (img.north west) [xshift=2mm, yshift=-106.5mm]{\large\bfseries E};
    \node at (img.north west) [xshift=85mm, yshift=-106.5mm]{\large\bfseries F};
        \node at (img.north west) [xshift=12mm, yshift=-145.5mm]{\large\bfseries G};
            \node at (img.north west) [xshift=85mm, yshift=-145.5mm]{\large\bfseries H};
\end{tikzpicture}
    \caption{Single cyst dynamics. (A) Temporal snapshots of the cavity (red asterisk) inflation in a epiblast of human induced pluripotent stem cells (hiPSCs) over a period of $t_{1}-t_{4}\sim4$ days (scale bar, $50\ \mu$m). (B) Quantification of Dextran, added in the whole culture media, intensity inside the lumen as a function of the lumen radius for different Dextran weights (sizes). Higher Dextran intensities is a readout of leaky lumens. We note that fluorescently labelled Dextran can also be microinjected directly into the lumen, which can be useful for determining the leakage kinetics by measuring the change in fluorescence over time \cite{hill2017real}, providing experimental access to the hydraulic permeability $\mathcal{L}_{w}$. Adapted from \cite{indana2024lumen}. Temporal snapshots (C) of the steady growth (D) of the luminal space in the otic vesicle after its birth through cavitation. hpf: Hours post-fertilization. Adapted from \cite{mosaliganti2019size}. Oscillatory growth of a human blastocyst cavity (red asterisk)  illustrated by temporal snapshots (E) and by the radius $R$ as a function of time (F). 
    (scale bar, $100\ \mu$m). Panels G and H show similar oscillatory growth but now in Hydra vulgaris (scale bar, $200\ \mu$m). In both oscillatory cases (adapted from \cite{ruiz2017organ}), lumens are  created via cord hollowing.}
    \label{fig6}
\end{figure*}
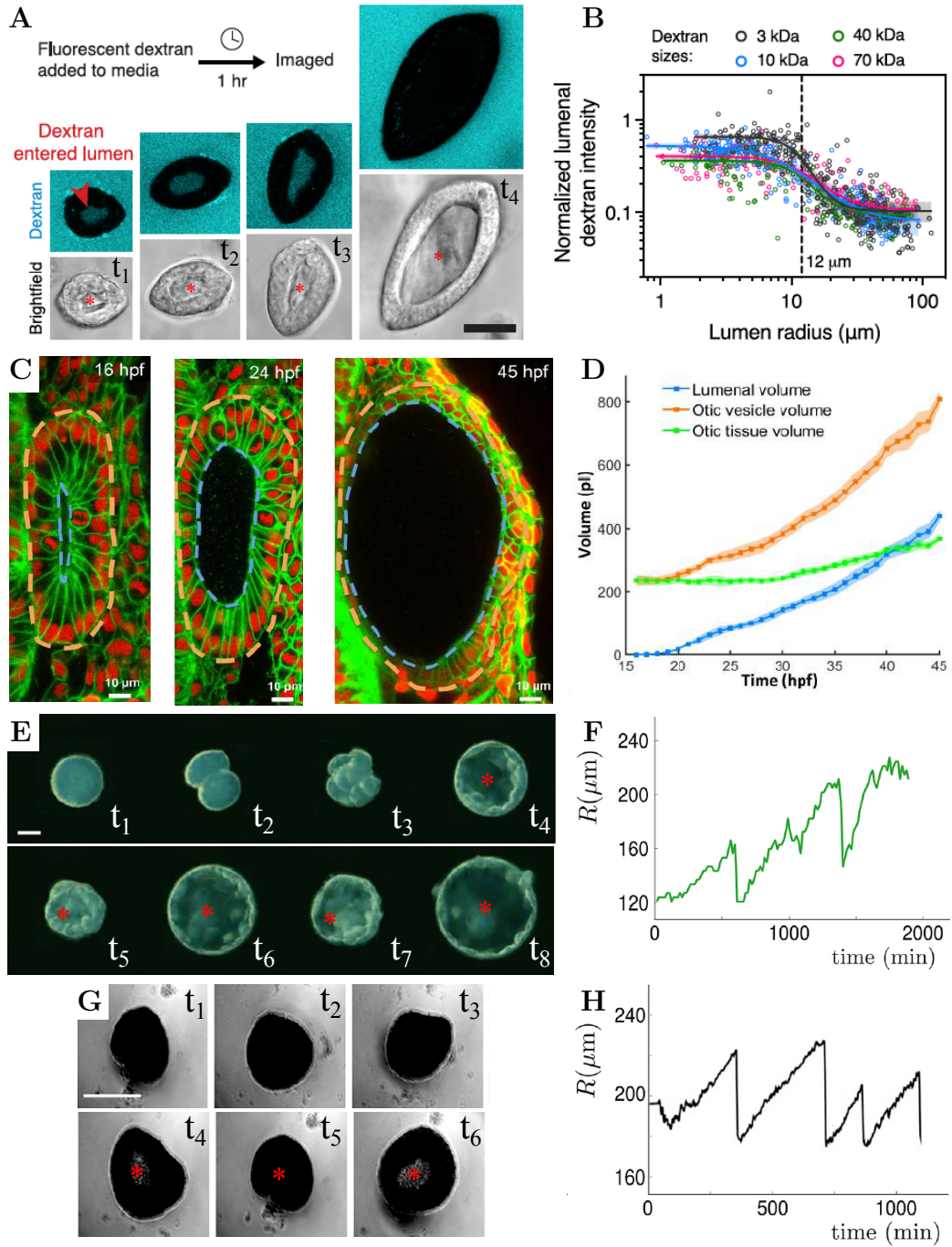

The hydraulic processes described above require sufficiently mature tight junctions to build the initial osmotic pressure. This condition is generally met in late lumenogenesis, but in early stages tight junctions can be highly permeable. Recently, Indana \textit{et al}.~\cite{indana2024lumen} demonstrated that lumens with a radius smaller than $\sim12\ \mu$m in epiblasts of hiPSC cells lack mature tight junctions. The authors showed that fluorescent dextran---a macromolecule used as a tracer for intercellular permeability---enters the small cavities through intercellular spaces (Figs.~\ref{fig6}A and~\ref{fig6}B).  In this leaky regime, FRAP measurements reveal that any ions pumped into the luminal space are lost via intercellular gaps \cite{indana2024lumen}, preventing the establishment of $\Delta\Pi$, and thus of $\Delta P$. Instead, lumen expansion is governed by an active mechanism: apical actin polymerization, in which the growth of a F-actin mesh at the apical surface drives outward inflation of the cavity. Once lumens reach a radius of $\sim12\ \mu$m, tight junctions mature and seal (fluorescent dextran is excluded from the cavity), and the osmotic engine takes over as the dominant growth mode. Consistently, inhibition of actin polymerization suppresses lumenogenesis in small epiblasts but has no effect on larger ones \cite{indana2024lumen}.

The leakage due to immature tight junctions during the early lumenogenesis in hiPSC epiblasts is intrinsic to the developmental stage. However, in systems with mature tight junctions, biologically-regulated leakage dynamics may modulate lumen inflation, contributing to lumen size selection and growth dynamics \cite{dasgupta2018physics}. In fact, this hydraulic effect can lead to inflation of the balloon at a non-uniform rate, in contrast to the approximately steady growth observed, for example, in the otic vesicle (Figs.~\ref{fig6}C and~\ref{fig6}D). 

In many biological systems, such as bile canaliculi, blastocysts, and regenerating Hydra \cite{ruiz2017organ,dasgupta2018physics}, lumen growth exhibits oscillatory behavior, with alternating phases of expansion and contraction (Figs.~\ref{fig6}E-~\ref{fig6}H). Although leakage, or transient tissue ruptures, have been proposed to contribute to these oscillations, the full physical picture indicates that the hydraulic machinery alone cannot account for such nonlinear dynamics \cite{ruiz2017organ,dasgupta2018physics}. Rather, the oscillatory inflation results from the mechanical coupling between the lumen and its surrounding epithelial layer under cortical tension, together with fluid flows. This sets the stage for our next discussion, which examines how the epithelial layer surrounding the cavity accommodates and resists mechanical deformations.

\subsection{Balloon sculpting: mechanics and tissue response}\label{sculpting}
Once a lumen has nucleated and the osmotic machinery has inflated the active balloon, the resulting hydrostatic pressure $\Delta P$ needs to be balanced by the surrounding epithelial layer \cite{cell_lecuit_2007,forces_heisenberg_2013} (Fig.~\ref{fig6}). The classical static description of this balance comes from the Young-Laplace equation, which relates the pressure difference across a curved interface to the surface tension and local curvature \cite{iii_young_1804,Laplace1805MecaniqueCelesteV4}. In the case of a spherical lumen with radius $R_{l}$, the relationship is
\begin{equation}
\Delta P  = \frac{2\gamma}{R_{l}}
\label{OGLaplace},
\end{equation}
Here $\gamma$ is a homogeneous surface tension which, in a biological context, is a phenomenological constant accounting for both in-plane membrane tension and actomyosin cortical tension \cite{furrow_turlier_2014,dasgupta2018physics,mechanical_hu_2019,dynamics_alsous_2021}.  However, experimental observations reveal that real lumens frequently deviate from the idealized spherical geometry, particularly during early stages of lumenogenesis, when they exhibit irregular, non-spherical morphologies with regions of negative curvature that are incompatible with uniform positive pressure \cite{bryant2010molecular,yang2021cell,vasquez2021physical,mukenhirn2024tight}. Underlying these incompatibilities and the resulting irregular lumen morphologies is the rich, active structure of the epithelial container. Here, we review two complementary studies in MDCK systems showing how active regulation of cortical tension can give rise to nonspherical lumen shapes \cite{vasquez2021physical,mukenhirn2024tight}.

\begin{figure}
\centering
\begin{tikzpicture}
  \node[inner sep=0] (img)
    {\includegraphics[width=0.9\linewidth]{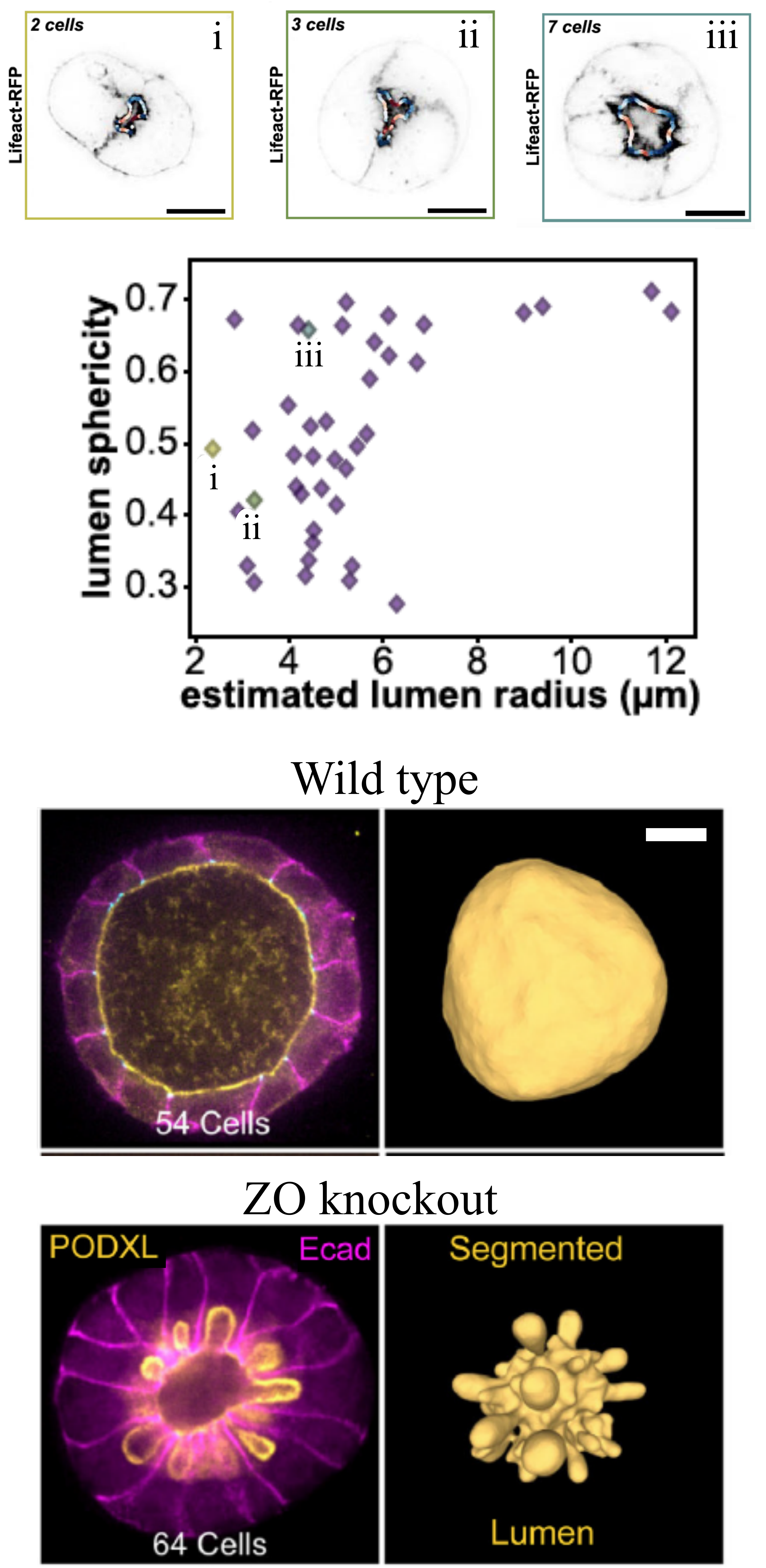}};
  \node at (img.north west) [xshift=0mm, yshift=0mm]{\large\bfseries A};
  \node at (img.north west) [xshift=0mm, yshift=-76mm]{\large\bfseries B};
\end{tikzpicture}
    \caption{Nonspherical lumen geometries. (A) Slices of MDCK spheroids showing irregular luminal shapes (i-iii), expressing Lifeact-RFP (polymerized actin). Scale bars are $10 \mu$m. The plot shows $\Psi$ vs the estimated lumen radius (assuming a spherical cavity). Adapted from \cite{vasquez2021physical}. (B) Luminal shapes in wild type and ZO knockout in MDCK-II cysts. The lateral membranes are stained by E-cadherin and apical membrane by podocalyxin (scale bar, $10\ \mu$m). Adapted from \cite{mukenhirn2024tight}.}
    \label{fig7p2}
\end{figure}

First, Vasquez and colleagues \cite{vasquez2021physical} showed experimentally that small to intermediate lumens ($R_{l}<6\mu$m) within MDCK clusters deviate  strongly from a spherical shape, as quantified by the sphericity:
\begin{equation}
\Psi = \frac{\pi^{1/3}(6V)^{2/3}}{A},
\label{sphericity}
\end{equation}
which compares the actual surface luminal area $A$ to that of a sphere with equivalent volume $V$ (Fig.~\ref{fig7p2}A). Irregular lumens yielded $\Psi\approx0.3$--$0.7$, deviating from a perfect sphere ($\Psi=1$) \cite{vasquez2021physical}. The failure of Young-Laplace predictions at early stages of lumenogenesis reflects the complex mechanical and active properties of the epithelial layer surrounding the lumen \cite{cell_lecuit_2007,mechanics_guillot_2013,forces_heisenberg_2013,indana2024lumen}. Unlike a simple liquid or solid surface, the multicellular layer has a richer internal structure, including the actomyosin cortex, intercellular junctions, and connections to the ECM, each contributing unique and 
distinct mechanical properties \cite{tissue_discher_2005,salbreux2012actin,cell_moeendarbary_2014,shen2017mechanical,adherens_rbsam_2018}. In particular, the authors of Ref.~\cite{vasquez2021physical} argue that non-spherical luminal shapes arise from a geometrical constraint at the level of individual epithelial cells, imposed by the actomyosin cortex (see Section~\ref{cellcon}). Because each cell has a preferred apical area, a lumen that is too small to accommodate the total apical area forces the apical surface to fold.

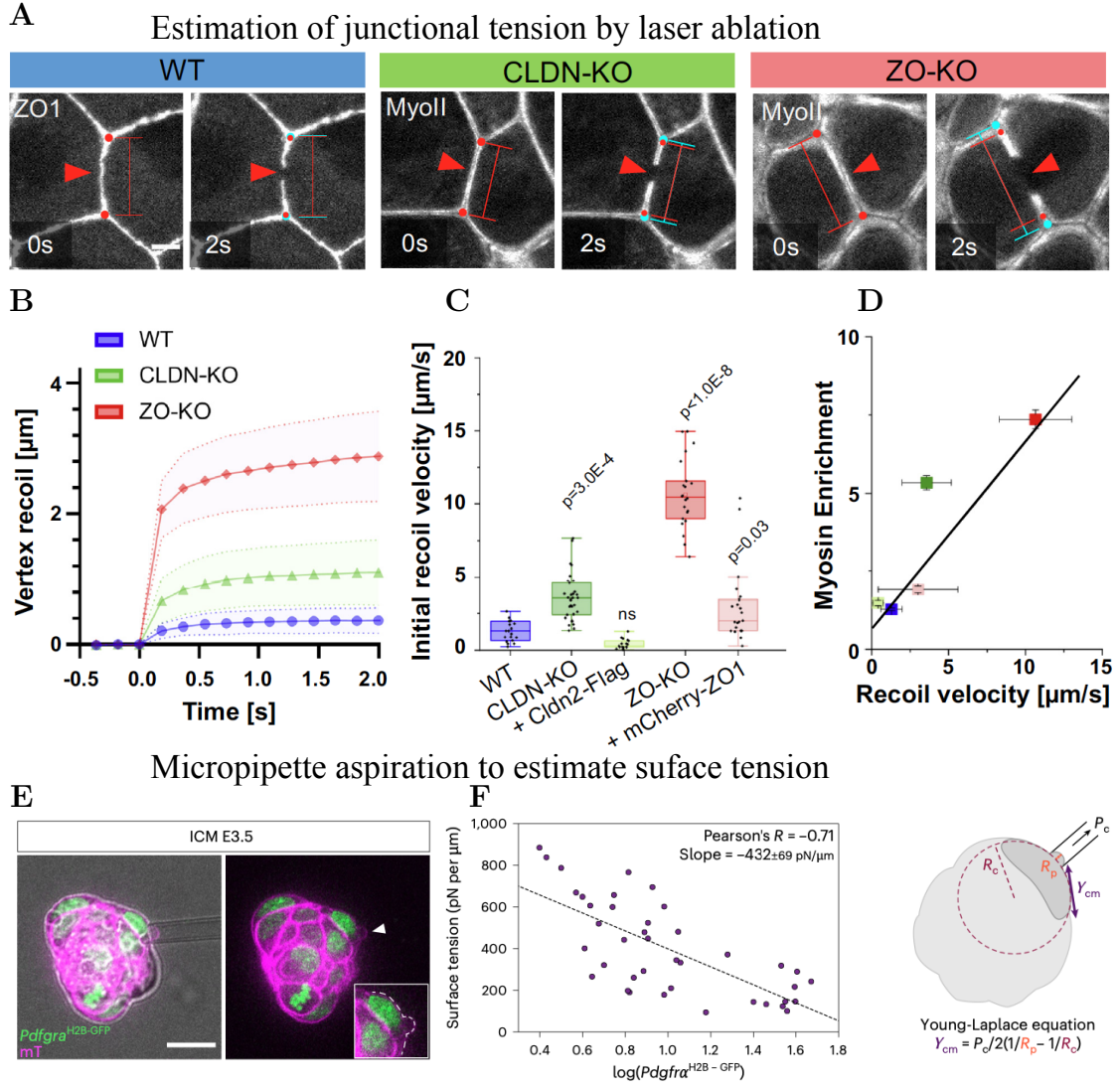
\begin{figure*}
\centering
\begin{tikzpicture}
  \node[inner sep=0] (img)
    {\includegraphics[width=0.86\linewidth]{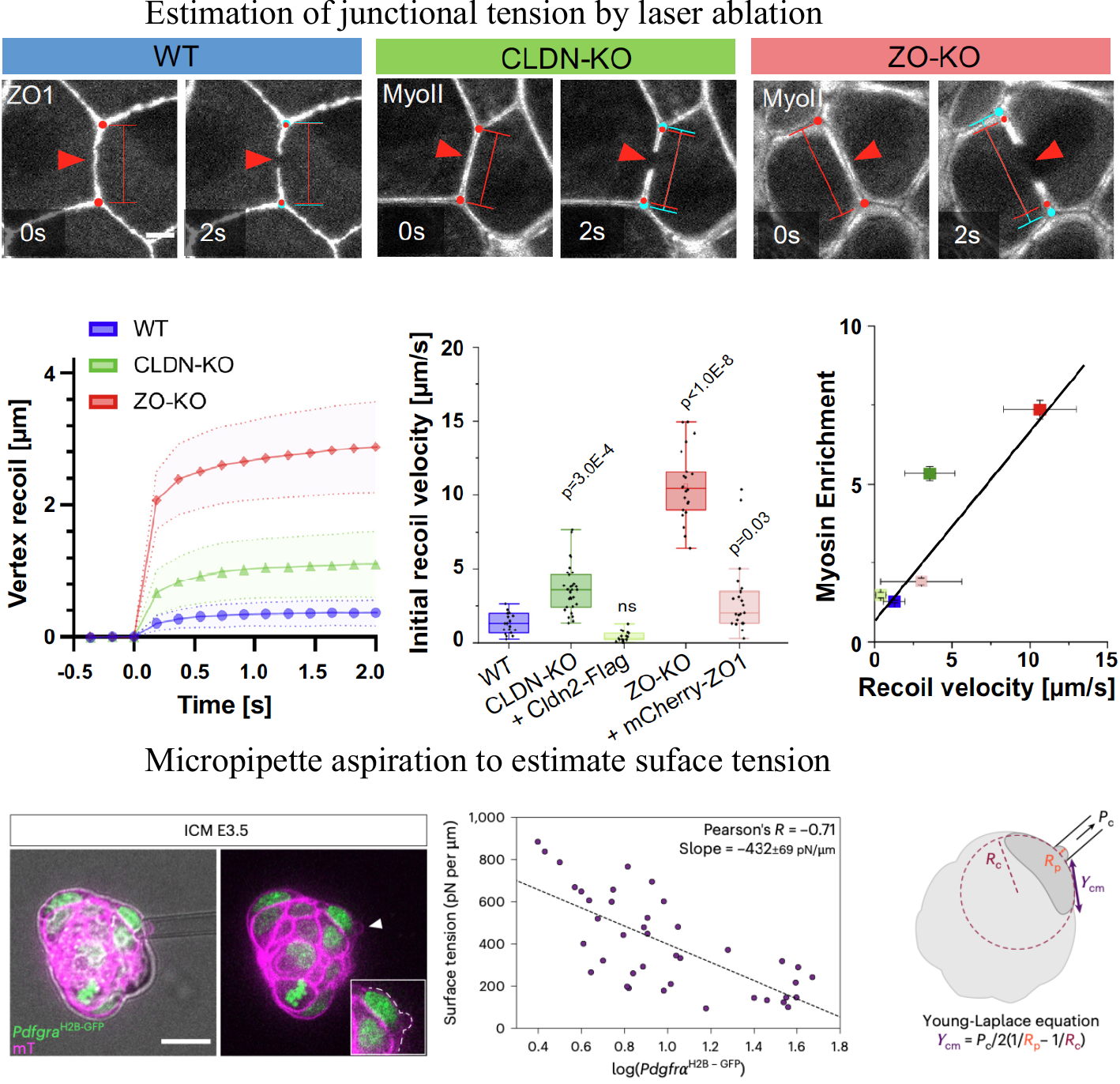}};
  \node at (img.north west) [xshift=2mm, yshift=0mm]{\large\bfseries A};
   \node at (img.north west) [xshift=2mm, yshift=-38mm]{\large\bfseries B};
 \node at (img.north west) [xshift=60mm, yshift=-38mm]{\large\bfseries C};
   \node at (img.north west) [xshift=115mm, yshift=-38mm]{\large\bfseries D};
\node at (img.north west) [xshift=2mm, yshift=-104mm]{\large\bfseries E};
\node at (img.north west) [xshift=63mm, yshift=-104mm]{\large\bfseries F};
\end{tikzpicture}
\caption{ Experimental approaches to characterize surface tension in lumens. (A) Laser ablation of cell-cell junctions of MDCK-II cysts with claudin and ZO knockouts. Red arrow - position of cut. (B) Dynamics of junctional recoil after laser ablation of MDCK-II and knockouts. Laser cut was done at $0$ s. (C) Mean initial recoil velocities of MDCK-II and knockouts show that wild type (WT) tissue is under low mechanical tension. Claudin knockout tissue has increased tension and ZO knockout tissue builds up strong junctional tension. (D) Correlation between myosin junctional enrichment and initial recoil velocity. Adapted from \cite{mukenhirn2024tight}. (E) Micropipette aspiration of ICMs expressing PdgfraH2B-GFP (green) and membrane tdTomato (mT, magenta) in a blastocyst. White arrowhead marks the site of cell aspiration and the white dotted line indicates cell surface contour. (F) Scatter-plot of measured surface tension of outer cells versus logarithm of PdgfraH2B-GFP fluorescence intensity of the cell (left panel). Tension is calculated using Young-Laplace equation accounting for the curvature differences on the cell surface, where $\gamma_{cm}$ indicates cell-medium interfacial tension, $P_{c}$, aspiration pressure, $R_{p}$, radius of the pipette, and $R_{c}$ is the curvature radius of the cell surface near the aspiration site (right panel). Adapted from \cite{moghe2025coupling}.}
    \label{mergeb}
\end{figure*}

The second study, by Mukenhirn and colleagues \cite{mukenhirn2024tight}, identified a role for tight junctions in regulating apical tension in MDCK-II cysts through knockout of ZO and claudin proteins, and linked this regulation to the stabilization of convoluted luminal shapes in late lumenogenesis (Fig.~\ref{fig7p2}B). Before discussing these results in detail, we briefly introduce the key experimental technique used to measure apical tension: laser ablation \cite{mukenhirn2024tight}. In this technique, a high power laser physically disrupts the actomyosin cortex locally, at the sites of cell-cell adhesion, releasing tensile forces (Fig.~\ref{mergeb}A). The adjacent junctions experience this sudden imbalance of forces and recoil away from the ablated junction (Fig.~\ref{mergeb}B), with an initial recoil velocity which is a measure of cortical tension before ablation, and a relaxation time which signifies the viscoelastic dissipation of the tissue. Laser ablation on apical cell-cell junctions of ZO-knockouts resulted in higher apical junctional tension and faster initial recoil velocities when compared to the claudin-knockouts and MDCK-II wild type controls (Fig.~\ref{mergeb}C). This increase in recoil correlated well with the myosin enrichment at the junctions (Fig.~\ref{mergeb}D) \cite{mukenhirn2024tight}. Laser ablation was also used to cut open the MDCK cysts and measure $\Delta P$ by assuming a Hagen-Poiseuille law for the luminal fluid flow \cite{mukenhirn2024tight}. Although laser ablation provides only relative measures of tension rather than absolute force values---since the recoil velocity is also affected by the viscoelasticity of the cortex and the cytoplasm---it can be combined with micropipette aspiration \cite{villeneuve2019apkci} (Figs.~\ref{mergeb}E~and~\ref{mergeb}F) or atomic force microscopy (AFM) measurements to obtain forces quantitatively.

Mukenhirn \textit{et al}.~\cite{mukenhirn2024tight} showed experimentally that in MDCK-II wild-types, hydrostatic pressure differences of $\Delta P\sim 65$ Pa together with relatively low junctional line tensions ($\gamma\sim50$--$100$ pN) place the system in a growth-dominated regime that  readily stabilizes lumens.
However, when ZO-1 and ZO-2 proteins are depleted---producing  a $\sim10$-fold increase in junctional tension through elevated myosin-IIa accumulation at apical junctions, and a $\sim6$-fold decrease in $\Delta P$---the system shifts to a tension-dominated regime. In this ZO-knockout case, lumens collapse into a folded morphology where apical surface regions buckle inward with a characteristic wavelength, despite the osmotic machinery being intact. The authors suggested that lumen expansion requires not only a
hydrostatic pressure difference but also an active suppression of cortical actomyosin contractility; this effect is mediated by the tight-junction scaffold proteins ZO-$1$ and ZO-$2$, which couple junctions to the perijunctional actomyosin network and can tune myosin-dependent tension \cite{jama_monteiro_2013,zo1_tornavaca_2015,zo2_gonzlezmariscal_2019}. In fact, the authors checked that ROCK---a protein kinase that controls the cell's cytoskeleton---inhibition resulted in no significant change in MDCK-II cyst wild types, while in ZO-knockouts the lumen volume increased significantly, confirming that release of apical junctional tension promotes lumen inflation in tight junction deficient cysts. We note that cortical tension can also be inhibited using  inhibitors of actin depolymerization (Latrunculin A), microtubules (nocodazole) \cite{vasquez2021physical} and myosin light chain kinase inhibitor (ML-7) \cite{indana2024lumen}.

The key observation linking these two studies is the conserved role of the apical area. In ZO-knockouts, the convoluted pattern of the collapsed lumens indicates that, even when the luminal volume decreases significantly, the total apical surface area remains essentially unchanged \cite{mukenhirn2024tight}. This is the same geometrical principle identified by Vasquez \textit{et al}.~\cite{vasquez2021physical} for small lumens in  wild-type MDCK cysts:
each cell has a preferred apical area that must be accommodated, largely independently of luminal volume. When the lumen is too small for this area to be realized on a spherical surface, nonspherical morphologies emerge. In summary,  apical area can act as a geometrical constraint on lumen sculpting at both the cell and tissue levels in MDCK cysts. The incorporation and consequences of this constraint in mathematical models are discussed in  Section~\ref{cellcon}.

The results of Vasquez \textit{et al}.~\cite{vasquez2021physical} and Mukenhirn \textit{et al}.~\cite{mukenhirn2024tight} highlight the active balance between luminal hydrostatic pressure and apical epithelial tension.
In light of the folded morphologies in Figs.~\ref{fig7p2}A and~\ref{fig7p2}B, a natural question is which bulk mechanical properties control the ability of the epithelial tissue to deform.
Experimental findings from Shen \textit{et al}.~\cite{shen2017mechanical} have helped to elucidate this question. Specifically, AFM indentation of MDCK-II cysts have shown hysteresis in the force-distance approach and retraction curves, characteristic of a viscoelastic behavior. By analyzing only the approach  curve, the elastic modulus of the cyst, modeled as a pressurized shell, was estimated as $E_{s}=8.2\pm4.2$ kPa using a shell model \cite{fery2004mechanics}:
\begin{equation}
F = \frac{4}{3}E_{s}h\epsilon\left(h+4\pi R\epsilon^{2}\right).
    \label{shell}
\end{equation}
Here $F$ is the applied force, $h$ is the thickness of the epithelial layer ($5-8 \mu$m \cite{shen2017mechanical}), $R$ is the radius of the cyst, and $\epsilon=\delta/2R$ is the relative deformation, with $\delta$ the indentation length.  Beyond this elastic response, measurements at different loading speeds revealed an early exponential relaxation followed by a power-law decay \cite{shen2017mechanical}, indicating that the cyst mechanical behavior cannot be captured purely by an elastic description. More generally, the mechanical response of a multicellular aggregate to pressure-induced deformations depends critically on its constitutive properties, which are far more complex than those of a simple linear elastic material. \cite{viscoelastic_forgcs_1998,role_marmottant_2009,physical_trepat_2009,shen2017mechanical}. As we review in Section~\ref{tirupture}, the choice of a constitutive model for a epithelial shell, whether elastic, viscoelastic or plastic, affects the theoretical predictions of lumen dynamics.

The hydrostatic pressure within lumens is not only resisted by the epithelial container, but also by the ECM, specifically the basement membrane. Studies of MDCK cells cultured in synthetic hydrogels suggest that lumen formation occurs in hydrogels with an optimal stiffness of $\sim 4$ kPa \cite{enemchukwu2016synthetic}. Matrices that are too rigid disrupt lumenogenesis by preventing lumen inflation (see Section~\ref{matrices}). Beyond acting as a mechanical constraint, the ECM also has been reported to have a sculpting role after lumen creation in hepatocytes \cite{extracellular_li_2016}. The spatial distribution of ECM adhesion sites generates anisotropic intercellular tension that guides lumen elongation. In particular, Li and collaborators \cite{extracellular_li_2016} demonstrated that lumens preferentially extend toward regions of minimal intercellular tension, i.e., away from the ECM. 

\subsection{Mechanochemical coupling between the lumen and its cellular boundary}\label{mechanoprocess}

A common conceptual framework emerges from the three previous subsections: fluid-filled active balloons are shaped by a tissue-scale mechanical balance among luminal pressure, cortical tension, and ECM confinement. This force balance, however, is not the whole story. The epithelial container can also respond biochemically to mechanical cues from the lumen by modifying tissue contractility, remodeling tight junctions, or altering gene expression \cite{hannezo2019mechanochemical,chan2020integration}. Next, we review this mechanochemical coupling between the luminal fluid and the epithelial shell.

One example of mechanochemical coupling is the direct effect of luminal pressure on epithelial mechanics and cell fate: hydrostatic pressure does not merely provide passive mechanical support, but can actively regulate the behavior of the surrounding cellular layer. The regulatory task of luminal pressure of cell function in particular and morphogenesis in general has been reviewed by Chan and Hiiragi \cite{chan2020integration} and by Bagnat \textit{et al}.~\cite{bagnat2022morphogenetic}, and here we briefly highlight key examples. In the mouse blastocyst, lumen inflation stretches the TE cells enclosing the cavity. This stretching generates an increase in actomyosin contractility, promoting tight junction maturation, which is necessary for the subsequent luminal growth \cite{chan2019hydraulic}. This positive feedback can be impaired through transient tight junctions leakage events when the lumen is too pressurized \cite{chan2019hydraulic,chan2020integration}. Luminal pressure also dictates cell fate in the mouse blastocyst, where small cavity sizes have been shown to stimulate asymmetric division of TE cells, creating daughter cells that face the interior and become ICM cells \cite{chan2019hydraulic}. 

Beyond the blastocyst, luminal pressure also orchestrates cell fate in other developing systems. During lung morphogenesis the increase in amniotic (luminal) fluid together with a heterogeneous resistance of alveolar cells dictates the fate of these cells. Specifically, cells that cannot resist the pressure from the luminal fluid are flattened into thin gas-exchanging cells, while cells that do resist the luminal pressure---via stiffening of their apical cortex---adopt a cuboidal secretory identity \cite{li2018strength}. 

Apart from its ability to retain fluid and exert pressure, the lumen also serves as a storage of signaling molecules as described in Section~\ref{hub}. The signaling hub function of lumens in diverse biological contexts have been discussed in both Chan and Hiiragi \cite{chan2020integration} and Hannezo and Heisenberg \cite{hannezo2019mechanochemical} reviews, and here we highlight  one striking case studied by Durdu \textit{et al}.~\cite{durdu2014luminal}. In the lateral line of the zebrafish embryo,  groups of epithelial cells assemble into rosette-like shapes, each enclosing a shared lumen. Durdu and colleagues showed that these cavities localize fibroblast growth factor, a key  developmental signaling molecule, which is absent from neighboring cells outside the rosettes. This finding demonstrates that lumens can act as spatial morphogenetic checkpoints, influencing which epithelial cells within a tissue polarize, organize, and ultimately differentiate.

A more integrated form of mechanochemical feedback emerges in intestinal organoid morphogenesis, where lumen dynamics, cell fate, and tissue mechanics are tightly coupled. These organoids, which initially consist of a spherical epithelial layer enclosing a fluid-filled lumen, break spherical symmetry during development into two mechanically distinct regions : crypt and villus. 
The crypt region forms outward-budding epithelial pockets that harbor stem cells, whereas the villus region is the remaining relatively flat epithelium, composed primarily of enterocytes.
Yang and colleagues \cite{yang2021cell} demonstrated that cell-fate specification drives this morphological transition by regulating the spatial distribution of  actomyosin along the lumen container, yielding distinct apical and basal myosin patterns in the crypt and villus regions. 
In addition, the emergence of enterocytes redistributes  luminal fluid via a specific osmotic machinery that promotes enterocyte swelling and  lumen deflation.
This process  accelerates  crypt budding by facilitating epithelial deformations. Interestingly, inflating the lumen after budding does not restore  spherical symmetry, while inflating the initially spherical organoids prevents budding formation \cite{yang2021cell,xue2025mechanochemical}.

Finally, a compelling example of the coupling between lumen mechanics and biochemical signaling is the regeneration of Hydra tissue fragments. 
After excision, the tissue folds into a spherical shape and undergoes  an osmotically-driven oscillation, alternating phases of lumen inflation and deflation \cite{CFutterer_2003,Soriano2006} (Figs.~\ref{fig6}G~and~\ref{fig6}H). 
These oscillations are not a simple reaction to excision or a side effect of the regenerating process, but  are fundamental for establishing a Wnt signaling center,
a molecular signaling pathway that regulates body-axis formation. 
Ferenc and colleagues \cite{ferenc2021mechanical} showed that Wnt3 expression correlates with the degree of tissue stretching during inflation, and that blocking the oscillatory dynamics inhibits regeneration. 
In a more recent study, Weevers \textit{et al}.~\cite{weevers2025mechanochemical} used micropipette aspiration to demonstrate that the mechanochemical coupling between Wnt signaling and tissue stretching forms a positive feedback in which Wnt signaling reduces tissue surface tension, while stretching promotes Wnt3 activation. 
This local self-amplification acts in synchrony with a long-range inhibition mediated by  luminal pressure: because the luminal fluid is effectively incompressible, stretching in one region must be compensated elsewhere to maintain hydrostatic pressure.
Beyond providing a mechanochemical basis for spatial patterning, different from classical Turing or Gierer-Meinhardt mechanisms \cite{turing1990chemical,gierer1972theory}, this activator-inhibitor process positions the lumen as a key regulator of tissue symmetry breaking.

Taken together, these examples highlight the lumen as a mechanical and biochemical actor that can influence tissue organization across biological systems (blastocyst, intestinal organoids, regenerating Hydra) through nontrivial feedback mechanisms. In Section~\ref{feedbackModels}, we will review the theoretical models---the main theme of our next section---introduced to explain some of the experimental observations described in this subsection.

\section{Theoretical and computational models of lumen formation}\label{theocomp}

Section~\ref{sec4} emphasized that lumenogenesis is a multiscale physical process involving cavity nucleation, osmotic inflation, tissue mechanics, and mechanochemical feedbacks. A natural next step is to ask how these ingredients can be distilled into mathematical models that are simple enough to analyze, yet rich enough to capture the main experimental phenomenology. In the absence of first-principles theories for morphogenesis in general, and lumenogenesis in particular, theoretical and computational approaches play a central role in identifying minimal ingredients, testing mechanistic hypotheses, and generating quantitative predictions \cite{FUJI2022173}. Rather than attempting a single unified description, the literature has developed a diverse set of modeling frameworks, each tailored to a particular aspect of lumen formation and a specific biological system, operating at its own level of coarse graining, and 
utilizing distinct computational tools.

The purpose of this section is to review these frameworks and show how they illuminate several of the physical scenarios discussed in the previous section. We begin with models for the emergence of a single lumen from many nascent cavities, focusing on coarsening-like dynamics driven by hydraulic exchange, active pumping, cell rearrangements, and cell proliferation (Section~\ref{coarseningModels}). We then turn to oscillatory lumen growth, where theory has been used to formalize the competition between inflationary driving forces and the negative feedbacks that arrest or reverse expansion (Section~\ref{osci}). A third theme concerns lumen shape itself: why many biological lumens deviate strongly from spheres, and how such morphologies can emerge from cell-level geometric constraints, collective cell dynamics, differential growth, and active rearrangements (Section~\ref{shapesModel}). Finally, we discuss recent models in which the lumen is no longer treated merely as an output of cellular activity, but as an active participant in development through mechanochemical feedbacks on tissue contractility, signaling, and fate specification (Section~\ref{feedbackModels}).

Taken together, these studies illustrate both the breadth of current theoretical approaches to lumenogenesis and the complementary roles they play. Some models are intentionally minimal and analytically tractable, making it possible to isolate generic principles such as hydraulic screening, coarsening laws, or instability thresholds. Others are more phenomenological and computational, incorporating cell division, cell motion, topology changes, or tissue-scale mechanics to connect more directly with specific experiments. What unifies them is a common goal: to translate the complex biological picture of lumen formation into a physical language of transport, force balance, elasticity, active stresses, and feedback. In that sense, this section is not only a survey of models, but also a map of how biological questions about lumenogenesis have been recast into problems in soft matter, nonequilibrium physics, and mechanochemical pattern formation.

\subsection{Coarsening-like dynamics and lumen selection}\label{coarseningModels}
In Section~\ref{routes}, we discussed how in multicellular aggregates the  emergence of luminal space can be through the nucleation of multiple lumens (Fig.~\ref{fig4}). 
Although this nucleation mechanism has only recently begun to be addressed theoretically \cite{dinet2023patterning}, the subsequent dynamics that consolidate multiple lumens into a single lumen, or fail to do so, has received greater attention
\cite{cerruti2013,dumortier2019hydraulic,le2021hydro,LEVERGESERANDOUR202212,lu2025generic,lee2025permeability}.
\subsubsection{Passive and active fluxes as drivers of lumen coarsening}\label{hydralumen}
In the blastocyst, after hydraulic fracture of cell-cell adhesion between, multiple lumens nucleate throughout the system (top panel, Fig.~\ref{fig4}). The ensuing coarsening introduces a type of collective dynamics that is unique to biological systems: a network of actively maintained, pressurized compartments
that interact through fluid and solute exchange. 
From a purely physics perspective, this system belongs to the broad family of coarsening phenomena \cite{lifshitz1961kinetics}, but with biological-specific richness: long-time transport is predominantly hydraulic rather than diffusive, active ion pumps drive water flows, and intrinsic cellular heterogeneities bias the location at which the final lumen is selected and arrested \cite{dumortier2019hydraulic,le2021hydro}.
In what follows, we describe the modeling approaches used in the literature to explain the coarsening-like dynamics involved in blastocoel formation. 
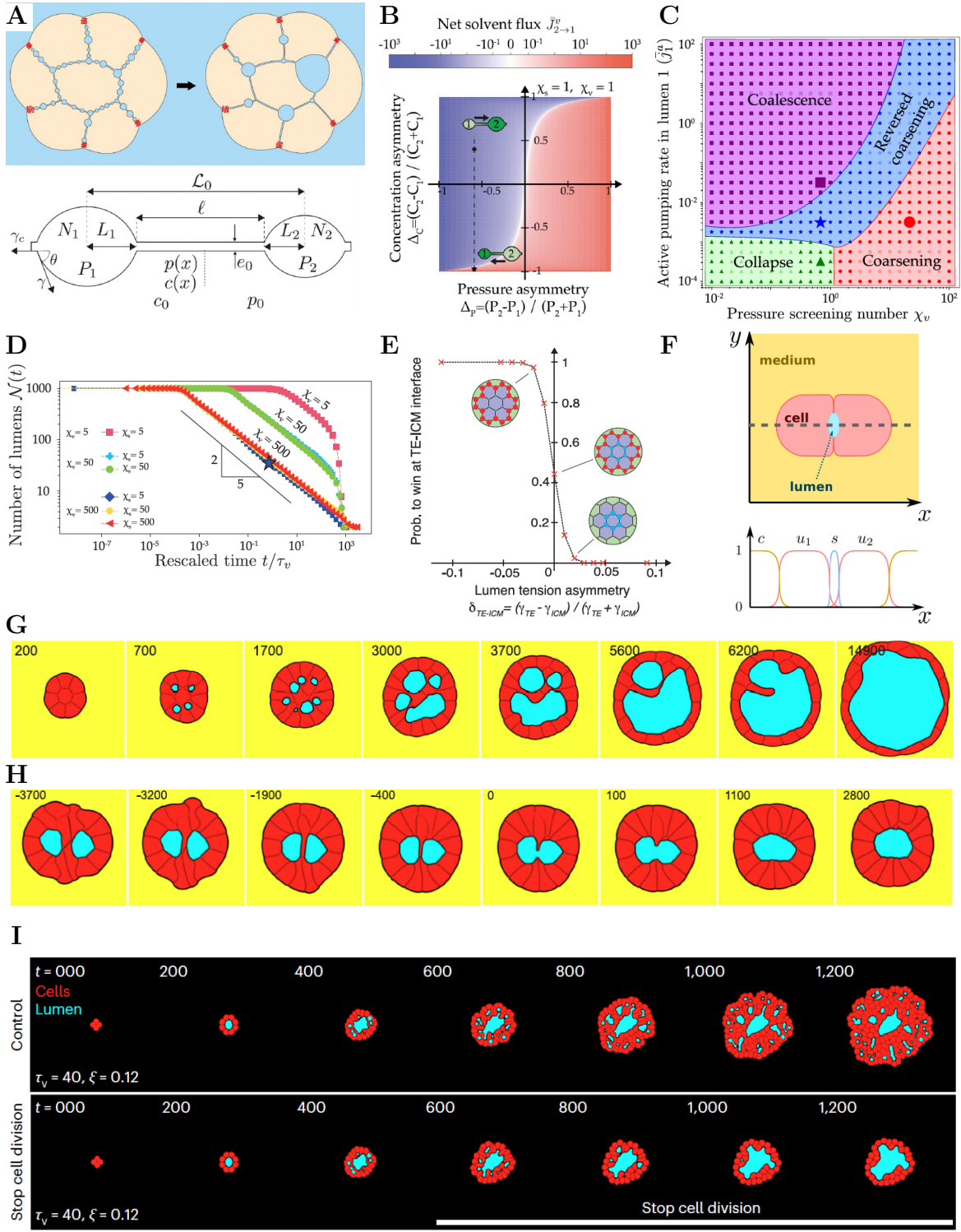
\begin{figure*}
\centering
\begin{tikzpicture}
\node[inner sep=0] (img)
{\includegraphics[width=0.9\linewidth]{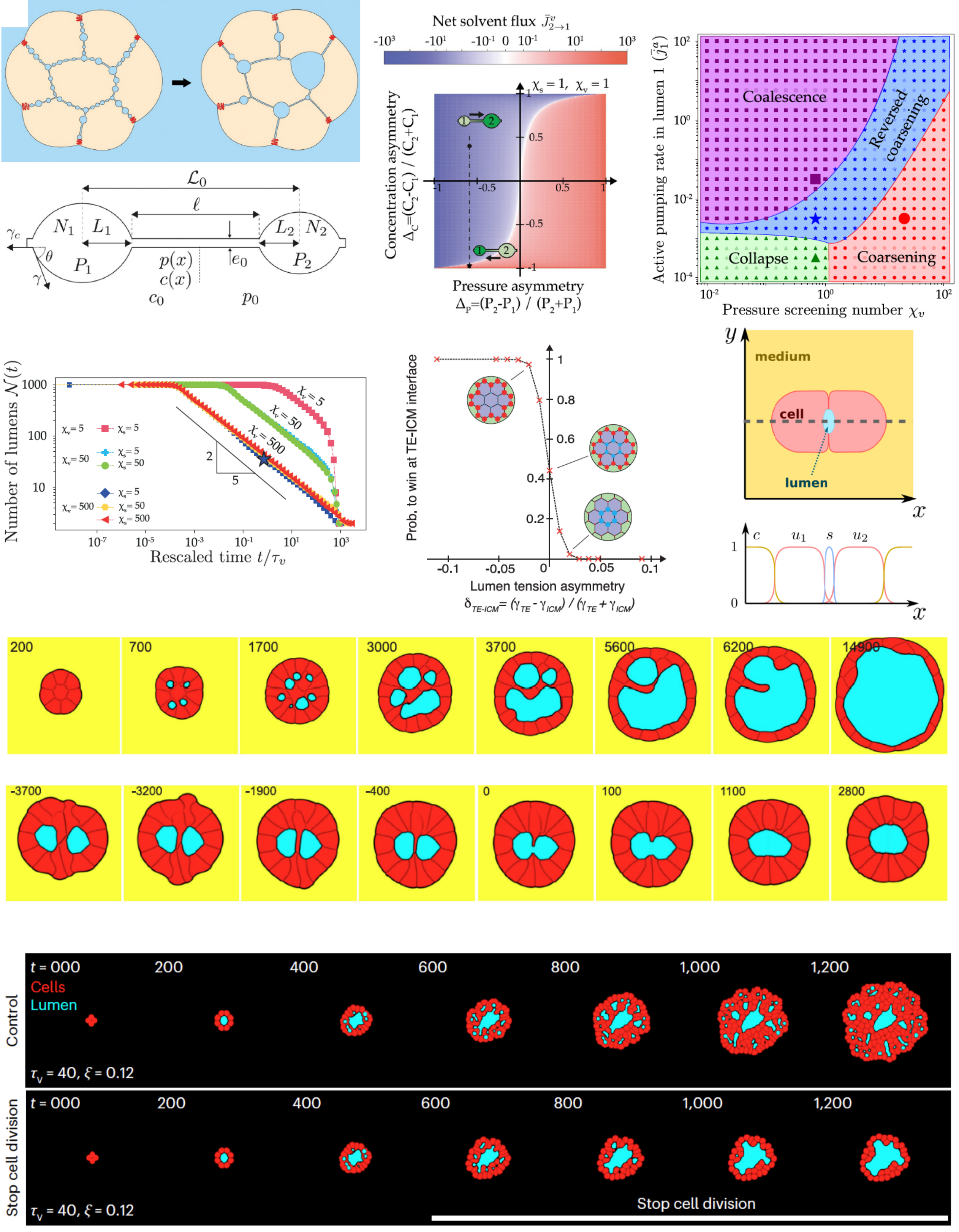}};
  \node at (img.north west) [xshift=2mm, yshift=-2mm]{\large\bfseries A};
   \node at (img.north west) [xshift=62mm, yshift=-2mm]{\large\bfseries B};
  \node at (img.north west) [xshift=107mm, yshift=-2mm]{\large\bfseries C};
   \node at (img.north west) [xshift=2mm, yshift=-55mm]{\large\bfseries D};
\node at (img.north west) [xshift=62mm, yshift=-55mm]{\large\bfseries E};
\node at (img.north west) [xshift=107mm, yshift=-55mm]{\large\bfseries F};
\node at (img.north west) [xshift=2mm, yshift=-100mm]{\large\bfseries G};
\node at (img.north west) [xshift=2mm, yshift=-124mm]{\large\bfseries H};
\node at (img.north west) [xshift=2mm, yshift=-150mm]{\large\bfseries I};
\end{tikzpicture}
    \caption{Modeling approaches for the emergence of a single lumen from multiple ones. (A) Top panel: Schematic drawing of two instances of the coarsening dynamics in the blastocyst. The red lines illustrate the tight junctions. Bottom panel: theoretical description of two lumens connected by an intercellular channel. (B) Phase diagram in $\Delta_{p}-\Delta_{c}$ space of the pair-lumen dynamics. (C) Phase diagram in $\chi_{v}-\bar{j}_{1}^{a}$ space of the pair-lumen dynamics when one active pump is included. The upper bars in the flux quantities in (B) and (C) indicate that they are dimensionless (see details in Ref.~\cite{le2021hydro}). (D) Evolution of the number of lumens over time on a 1D chain. The time constant $\tau_{v}$ is a fluid equilibration time scale given by $L_{o}/2RT\lambda_{v}c_{o}$ with $L{o}$ ($c_{o}$) the mean initial lumen size (solute concentration). (A-D) are adapted from \cite{le2021hydro}. (E) Plot illustrating the most probable final lumen position (red and light blue shaded areas) after coarsening dynamics in a 2D hexagonal network model, as a function of lumen tension asymmetry. Adapted from \cite{dumortier2019hydraulic}. (F) Phase field representation of two cells, one lumen and the surrounding medium. (G) Temporal evolution of a multicellular phase field model involving cell division and lumen fusion. (H) Temporal evolution of a multicellular phase field model involving cell motion and lumen fusion. (F-H) are adapted from \cite{lu2025generic}. (I) Temporal snapshots of multicellular phase field simulations where cell proliferation is active (top panel) and inactive (bottom panel). Adapted from \cite{lee2025permeability}.}
    \label{fig8}
\end{figure*}

Irrespective of whether a one-or two-dimensional approach is used to study  coarsening dynamics, the basic modeling unit  is a pair of lumens joined by a thin intracellular channel of width $e_{o}$ and length $\ell(t)$. The lumens are assumed to be embedded in an infinite cellular medium acting as a chemostat of solute concentration $c_{o}$ and as barostat of pressure $p_{o}$ (Fig.~\ref{fig8}A). Each lumen, $i=1,2$, is characterized by its cross-sectional area $A_{i}$ and its molar ion content $N_i$. 
Lumens are under pressure, which is balanced by a luminal tension $\gamma_{i}$ via the Young-Laplace law
\begin{equation}
\delta P_{i}=P_{i}-p_{o}=\frac{\gamma_{i}\sin(\theta)}{L_{i}},
    \label{balanceGeom}
\end{equation}
where $L_{i}=\sqrt{\mu(\theta) A_{i}}$ is the half length of the lumen $i$ (Fig.~\ref{fig8}A), and $\mu(\theta)$ is a geometric factor that depends on $\theta$, the contact angle between the lumens and the channel. This angle is determined by a tension balance at the lumen-channel junction \cite{dumortier2019hydraulic,le2021hydro}; $2\gamma_{i}\cos(\theta)=\gamma_{c}$, where $\gamma_{c}$ is the tension at the channel.

The volume dynamics of the lumen $i$ are parameterized through the length $L_i$ \cite{le2021hydro}:
\begin{equation}
\frac{dL_{i}}{dt}=2\mu(\theta)\nu(\theta)\lambda_{v}(RT\delta C_{i}-\delta P_{i})-\frac{\mu}{2L_{i}}J_{i}^{v},
    \label{VolChange}
\end{equation}
where $\delta C_{i}=C_{i}-c_{o}$ is the concentration gradient across the membrane ($C_{i}=N_{i}/A_{i}$), $\nu(\theta)$ is a geometric factor, $J_{i}^{v}$ is the hydraulic flow at the channel boundary, and $\lambda_{v}$ is the hydraulic permeability. Similarly, an equation for the evolution of ions in the lumen $i$ reads:
\begin{equation}
\frac{dN_{i}}{dt}=2\nu L_{i}\left(\lambda_{s}RT\delta\log\left(\frac{c_{o}}{C_{i}}\right)+j_{i}^{a}\right)-J_{i}^{s},
    \label{SolChange}
\end{equation}
where $j_{i}^{a}$ is an active pumping rate by unit length of cell membrane, 
a key mechanism for lumen growth as we discussed in Section~\ref{pumping}, $J_{i}^{s}$ is the ionic flux of solute $i$ at the channel boundary, and  $\lambda_{s}$ is the solute permeability. Both the hydraulic and ionic fluxes at the channel boundary can be obtained by modeling the respective flows along the thin channel. For the fluid flow, a rapid equilibration of hydrostatic pressure is assumed across the channel thickness (lubrication limit), allowing a Poiseulle flow description;  $-\kappa_{v}\partial\delta p/\partial x$, with $\delta p=p(x)-p_{o}$ the pressure difference across the channel membrane and $\kappa_{v}\approx e_{o}^{3}/12\eta$ the hydraulic conductance, with $\eta$ the fluid viscosity. Then, the volume conservation in the thin region is
\begin{equation}
\kappa_{v}\frac{\partial^{2}\delta p}{\partial x^{2}}+2\lambda_{v}(RT\delta c-\delta p)=0,
\label{channelV}
\end{equation}
where $\delta c=c(x)-c_{o}$ is the concentration gradient across the channel membrane. For the solute flow, assuming a fast mixing and diffusion-dominated transport along the channel, the longitudinal flow of solute can be described by Fick's law: $-D\partial\delta c/\partial x$, where $D$ is a diffusion coefficient. Then, the balance of ions in the channel yields \cite{le2021hydro}
\begin{equation}
e_{o}D\frac{\partial^{2}\delta c}{\partial x^{2}}+2\lambda_{s}RT\log\left(1+\frac{\delta c}{c_{o}}\right)-2j^{a}=0.
\label{channelI}
\end{equation}
The exchanges of fluid and solute along the channel are controlled by two characteristic length scales (screening lengths): $\xi_{v}=\sqrt{\kappa_{v}/2\lambda_{v}}$ and $\xi_{S}=\sqrt{De_{o}c_{o}/2\lambda_{s}RT}$, respectively, which measure the typical distance over which a pressure or concentration difference is dissipated by lateral permeation. When $\ell\gg\xi_{v}$ ($\ell\gg\xi_{s}$), the two lumens are hydraulically (osmotically) uncoupled, while when $\ell\ll\xi_{v}$ ($\ell\ll\xi_{s}$), they exchange fluid (solute) freely.

In the double limit $\delta C_{i},\delta c\ll c_{o}$, Eqs.~(\ref{channelV}) and~(\ref{channelI}) can be solved exactly, closing the system together with size conservation $L_{1}+L_{2}+\ell=\mathcal{L}_{o}$ \cite{le2021hydro}.  The competition between pressure and concentration differences is summarized in the phase diagram of  Figure~\ref{fig8}B, where the direction of the net hydraulic flow $J_{2\rightarrow 1}^{v}=J_{2}^{v}-J_{1}^{v}$ is measured as a function of the pressure asymmetry $\Delta_{p}=(P_{2}-P_{1})/(P_{2}+P_{1})$ and the concentration asymmetry $\Delta_{c}=(C_{2}-C_{1})/(C_{2}+C_{1})$. The phase diagram shows that normal coarsening---drainage from small to large lumen---happens for a wide range of $\Delta_{c}$ values. 
However, if the larger lumen is sufficiently depleted of ions, reverse drainage---from the larger lumen to the smaller one---can instead dominate the dynamics (Fig.~\ref{fig8}B).

Active pumping qualitatively expands the range of possible outcomes. Imposing a non-zero active pump in one lumen ($j_{1}^{a}\neq0$, $j_{2}^{a}=0$) introduces a sustained ion source that competes with the pressure-driven hydraulic discharge. Depending on the values of $j_{1}^{a}$ and the pressure screening number $\chi_{v}=\xi_{v}/\ell_{o}$ (with $\ell_{o}$ the initial channel length), four possible dynamical regimes are revealed (Fig.~\ref{fig8}C). In addition to coarsening and reversed coarsening, both lumens can collapse when  active pumps are negligible and the lumens are hydraulically isolated. Conversely, when the pump is strong and $\chi_v \le 1$, the two lumens may coalesce:  their growth overcomes the resistance of the connecting the channel. Such coalescence  has been observed in the late-stage  mouse blastocoel formation \cite{dumortier2019hydraulic}, and in experiments and models of fluid droplet coarsening \cite{meakin1992droplet,stricker2015impact}.

Once the model for the pair of lumens is established, extending it to $\mathcal{N}$ lumens allows the study of  coarsening dynamics \cite{le2021hydro}.
In a 1D chain with sealed boundary conditions and no active pump, the number of lumens decays as a power-law in time: $\mathcal{N}\sim t^{-2/5}$ (Fig.~\ref{fig8}D). This decay law is largely independent of $\chi_{v}$ and $\chi_{s}=\xi_{s}/\ell_{o}$ over a large range of values, except for small pressure screening numbers ($\chi_{v}<10$). The scaling exponent is similar to that predicted for the coarsening  of hydraulically coupled droplets in thin dewetting films \cite{glasner2003coarsening,pismen2004mobility,glasner2005collision}. Indeed, in the hydraulic limit---$\chi_v\gg 1$, $\chi_s\ll 1$, and $\tau_s\ll \tau_v$, so that solute dynamics relax rapidly---the model Eq.~(\ref{VolChange}) becomes analytically equivalent to a mean-field model introduced for thin dewetting films. Incorporating active pumps along the  chain yields a new self-similar behavior for the number of lumens for late coarsening times: $N(t)\sim t^{-1}$. This faster decay is driven by lumen coalescence, which occurs earlier or later depending on the pump strength $j^{a}$ \cite{le2021hydro}.

Dumortier \textit{et al}.~\cite{dumortier2019hydraulic} considered a two-dimensional network topology in the hydraulic limit to explain blastocoel positioning  in a more realistic geometric setting. In their model, the evolution of the lumen area is given by 
\begin{equation}
\frac{dA_{i}}{dt}=\lambda_{v}\omega(\theta_{i})\sqrt{A_{i}}+\sum_{j\in W_{i}}\frac{\kappa_{v}(P_{i}-P_{j})}{\ell_{ij}},
\label{networkLumen}
\end{equation}
where $\omega(\theta_{i})$ is a geometric factor encoding the tension balance at the edge of the lumen and is a function of the lumen-channel contact angle $\theta_{i}$, $\ell_{i,j}$ is the length of the channel between lumens $i$ and $j$, and $W_{i}$ is the set of neighbors of lumen $i$ given by the topology of the network. 
Neglecting solute-concentration dynamics allowed them to systematically explore how 
spatial variations in lumen tension $\gamma_{i}$, captured through the cell-type dependent geometric factor $\omega(\theta)$,  bias 
lumen positioning.
Because ICM cells are more contractile \cite{maitre2016asymmetric,dumortier2019hydraulic}, a property that can be quantified, for example, by micropipette aspiration \cite{moghe2025coupling}, the tension of a lumen surrounded by ICM cells is larger than that of a 
lumen  sandwiched between a TE cell and an ICM cell. This heterogeneity, quantified by the lumen tension asymmetry
\begin{equation}
\delta_{TE-ICM}=\frac{\gamma_{TE}-\gamma_{ICM}}{\gamma_{TE}+\gamma_{ICM}},
\end{equation}
selects the blastocoel position at the TE-ICM interface (Fig.~\ref{fig8}E) \cite{dumortier2019hydraulic}. This selection follows from the Young-Laplace law: larger tensions produce larger hydrostatic pressures, which drive fluid discharge from higher-pressure luminens (ICM-associated) toward lower-pressure lumens (TE-associated).

In Ref.~\cite{le2021hydro}, a heterogeneous distribution of active pumps, motivated by a higher abundance at TE cells than at inner ICM cells, was also shown to favor positioning the final lumen at the TE-ICM interface.  Beyond the coarsening dynamics of the lumens, other theoretical studies have explored the positioning of the blastocoel  from purely mechanical considerations, showing that the geometry of the confining space can bias the final lumen position \cite{honda2008computer}.

The network models described above show how redistribution of fluid due to pressure differences, active pumping, and inhomogeneous cell properties determines the position of the final single lumen in the blastocyst. In these models, lumen centroids remain fixed during discharge, consistent with experiments \cite{dumortier2019hydraulic}, and the surrounding tissue is therefore treated as static as well. This absence of motion, however, is not a general feature of all  biological systems exhibiting coarsening-like lumen dynamics. Next, we review modeling approaches that explicitly incorporate  cell dynamics into the study of lumen fusion.

\subsubsection{Lumen coalescence in agent-based models}\label{phasefieldS}

In modeling studies of lumen dynamics in MDCK cysts, pancreatic spheres, and epiblasts, attention has also focused on how intrinsic cellular dynamics  during development---including cell deformation, cell-cell interactions, cell division and cell arrangement---play a role in lumen formation and stabilization \cite{lu2025generic,tanida2025predicting,lee2025permeability}. 
This is particularly relevant in the case of epiblasts \cite{lu2025generic}, where experimental observations suggest that unpressurized cavities coarsen through cell movements (bottom panel, Fig.~\ref{fig4}). Lu, Fuji and colleagues \cite{lu2025generic} employed a two-dimensional multicellular phase field approach to capture cell and lumen dynamics in epithelial organoids. This phase field approach is a flexible continuum method that can couple cell boundary deformations to cell-cell and cell-medium interactions while keeping the discrete identity of individual cells \cite{shao2010computational,nonomura2012study,camley2014polarity,lober2015collisions,cao2019cell,hopkins2022local,tanida2025predicting,echeverria2025single}. This method has been used broadly by many groups working in theoretical and computational biophysics, and thus its motivation and implementation can vary in the literature. Here, we describe the framework used in Ref.~\cite{lu2025generic}.

The model consists in three components: cells, lumens and the ECM. Each of these is described by a scalar phase field variable that it is equal to $1$ inside the corresponding domain and $0$ outside, with a smooth interface connecting them (Fig.~\ref{fig8}F). The authors use $u$ for the cellular, $s$ for the lumens and $c$ for the ECM phase fields. The evolution of these phase fields follow variational dynamics, i.e., they relax to a steady state given an energy functional, which generally reads
\begin{equation}
E=\sum_{m}^{M}E_{u_{m}}+E_{s}+E_{c}+E_{int}
\label{totenergy},
\end{equation}
where the first three terms are the energetic contribution of $M$ cells, lumen and ECM, respectively, and the last term involves the interactions between each constituent. The phase field framework requires that each entity has an energetic contribution dictated by a Ginzburg-Landau potential. For example, for a single cell:
\begin{equation}
E_{pf}[u_{m}]=\int\left[\frac{D_{u}}{2}|\nabla u_{m}|^{2}+\frac{1}{4}u_{m}^{2}(1-u_{m})^{2}\right]d\mathbf{r}
\label{epf}.
\end{equation}
This energy introduces the interface, which is steady in 1D when both minima of the potential ($u_{m}=0$ and $u_{m}=1$) are equally favorable. The coefficient $D_{u}$ sets the width of the interface. One can write similar terms $E_{pf}[s]$ and $E_{pf}[c]$ for the interfaces of the lumen and ECM phase fields \cite{lu2025generic}. An extra energetic term for a single cell is area conservation, which prescribes the cell size $V$; $E_{v}[u_{m}]=\alpha_{m}(V-\int h(u_{m})d\mathbf{r})^{2}$ where $\alpha_{m}$ represents a elastic bulk modulus for the cell and $h(u_{m})=u_{m}^{2}(3-2u_{m})$. 
The role of the function $h(u_{m})$ is to minimize the contribution of the phase field interface when integrating of the whole phase field domain. The authors in Ref.~\cite{lu2025generic} also include an explicit term to control the surface of the cell: $E_{\gamma}[u_{m}]=\gamma\int|\nabla h(u_{m})|^{2} d\mathbf{r}$, where $\gamma$ is the surface tension. We note that the term proportional to $D_{u}$ in Eq.~(\ref{epf})  implicitly includes a surface tension contribution \cite{collins1985diffuse,shao2010computational}. In summary, the energy of a single cell is $E_{u_{m}}=E_{pf}[u_{m}]+E_{v}[u_{m}]+E_{\gamma}[u_{m}]$.

For the ECM, a similar term to $E_{v}[u_{m}]$ is employed to prescribe the size of the ECM, $V_c$; $E_{v}[c]=\alpha_{c}(V_{c}-\int h(c)d\mathbf{r})^{2}$, where $\alpha_{c}$ is the elastic modulus of the ECM. At the same time, this term models the elastic resistance of the ECM to deformations from its relaxed size $V_{c}$. The ECM energetic contribution also considers a pressure term; $E_{p}[c]=-\xi_{c}\int h(c)d\mathbf{r}$, modeling the swelling of a hydrated gel. An equivalent pressure contribution is used in the lumen description; $E_{p}[s]=-\xi_{s}\int h(s)d\mathbf{r}$. Mathematically, the role of these energies, controlled by the pressure coefficients $\{\xi_{c},\xi_{s}\}$, is to tilt the double-well potentials of the Ginzburg-Landau functionals and generate outward motion of the respective interfaces. The total energy of the ECM is given by $E_{c}=E_{pf}[c]+E_{v}[c]+E_{p}[c]$, while the total energy of the lumen is $E_{s}=E_{pf}[s]+E_{p}[s]$.

The energy term $E_{int}$ includes  cell-cell and cell-ECM interactions common in modeling approaches \cite{camley2014polarity,jain2024cell,loewe2020solid}, such as adhesion (interface sharing) and steric repulsion (volume exclusion), $E_{int}=E_{adh}+E_{rep}$. The incorporation of the lumen as a phase field variable requires the incorporation of an interaction energy with the cells and the ECM, which in Ref.~\cite{lu2025generic} has the form of volume exclusion. 

The temporal evolution of  the phase fields $\Phi=\{u_{m},s,c\}$ follows gradient dynamics, obtained by taking the functional derivative of Eq.~(\ref{totenergy}) with respect to each corresponding  field:
\begin{equation}
\partial_{t}\Phi=-\delta E/\delta\Phi
\label{evolPhi}.
\end{equation} 
The evolution of the multicellular phase field model can be coupled to out-of-equilibrium processes, such as cell-cell division, cell proliferation and cell motion, which have been implemented in many multicellular models \cite{shao2010computational,nonomura2012study,belmonte2016virtual,akiyama2019numerical,cao2019plasticity,camacho20223d,kuang2023morphosim,ranamukhaarachchi2025global}. In the context of lumen dynamics, Lu, Fuji and colleagues \cite{lu2025generic} introduced cell growth and division to account for the dynamics observed in MDCK cysts (middle panel, Fig.~\ref{fig4}). 
Specifically, they allowed the target cell volume to evolve in time, $V\rightarrow V(t)$, and introduced a size threshold beyond which a cell divides into
two daughter halves. Most importantly, they incorporated an experimetally-motivated rule for lumen creation: the $s$-phase field is initialized between two cells after cell division (top panel, Fig.~\ref{fig8}F). 
The numerical integration of Eq.~(\ref{evolPhi}) for the MDCK cyst case is shown in Fig.~\ref{fig8}G, where, following cell division, multiple lumens nucleate. 
Driven by their dynamics---including internal pressure and volume exclusion interaction with the other phase fields---these small lumens  eventually converge
into a large, single lumen in a similar fashion to the coalescence mechanism discussed in Section~\ref{hydralumen}. Additionally, the epiblast dynamics (bottom panel, Fig.~\ref{fig4}) can be reproduced if individual cell motion is coupled to Eq.~(\ref{evolPhi}). By prescribing experimental single cell trajectories at the $16$-cell stage into their model, and without cell division and proliferation, the authors showed that the lumen coalescence dynamics driven by cell rearrangement is also possible (Fig.~\ref{fig8}H). For the latter case, an additional constraint of lumen volume ($V_{s}$) conservations was implemented, $E_{v}[s_{i}]=\alpha_{s}(V_{s}-\int h(s_{i})d\mathbf{r})^{2}$ with $i=1,2$, which assumes that lumen fusion is faster than lumen growth \cite{lu2025generic}.

The phase field framework used by Lu, Fuji \textit{et al}.~\cite{lu2025generic} incorporates several experimentally motivated rules that qualitatively reproduce lumen coalescence in MDCK cysts and epiblasts. A natural question, motivated by the rule that nucleates a new lumen after cell division  (top panel, Fig.~\ref{fig8}F), is how cell proliferation---either physiological or abnormal---influences the selection of a single lumen
by continuously generating new small cavities. In the following, we discuss the role of the cell proliferation rate in lumen coalescence.

Cerruti, Puliafito \textit{et al}.~\cite{cerruti2013} studied the effects of cell proliferation rate and cell division orientation in the coarsening lumen dynamics within MDCK cysts by means of a three dimensional Cellular Potts model (CPM) \cite{potts1952some,graner1992simulation,glazier1993simulation}. In the CPM,  cells, ECM and lumen are represented as extended domains of discrete sites on a cubic lattice, where each site carries a domain index $\sigma$ indicating a unique entity, and also a label $\tau(\sigma)$ that prescribes the physical identity of the domain (cell, ECM or lumen). The evolution of the system follows a Monte Carlo update: at each step one site, $\sigma'$, is randomly selected and an attempt of overwriting a neighboring site, $\sigma''$, from a different domain, is made. The attempt is accepted if the change in the discretized energy of the system (see below) is negative, $\Delta H\leq0$, and rejected otherwise with probability $1-e^{\Delta H/T_{m}}$, where $T_{m}$ specifies the amplitude of the fluctuations for  attempts $\sigma'\rightarrow\sigma''$. The mechanical energy of the MDCK cyst model used by in Ref.~\cite{cerruti2013} is
\begin{eqnarray}
E &=& \sum_{cells}\bigg\{( \alpha S_{cell-cell} + \beta S_{cell-ECM} + \gamma S_{cell-lumen}  \nonumber\\
&& +C_{V}(V_{cell}) + C_{A}(S_{cell-lumen})\bigg\},
\label{potts1}
\end{eqnarray}
where $S_{i-j}$ indicates the shared surface between the respective physical domains. The surface tensions obey $\alpha>\beta>\gamma>0$, and thus cell-cell contacts are favored over cell-ECM and cell-lumen contacts. The term proportional to $C_{V}$ enforces cell volume conservation, while the last term models the conservation of the apical surface. The model includes cell division to tile the spherical acinar structure and, similar to Ref.~\cite{lu2025generic}, each division generates a small lumen domain  between neighboring cells.

The model predicts that the orientation of the division plane, parameterized by 
an angle $\phi$ affects the lumen selection. The baseline $\phi=0$ is defined as a plane orthogonal to the apical surface---following the typical division pattern observed in MDCK cysts with a strong apicobasal polarity  \cite{martin2007pten,zheng2010lgn}. Below $\phi=20^{\circ}$, on average, the outcome is physiological: only one lumen is selected after the coalescence dynamics. However, for angles $\phi>20^{\circ}$, the average number of lumens present after long simulation times increases linearly with $\phi$ up to $\phi=90^{\circ}$ \cite{cerruti2013}. This $\phi$-mediated transition reflects frustration of the coalescence process due to misaligned proliferation, but is not the full picture. Cerruti, Puliafito and colleagues \cite{cerruti2013} showed that the disruption of single-lumen selection also depends on the comparison ratio between the proliferation rate and the time scale of the coalescence events. Specifically, if the cell division rate is relatively slow compared to lumen fusion events  coalescence can be completed and a single lumen is formed, even when $\phi=90^{\circ}$ \cite{cerruti2013}. 

Lee \textit{et al}.~\cite{lee2025permeability} further explored the consequences of cell proliferation in lumen coarsening by investigating how the interplay between luminal pressure ($\xi$) and cell proliferation rate ($\tau_{V}$) controls lumen shapes in pancreatic organoids, using the phase field approach described above. The simulations showed that after cell and lumen growth, the final geometry can be either a spherical lumen or a fragmented multilumen network depending on the values of $\xi$ and $\tau_{V}$. Specifically, a  fragmented luminal geometry arises when the pressure is low and the proliferation rate is high. Experiments in MDCK cysts, together with phase field modeling (Fig.~\ref{fig8}I), confirmed that arresting proliferation in branching organoids allows  coarsening to dominate, driving  the system toward a spherical shape, consistent with the findings of Cerruti \textit{et al}.~\cite{cerruti2013}.

In summary, the agent-based models reviewed here, namely phase field and cellular Potts models, illustrate that is possible to investigate the coarsening-like dynamics in the lumenogenesis of MDCK cysts, epiblasts and pancreatic organoids, using experimentally-motivated physics rules. The models further demonstrate that cell dynamics---motion, division and proliferation---act as  active boundary conditions on  luminal spaces, thereby determining whether a single dominant lumen is successfully established.

\subsection{Oscillatory dynamics of lumen growth}\label{osci}
Whether a single lumen is created through coarsening-like dynamics or nucleates as a unique cavity during lumenogenesis, its subsequent growth can display rich nonlinear behavior. As we discussed at the end of Section~\ref{pumping}, lumen growth is not always uniform, but can exhibit oscillations (Figs.~\ref{fig6}E-H). This type of nonlinear dynamics has been observed in several systems \cite{CFutterer_2003,niimura2003time,Soriano2006,ruiz2017organ,dasgupta2018physics,chan2019hydraulic}, suggesting it is a robust phenomenon in lumenogenesis. In the literature, a common unifying view of oscillatory lumen dynamics is that they arise from competition between a driving mechanism of lumen inflation and a negative feedback that opposes growth.

\subsubsection{Leakage-driven oscillations}\label{dasgupta}
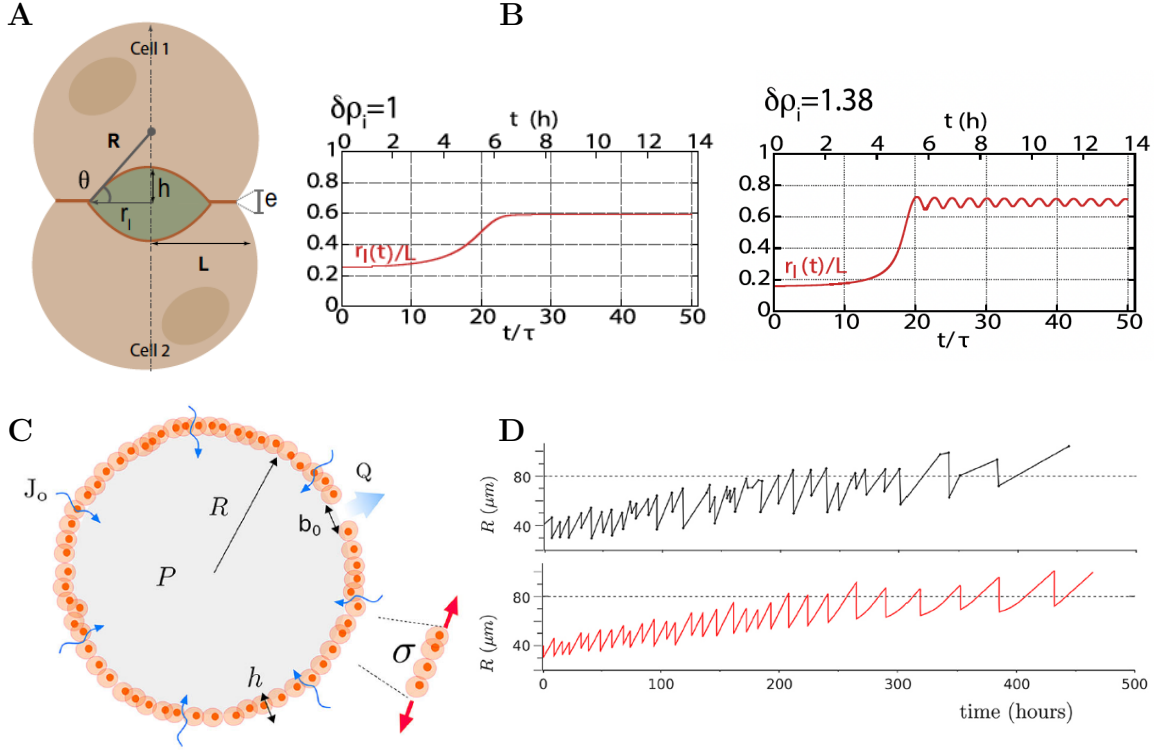
\begin{figure*}
\centering
\begin{tikzpicture}
\node[inner sep=0] (img)
{\includegraphics[width=0.88\linewidth]{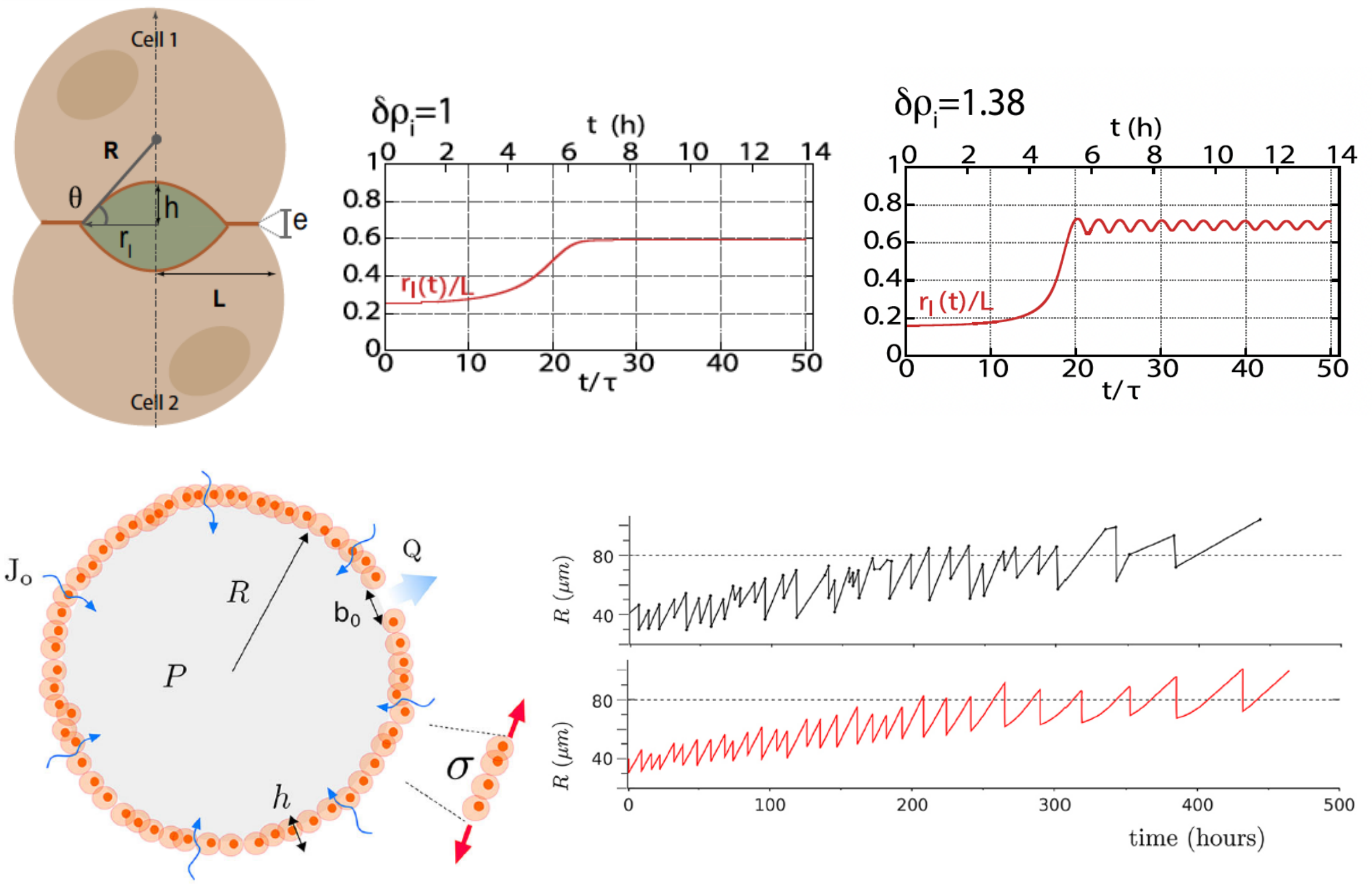}};
\node at (img.north west) [xshift=0mm, yshift=0mm]{\large\bfseries A};
\node at (img.north west) [xshift=65mm, yshift=0mm]{\large\bfseries B};
\node at (img.north west) [xshift=0mm, yshift=-55mm]{\large\bfseries C};
\node at (img.north west) [xshift=65mm, yshift=-55mm]{\large\bfseries D};
  \end{tikzpicture}
  \caption{Oscillatory behavior in models of lumen dynamics. (A) Schematic drawing of a lumen within two cells. (B) Temporal evolution of the normalized lumen radius over time for different pumping efficiencies $\delta p_{i}$. (A) and (B) are adapted from \cite{dasgupta2018physics}. (C) Schematic representation of a cyst undergoing a localized rupture of size $b_{o}$. (D) Temporal dynamics of lumen radii within a multicellular layer of MCF10-DCIS.com cells (top panel), and in a theoretical model including rupture-healing dynamics (bottom panel). Both (C) and (D) are adapted from \cite{ruiz2017organ}.}
    \label{fig9}
\end{figure*}

In Dasgupta \textit{et al}.~\cite{dasgupta2018physics}, the authors modeled leakage effects to study oscillatory dynamics of the lumen between two hepatocytes in the bile canaliculi, where a thin intercellular cleft (or gap) connects the lumen with the outside (Fig.~\ref{fig9}A), providing a leakage pathway. The volume and ion dynamics in the lumen and in the cleft are described by equations equivalent to Eqs.~(\ref{VolChange})-~(\ref{channelI}). In this case, the flux at the cleft boundary represents the leakage, which is geometry-dependent: as the lumen expands ($r_{l}\rightarrow L$) the length of the neighboring cleft shortens, offering  a path of less resistance for the leakage. In the limit $L-r_{l}\ll \xi_{v}$, the volume leakage scales as $\sim(L-r_l)^{-1}$ \cite{dasgupta2018physics}, which diverges as the lumen approaches to the cell basal boundary.

The geometry-dependent leakage is the restoring mechanism that saturates growth, which above a critical threshold of active ion pumping stabilize a steady-state lumen. In order to obtain out-of-equilibrium dynamics, Dasgupta \textit{et al}.~\cite{dasgupta2018physics} argued that is necessary to make the cortical tension $\sigma$ a temporal variable to account for the lumen expansion. In particular, the authors motivate the introduction of a viscous correction as a natural consequence of large cortical expansions due to lumen inflation, resulting in a strain rate-dependent cortical tension \cite{kruse2005generic}:
\begin{equation}
\sigma(t)=\sigma_{o}\left[1+\tau_{c}\left(\frac{1}{R}\frac{dR}{dt}+\frac{\sin(\theta)}{2(1-\cos(\theta))\frac{d\theta}{dt}}\right)\right].
\label{sigmaT}
\end{equation}
Here, $\tau_{c}$ is a relaxation time characterizing the viscous response, and the term in round brackets is a measure of the deformation rate of the lumen area, with $\theta$ the lumen-cleft contact angle and $R$ the curvature radius of the lumen (Fig.~\ref{fig9}A). The constant value $\sigma_{o}$ is obtained from solving the tension balance at the lumen-cleft edge in steady state, $\sigma_{o}(1-\cos(\theta_{s}))=E$, where $E$ is an E-cadherin correction and $\theta_{s}$ the contact angle in steady state. At finite $\tau_{c}$, the model of Dasgupta \textit{et al}.~\cite{dasgupta2018physics} exhibits a transition from overdamped to sustained oscillations of the lumen growth when the pumping efficiency---ratio of active pumping versus passive ion transport---increases (Fig.~\ref{fig9}B).

This model demonstrates that it is possible to build a theory for lumen oscillations, such as those observed in bile canaliculi \cite{dasgupta2018physics}, solely based on mechanical, geometric, and transport arguments. In this framework, the oscillations emerge as an interplay between a viscous cortical response and a geometry-dependent leakage. In principle, the same conceptual framework could be extended to other lumen systems in which leakage responds to lumen expansion.

\subsubsection{Hydraulically gated oscillations from tissue rupture}\label{tirupture}

In the context of multicellular cysts and regenerating organisms, oscillatory dynamics have been experimentally described as a result of alternations in rupture and healing of the tissue surrounding the cavity \cite{kucken2008osmoregulatory,wang2019mouth}. Ruiz-Herrero \textit{et al}.~\cite{ruiz2017organ} studied lumen dynamics in MCF10-DCIS.com  cell spheroids, and theoretically identified the minimal ingredients necessary for sustaining  oscillatory lumen growth. They modeled the active balloon as a spherical shell with wall stress $\sigma$, thickness $h$, and an inner cavity radius $R$. The cellular wall can modify its volume through cell proliferation at rate $J_{c}$;
\begin{equation}
4\pi d(R^{2}h)/dt=J_{c}.
\label{cellflow}
\end{equation}
The usual osmotic machinery inflates the lumen, thus increasing the stress: $\sigma=PR/2h$. The lumen  can rupture when this stress exceeds a threshold ($\sigma>\sigma_{2}$). The consequence of the rupture can be captured by a leak flux $Q$ (Fig.~\ref{fig9}C)---modeled as a Poiseulle flow---in the lumen volume conservation equation;
\begin{equation}
d{R}/dt\sim(\Pi-P)-Q\Theta(\sigma-\sigma_{2}),
\label{volH}
\end{equation}
where $\Theta$ is a sigmoid function. The outflow eventually stops when the wall stress is relaxed to $\sigma_{1}$, modeling tissue healing. In a first attempt, the authors closed their model by assuming a purely elastic response of the tissue wall:
\begin{equation}
\frac{d\sigma}{dt} = \frac{E}{R}\frac{dR}{dt},
\label{cons1}
\end{equation}
where $E\sim20$ kPA is the elastic modulus of the tissue.

The system of Eqs.~(\ref{cellflow})-(\ref{cons1}), including the tangential force balance in the cavity wall has oscillatory solutions, including the lumen size $R$ \cite{ruiz2017organ}, thanks to the separation of stress thresholds ($\sigma_{2}>\sigma_{1}$). However, the oscillations were about constant mean values, failing to reproduce the progressive increase of $R$ observed experimentally in MCF10-DCIS.com  cell spheroids (top panel, Fig.~\ref{fig9}D). To address this limitation,  Ruiz-Herrero \textit{et al}.~\cite{ruiz2017organ} explored more realistic tissue mechanics models, compared to the simplistic elastic view, in a second modeling attempt.

After  a viscoelatic description was also unsuccessful (see details in Ref.~\cite{ruiz2017organ}), the authors opted for a threshold-based elastic-plastic constitutive law. In this case, the cavity wall behaves elastically below a yield stress $\sigma<\sigma_{y}$ and undergoes an irreversible plastic deformation when $\sigma>\sigma_y$ with a  plastic modulus $H$ that is smaller than $E$. As $\sigma_{y}\leq\sigma_{2}$, after every rupture-healing cycle $i$ a plastic deformation occurs, modifying the lumen radius
\begin{equation}
R_{i}=R_{max}\left(1+\frac{\sigma_{2}-\sigma_{y}}{\frac{1}{H}-\frac{1}{E}}\right)^{i-1},
    \label{cycles}
\end{equation}
where $R_{max}$ is the maximum lumen radius. This approach captures both the oscillatory dynamics and the progressive size increase observed in growing cysts in MCF10-DCIS.com cell spheroids (bottom panel, Fig.~\ref{fig9}D).

The theoretical framework proposed by Ruiz-Herrero and colleagues \cite{ruiz2017organ} suggests that a mechanical phase transition, triggered by tissue rupture, is key to reproducing oscillatory lumen growth dynamics in a mechanics-based  model. Similar to Section~\ref{dasgupta}, this underscores the importance of choosing an appropriate constitutive law to capture nonlinear lumen dynamics observed experimentally.

\subsubsection{Active flexoelectricity as a control of lumen oscillations}
Duclut and colleagues \cite{duclut2019fluid} introduced a different theoretical mechanism for lumen oscillations, by coupling hydraulics, mechanics and electricity of a polarized spherical tissue with a cavity (see also \cite{duclut2021hydraulic}). The authors followed the work of Sarkar \textit{et al}.~\cite{sarkar2019field} and introduced a coarse-grained theory in which radially polarized cells exchange momentum and charge with the interstitial fluid, and flexoelectric effects---ionic currents generated by polarity gradients---modify the surface tension at the cavity boundary. Notably, flexoelectricity allows surface tension to become negative, acting as a mechanism for overcoming the lumen nucleation barrier. The authors showed that in the small lumen limit, oscillations of the lumen radius can be sustained when the cavity is inflated by flexoelectric effects while  it is deflated by ion pumping. If the time scales of these mechanisms are sufficiently separated, it is possible to show that oscillations emerge via Hopf bifurcation \cite{duclut2019fluid}.

The contribution of Duclut \textit{et al}.~\cite{duclut2019fluid} makes explicit a general principle shared by all the lumen models reviewed in this subsection: oscillations emerge through a Hopf bifurcation from the competition between two characteristic time scales. In Dasgupta \textit{et al}.~\cite{dasgupta2018physics}, this is a competition between  osmotic inflation (limited by leakage) and  viscous relaxation of the tissue cortex. In Ruiz-Herrero \textit{et al}.~\cite{ruiz2017organ}, it is between  slow osmotic lumen growth and  fast discharge following tissue rupture. This commonality suggests that oscillatory lumen dynamics can be framed  within the general language of nonlinear dynamics, independently of the
specific biological system being modeled.

\subsection{Emergence of non-spherical luminal morphologies}\label{shapesModel}
We have motivated lumenogenesis from the point of view of morphogenesis, where a key signature of the process is the creation of unique and complex forms. As described in Section~\ref{sculpting}, the lumen is not always a simple symmetrical object  and in many physiological contexts the "spherical cow" \cite{bovyn2024shaping} of lumen formation, governed by the Young-Laplace Eq.~(\ref{OGLaplace}), is the exception rather than the rule (Figs.~\ref{fig7p2}A and~\ref{fig7p2}B). Understanding the emergence of complex active balloon shapes requires a mathematical framework that links single-cell  properties to lumen shape \cite{vasquez2021physical,mukenhirn2024tight,indana2024lumen,guha2026control}, and that captures how active interactions between cells and their environment can drive a spherical lumen out of equilibrium \cite{belmonte2016virtual,rozman2020collective,gill2024developmental}.

\subsubsection{Cell-level constraints on lumen geometry}\label{cellcon}

In Section~\ref{sculpting}, we reviewed two scenarios where lumens in MDCK cysts deviate from spherical shapes (see Figs.~\ref{fig7p2}A and~\ref{fig7p2}B) due to a cell-level restriction, a preferred apical length \cite{vasquez2021physical,mukenhirn2024tight}. Here, we focus on modeling frameworks that explain how single-cell constraints allow the emergence of nonspherical lumen geometries \cite{honda2008computer,vasquez2021physical,mukenhirn2024tight,indana2024lumen,guha2026control}.

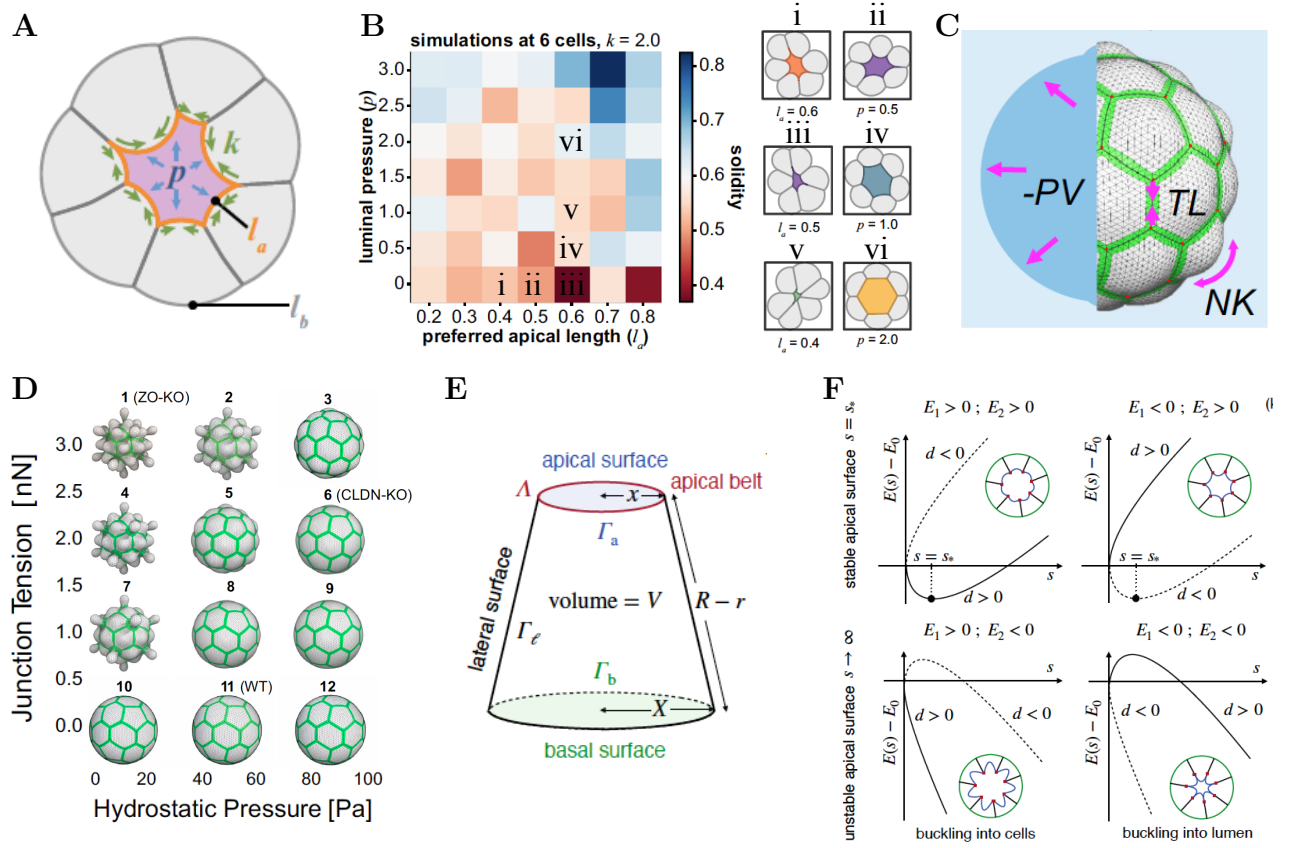
\begin{figure*}
\centering
\begin{tikzpicture}
\node[inner sep=0] (img)
{\includegraphics[width=0.98\linewidth]{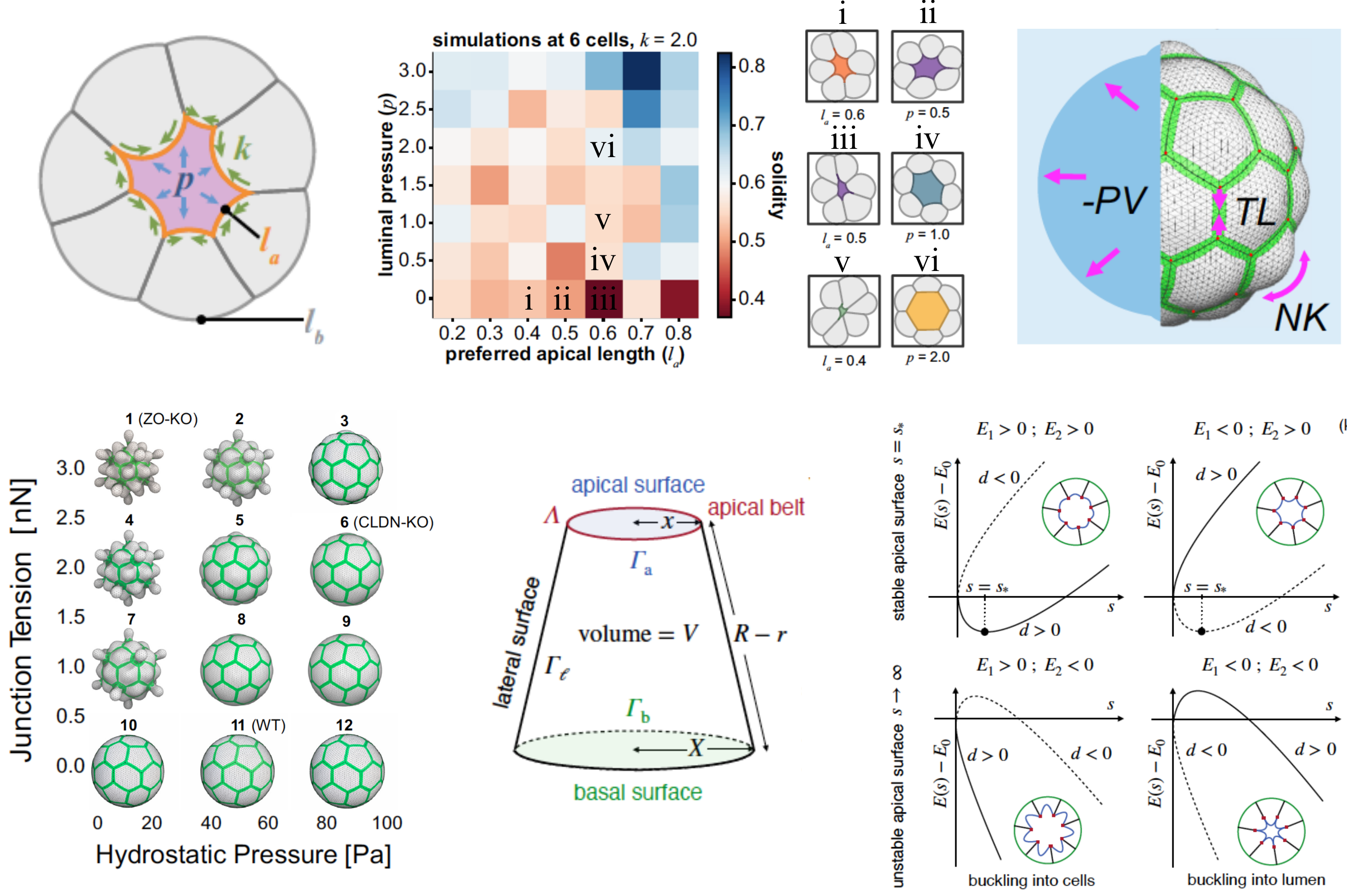}};
\node at (img.north west) [xshift=3mm, yshift=-3mm]{\large\bfseries A};
\node at (img.north west) [xshift=49mm, yshift=-3mm]{\large\bfseries B};
\node at (img.north west) [xshift=125mm, yshift=-3mm]{\large\bfseries C};
\node at (img.north west) [xshift=3mm, yshift=-51mm]{\large\bfseries D};
\node at (img.north west) [xshift=60mm, yshift=-51mm]{\large\bfseries E};
\node at (img.north west) [xshift=110mm, yshift=-51mm]{\large\bfseries F};
  \end{tikzpicture}
  \caption{Single-cell mechanics stabilize complex lumen morphologies. (A) Schematic of a lumen exhibiting an irregular shape. $p$ and $k$ are the dimensionless versions of the lumen pressure ($P_{L}$) and the stringency ($k_{l}$), respectively. (B) Phase diagram in $l_{a}-p$ space illustrating the differences in solidity of the shapes that minimize the energy in Eq.~(\ref{hamiltonian}). Adapted from \cite{vasquez2021physical}. (C) Schematic of the energy $\mathcal{F}$ at the lumen surface. (D) Equilibrium shapes in $P$-$T$ space after minimizing Eq.~(\ref{Ecurvature}) subjected to a constant apical area per cell. The knockout (KO) cases are reviewed in Section~\ref{sculpting}. Adapted from \cite{mukenhirn2024tight}. (E) Geometry of a single cell in the mean field vertex model. (F) Distinct apical surface modes as a functions of the energy coefficients $E_{1}$ and $E_{2}$ from the expansion in Eq.~(\ref{expansion}). Adapted from \cite{guha2026control}.
  }
    \label{fig10}
\end{figure*}

Vasquez \textit{et al}.~\cite{vasquez2021physical} used a 2D vertex model, representing a cross section of the MDCK cyst, to investigate the competition between  luminal pressure $P_{L}$ and  preferred apical length $l_{a}$ (Fig.~\ref{fig10}A). In the model, each cell is represented by four boundaries: two straight lateral sides with fixed length, and two curved sides, one apical and one basal, with preferred length $l_a$ and $l_{b}$, respectively. In cellular-mechanics applications, spatiotemporal dynamics in vertex models typically follow from the assumption of an overdamped regime \cite{farhadifar2007influence,fletcher2014vertex,bi2016motility,alt2017vertex,perez2020vertex}:
\begin{equation}
\xi\frac{d\mathbf{r}_{i}}{dt}=\mathbf{F}_{i},
    \label{vertex}
\end{equation}
where $\xi$ is a friction coefficient, the vector $\mathbf{r}_{i}$ encodes the vertices of the $i$-th computational cell, and $\mathbf{F}_{i}$ represent the forces at each vertex. In this particular case, the evolution is variational, i.e.,  $\mathbf{F}_{i}=-\partial H/\partial \mathbf{r}_{i}$, where
\begin{eqnarray}
H &=& -P_{L}A_{L}+ \sum_{i=1}^{N} \bigg\{ k_{A}(A_{i} - A_{0})^2 + 
k_{l}\left( (l_{a,i} - l_{a})^2 \right. \nonumber\\
&& \left. + (l_{b,i} - l_{b})^2 + l_{l,i}^2 \right) \bigg\}
\label{hamiltonian}
\end{eqnarray}
is the energy (or Hamiltonian) to be minimized. The first term accounts for the work done along the lumen surface, the term proportional to $k_{A}$ enforces cell area conservation, while the term proportional to $k_{l}$ prescribe the side lengths. The authors in Ref.~\cite{vasquez2021physical} introduces $k_{l}$ as a parameter that controls how stringently cells regulate their membrane size, reflecting multiple homeostatic processes and not only cortical tension. Upon energy minimization, and including cell division, the model predicts that increasing $l_{a}$ decreases lumen solidity---the ratio between the lumen area and its convex hull---at low luminal pressures (Fig.~\ref{fig9}B), producing noncircular shapes. In contrast, high luminal pressures restore circular shapes, in agreement with pressure-dominated regimes at large lumen radius in the experiments (Fig.~\ref{fig7p2}A).
Mukenhirn and colleagues~\cite{mukenhirn2024tight} studied the folded luminal shapes that emerges in MDCK cyst mutants (ZO knockout, Fig.~\ref{fig7p2}B) as an energy minimization problem. In their modeling approach, the luminal cavity is tessellated by an hexagonal array of cells, and each cell's apical surface is modeled as a paraboloid of base radius $r_{a}$ and depth $h$. Then, the free energy of the lumen surface is given by
\begin{equation}
\mathcal{F} = -PV + TL + N\kappa \int H(\mathbf{r})^2 \, dA.
\label{Ecurvature}
\end{equation}
The first term involves the luminal pressure effects similar to Eq.~(\ref{hamiltonian}), while the second term is an energy representing the line tension $T$ at the apical junctions of all neighboring cell pairs, where $L$ is the total apical length. The last term models the apical bending energy, with $N$ the number of cells, $\kappa$ the effective bending energy of the apical membrane, and $H(\mathbf{r})$ the mean curvature of this membrane (Fig.~\ref{fig10}C). Similar to Ref.~\cite{vasquez2021physical}, the authors impose a constraint on the total apical area, $S$, computed by assuming each cell has a paraboloid geometry \cite{harris1998handbook,mukenhirn2024tight}:
\begin{equation}
\label{SN}
S=N\frac{\pi r_{a}}{6h^{2}}\left[(r_{a}^{2}+4h^{2})^{3/2}-r_{a}^{3}\right].
\end{equation}
The geometric parameterization of the cell shape also offers an analytic calculation of the free energy in Eq.~\ref{Ecurvature}, from which the authors conclude that the apical bending rigidity opposes  lumen collapse when  hydrostatic pressure is relatively small. Numerical simulations of the minimization of $\mathcal{F}$ via a relaxational vertex model of the surface \cite{brakke1992surface}, subjected to the constrain of constant surface apical (luminal) area $S$, reproduces the irregular and spherical luminal shapes observed in their experiments across a physiological range of hydrostatic pressures and apical tension (Fig.~\ref{fig10}D). Together, the models of Vasquez \textit{et al}.~\cite{vasquez2021physical} and Mukenhirn \textit{et al}.~\cite{mukenhirn2024tight} demonstrate that a geometrical restriction in the apical size can explain a broad range of irregular lumens in normal and mutant MDCK cysts, respectively.

Guha Ray \textit{et al}.~\cite{guha2026control} extended the mean field approximation introduced by Mukenhirn and colleagues \cite{mukenhirn2024tight} by including  mechanical contributions of the basal and lateral domains. Physically, this means that they assume that cell mechanics are dictated solely by cell-cell and cell-ECM adhesion, together with cortical tension. 
The geometry consists of a cyst of $N$ identical, incompressible cells surrounding a central cavity, with each cell having a truncated conical morphology (Fig.~\ref{fig10}E) with apical area $A$. The total energy of the system is given by $E=Ne_{cell}-PV$, where 
\begin{equation}
e_{cell}=\Gamma_a A+ \Gamma_b\pi X^2 +\Gamma_\ell\pi(RX - rx) + \Lambda(2\pi x)
\label{energyPER}.
\end{equation}
The first three terms correspond to the membrane tensions at the apical, basal, and lateral sides, respectively. The last term is associated with the line tension generated by the apical belt enclosing the apical area (Fig.~\ref{fig10}E). 

Incompressibility of the cells and the imposed geometrical packing (similar to Ref.~\cite{mukenhirn2024tight}) allows one to write the total energy as a function of only one degree of freedom, the lumen radius $r$. In this reduction, it is possible to explore the apical instability of MDCK cyst mutants (ZO knockout, Fig.~\ref{fig7p2}B) when the apical area is fixed, by simply allowing apical bending with an amplitude $d$. When there is no bending ($d=0$), the dimensionless equilibrium lumen radius can be calculated and is equal to $r=r_{*}=\sqrt{N}/2$ \cite{guha2026control}. Near this equilibrium, the effect of apical bending can be studied through a stability analysis by introducing the small parameter $s=r-r_{*}\ll1$ in the total energy of the cyst, resulting in the following asymptotic expansion:
\begin{equation}
E = E_{0} \mp E_{1} s^{1/2} + E_{2} s + \mathcal{O}(s^{3/2}),
\label{expansion}
\end{equation}
valid in the limit $N\gg1$, in which
\begin{eqnarray}
E_{1} &=& \frac{N^{5/12}}{4\left(\sqrt{N} + 8v\right)^{1/3}} \left\{ p\left(\sqrt{N} + 8v\right)^{1/3} \sqrt{N}\right. \nonumber\\
&& \left. - 2\sqrt{N}\gamma - 4 \right\},
\label{E1}
\end{eqnarray}
\begin{eqnarray}
E_{2} &=& \frac{Np}{4} - \sqrt{N}\lambda - \frac{N^{1/6}}{\left(\sqrt{N}+8v\right)^{1/3}}\left(\sqrt{N} \right. \nonumber\\
&& \left.+ \frac{1}{\sqrt{N}+8v}\right) + \frac{N\gamma}{2}\left\{1 - \frac{1}{\left(\sqrt{N}+8v\right)^{1/3}\sqrt{N}}\right. \nonumber\\
&& \left.\left(\sqrt{N} + \frac{1}{\sqrt{N}+8v}\right)\right\}
\label{E2},
\end{eqnarray}
where $v=3V/4\pi a^{3/2}$, $p=\sqrt{a}P/\Gamma_{b}$, $\gamma=\Gamma_{l}/\Gamma_{b}$, $\lambda=\Lambda/\Gamma_{b}\sqrt{a}$, and $a=A/\pi$ with $A=\pi(x^{2}+d^{2})$ the apical area.

The signs of $E_{1}$ and $E_{2}$ determine the type and stability of the possible luminal morphologies (Fig.~\ref{fig10}F): if $E_{2}>0$, the total energy $E$ has a minimum independently of the sign of $E_{1}$. This minimum corresponds to a slight bend of the apical surfaces, either into the cells (for $E_{1}>0$) or into the lumen (for $E_{1}<0$). 
Conversely, if $E_{2}<0$, the energy has no minimum and the apical surfaces are unstable to a significant increase in the amplitude $d$, which is eventually saturated by nonlinear mechanisms (outwards when $E_{1}>0$  and inwards when $E_{1}<0$). This mathematical reasoning is related to the physics of the problem via Eqs.~(\ref{E1}) and~(\ref{E2}): high pressure and high lateral tension, relative to the basal tension, stabilize the lumen ($E_{2}>0$), while a large apical belt tension, relative to the apical area and the basal tension, tends to destabilize the surface of the cavity ($E_{2}<0$). Moreover, a key prediction of this extended model is that  the increase in lateral contractility observed in MDCK mutants \cite{mukenhirn2024tight,guha2026control} may represent a compensatory response to the destabilizing effect of the belt, as the contribution proportional to $\gamma$ ($\lambda$) in $E_{2}$ is positive (negative). In words, enhanced lateral tension penalizes cell shape deformations necessary for tissue folding, underscoring the coordinated regulation of all cell surfaces during luminal shape changes.

The three studies discussed here depict the value of vertex models and their mean-field approximations for the theoretical and computational study of lumenogenesis at the tissue level under cellular constraints. In particular, taken together, they demonstrate that the conservation of the apical area, combined with the differentiated tensions at the apical, basal and/or lateral cell surfaces, is sufficient to capture the instability of a smooth spherical lumen into a convoluted shape.

\subsubsection{Collective cell dynamics in lumen sculpting}

The models discussed above describe how static, cell-level constraints can select nonspherical cavities as equilibrium shapes. However, the balloon is active, with an epithelial layer that is out-of-equilibrium: cells grow, divide, and rearrange. The contribution of these collective processes to energy minimization theories enables new classes of lumen morphologies \cite{belmonte2016virtual,rozman2020collective,gill2024developmental}.


In Section~\ref{intro}, we referred to autosomal dominant polycystic kidney disease (ADPKD), the most common genetic kidney disorder, which profoundly affects  lumen morphology \cite{belmonte2016virtual}. 
Experimental evidence indicates altered cell-cell adhesion and increased cell proliferation in affected kidneys \cite{nadasdy1995proliferative,kher2011ectopic}.
In ADPKD, renal tubules develop two types of abnormal localized cysts: protruded (budded) spherical cavities connected to the tubule lumen by a narrow neck, and broader bulges in which the tubule wall swells outward  \cite{baert1978hereditary}. 
Belmonte \textit{et al}.~\cite{belmonte2016virtual} built a three-dimensional CPM, consisting of cells, lumens and  ECM, to study what triggers one abnormal morphology versus the other. Their model is similar to that of Cerruti \textit{et al}.~\cite{cerruti2013} (Section~\ref{phasefieldS}), but includes an important extension. In the new model, each cell surrounding the lumen is represented not as a homogeneous domain, but as four compartments: apical, basal, lateral and cytoplasmic. This detail at the subcellular level is required to isolate the adhesive mechanism mediated by cadherin, relevant in ADPKD, to the cell-cell interface.

In addition, a novel ingredient of the Belmonte \textit{et al}.~\cite{belmonte2016virtual} model is the phenomenologically motivated contact inhibition of proliferation \cite{martz1972role}, in which the evolution of the cell target volume $V_t$ is controlled by the contact fraction $\alpha = S_{contact}/S_{total}$, where $S_{contact}$ is the surface area shared with neighboring cells and $S_{total}$ is the total cell surface area. The idea is that fully surrounded cells cannot grow ($\alpha=1$; $dV_{t}/dt=0$), while cells with available space around them can proliferate ($\alpha<1$; $dV_{t}/dt>0$). Mathematically, the authors write 
\begin{equation}
    \frac{dV_t}{dt} = \kappa\, \frac{\alpha_c^n (1 - \alpha^n)}{\alpha_c^n + \alpha^n},
    \label{volgogo}
\end{equation}
where the right hand side is a Hill function that decays with $\alpha$. Here $\kappa$ is the maximum growth rate of the cell volume, $\alpha_c$ is a critical contact fraction for inhibition, and $n$ is a Hill coefficient. Once a volume threshold is reached, cells divide. In addition, the model includes explicit lumen dynamics \cite{engelberg2011mdck}:
\begin{equation}
    \frac{dV_{t,lumen}}{dt} = N_{cells} - \kappa_r S_{lumen},
    \label{lumenPotts}
\end{equation}
where $N_{cells}$ is the number of cells surrounding the cavity, $S_{lumen}$ is the lumen surface area, and $\kappa_r$ models the rate of lumen deflation due to leakage. Finally, a lumen is initiated at sites where three apical compartments are in contact.

To emulate the experimental observations of diseased kidneys, the authors simulate the case where a target cell (TC) on the surface of a tubular cavity has either a reduced cell-cell adhesion strength or a reduced contact inhibition. In the former case, the TC protrudes towards the ECM reducing its $S_{contact}$ ($\alpha$) and thus entering  a stage of proliferation via Eq.~(\ref{volgogo}). The resulting cyst buds outward from the tube and remains connected to it by a narrow neck, matching morphologies of previously observed cysts in human ADPKD nephrons and in-vitro induced cysts with reduced adhesion \cite{baert1977pathogenesis,kher2011ectopic}. On the contrary, decreasing the contact inhibition (increasing $\alpha_{c}$) caused TCs to proliferate within the plane of the initial tubular tissue, forming an expanding patch of cells that swells the epithelial tube wall outward. Thus, the model showed that the two ADPKD-associated cyst morphologies arise from two different perturbations: loss of cell-cell adhesion drives the budded shape, and loss of contact inhibition results in local swelling of the tissue.

From a purely theoretical perspective, Rozman and colleagues \cite{rozman2020collective} explored the role of collective cell dynamics in sculpting lumen shapes using a 3D vertex model with an energy structure similar to those  reviewed in Section~\ref{cellcon}, and with distinctive surface tensions for the apical ($\alpha$), basal ($\beta$), and lateral ($\gamma$) sides. A key addition is that cells surrounding the lumen are allowed to rearrange through active junctional noise, which mimics T1 transitions and fluidizes the tissue \cite{bi2015density,Krajnc2018}. 
These rearrangements are implemented with a threshold rule: if a cell-cell junction becomes shorter than a critical length, a T1 transition occurs with probability $1$; otherwise, it occurs with a small probability that is proportional to a transition rate $K_{T1}$, which decrease linearly in time.

\begin{figure*}
\centering
\begin{tikzpicture}
\node[inner sep=0] (img)
{\includegraphics[width=0.98\linewidth]{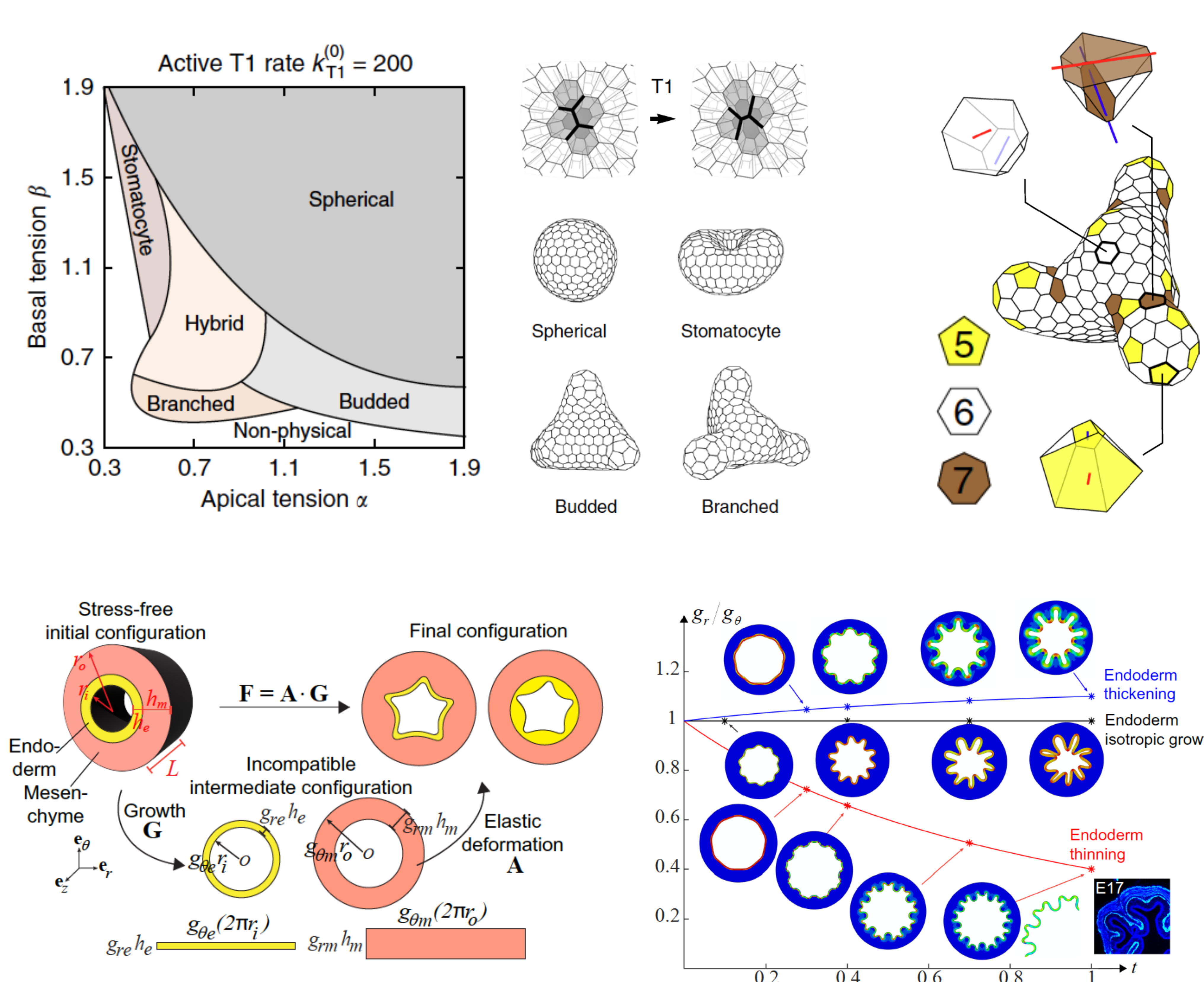}};
\node at (img.north west) [xshift=3mm, yshift=-3mm]{\large\bfseries A};
\node at (img.north west) [xshift=134mm, yshift=-3mm]{\large\bfseries B};
\node at (img.north west) [xshift=3mm, yshift=-80mm]{\large\bfseries C};
\node at (img.north west) [xshift=97mm, yshift=-80mm]{\large\bfseries D};
  \end{tikzpicture}
  \caption{Collective and tissue-scale control of luminal shapes. (A) Phase diagram of possible cyst morphologies in the $\alpha-\beta$ space. The initial active T1 rate in all the simulations is $k_{T1}^{(0)}=200$ (left). Schematic representation of a T1 transition and the four characteristic morphologies in the model (right). (B) Characterization of packing topology in the branched morphology. Adapted from \cite{rozman2020collective}. (C) Schematics of the 2D model geometry of the chick gut lumen, and the multiplicative decomposition of the deformation gradient $\mathbf{F}$. (D) Growth models for the foregut: Endoderm thickening ($g_{r}=1+1.2t$, $g_{\theta}=1+t$), isotropic growth ($g_{r}=g_{\theta}=1+t$), and endoderm thinning ($g_{r}=1-0.2t$, $g_{\theta}=1+t$). The later captures the period-doubling instability seen in E17 foregut (inset). Adapted from \cite{gill2024developmental}, where the reader can find the morphologies in the midgut and hindgut as well.}
    \label{fig11}
\end{figure*}
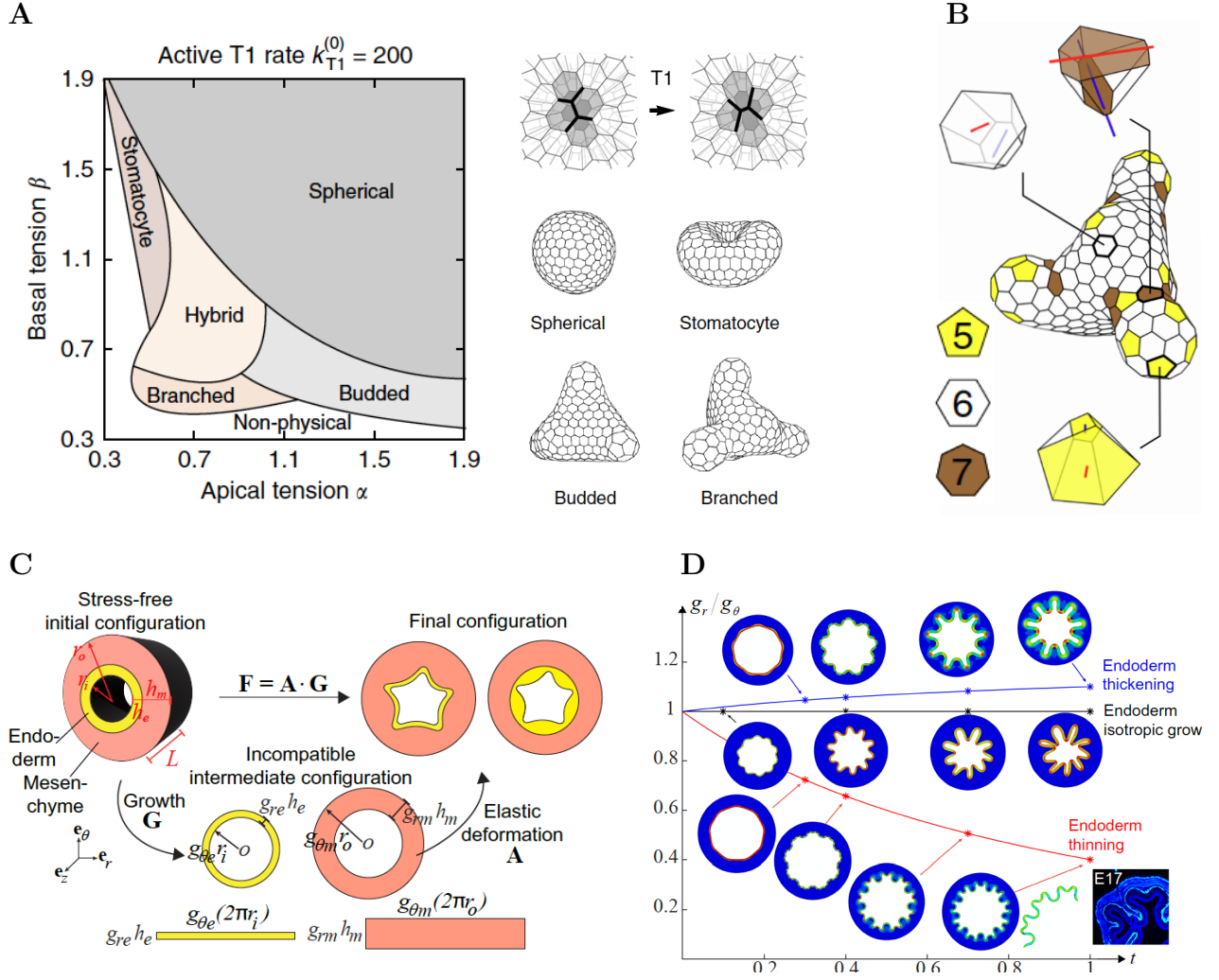

The model by Rozman \textit{et al}.~\cite{rozman2020collective} produces four different morphologies when the initial condition is a spherical cyst with $N_{c}$ cells and a fixed lumen volume: spherical, stomatocyte (cup-like), budded and branched (Fig.~\ref{fig11}A). The first three morphologies can exist with ($K_{T1}>0$) or without ($K_{T1}=0$) cell rearrangement. The origin of the budded and stomatocyte morphologies are the incompatibility of the preferred lumen volume with the total apical surface, and the apicobasal tension asymmetry ($\alpha\neq\beta$). Specifically, the cell width-to-height ratio scales inversely with the tissue tension, $\sim(\alpha+\beta)^{-1}$, so reduced  tension drives a columnar to squamous transition \cite{Krajnc2018}. This leads to a larger apical surface at a constant lumen volume, and the accommodation of this extra tissue surface generates nonspherical luminal geometries (similar to the convoluted shapes reviewed in Section~\ref{cellcon}). The selection, by energy minimization, between budded and stomatocyte shapes depends on the differential tension $\alpha-\beta$: budded forms are favored when $\alpha>\beta$, while stomatocyte morphologies are stable when $\alpha<\beta$.

Unlike the equilibrium morphologies discussed above, the branched morphology predicted by Rozman \textit{et al}.~\cite{rozman2020collective} requires out-of-equilibrium mechanisms to emerge. The authors found that a minimum $K_{T1}$ value drives the branching instability of the initial spherical cyst, where the mechanism involves the in-plane cell arrangement. Active T1 transitions promote the formation of pentagonal arrays of cells at branch tips and heptagonal arrays at the saddle-like branch bases (Fig.~\ref{fig11}B). These defects in arrangement, relative to the hexagonal array of cells, act as seeds of positive and negative Gaussian curvature, and the coupling between this curvature and the topology lock the system in a branched morphology \cite{nelson1987fluctuations,rozman2020collective}.


The modeling approaches discussed so far are cell-resolved, i.e., they carry cell-level information in their formalism. Complementary continuum perspectives have been implemented to address the emergence of luminal shapes from differential growth between tissue layers \cite{theory_hannezo_2014,buckling_nelson_2016,Goriely2017MathMechGrowth}. In the developing chick gut, the luminal boundary varies along the anterior-posterior axis, displaying wrinkles, creases and zigzags \cite{gill2024developmental}. These patterns are governed by the mismatch in growth rates of the endoderm (epithelial layer, $e$) and the surrounding mesenchyme ($m$) acting as a compliant substrate. Gill \textit{et al}.~\cite{gill2024developmental} modeled the chick gut as a growing two-layer tube, endoderm and mesenchyme of thicknesses $h_{e}$ and $h_{m}$ respectively, constrained by a muscular layer at the outer boundaries. The inner radius of the tube is $r_{i}$, delineated by the lumen boundary, and the outer radius is $r_{o}$ (Fig.~\ref{fig11}C). The thicknesses and the inner radius vary along the gut axis and during development \cite{gill2024developmental}.

The authors considered an in plane strain formulation of the system, restricting the analyses to 2D cross sections of the gut. They used differential growth theory \cite{rodriguez1994stress}, where deformation tissue gradients are decomposed in an elastic and a growing part: $\mathbf{F}=\mathbf{A}\cdot\mathbf{G}$. The elastic part, $\mathbf{A}$, is characterized as a neo-Hookean material, i.e., a hyperelastic solid whose strain energy density depends linearly on the first invariant of the Cauchy-Green tensor ($Tr(\mathbf{A}\cdot\mathbf{A}^T)$):
\begin{equation}
    W_{A} =\frac{\mu}{2}\left[J_{A}^{-2/3}Tr(\mathbf{A}\cdot\mathbf{A}^T) - 3\right] + \frac{K}{2}(J_{A} - 1)^{2}.
    \label{Wenergy}
\end{equation}
Here $J_{A}=\det\mathbf{A}$, $\mu$ is an elastic shear modulus, and $K$ is a bulk modulus ($K\gg\mu$) enforcing tissue incompressibility. The growth tensor $\mathbf{G}=diag(g_r, g_\theta, g_z)$ encodes spatiotemporal radial, circumferential, and axial growth ratios fitted to experimental measurements for both endoderm and mesenchyme layers. In a series of finite element simulations of the 2D version (cross sections) of the model (Fig.~\ref{fig11}C), where $g_{z}=1$, different luminal shapes were found along the gut (foregut, midgut, hindgut) as a function of the growth mode, anisotropic ($g_{r}\neq g_{\theta}$) vs isotropic ($g_{r}= g_{\theta}$), and the bulk ratio $\mu_e/\mu_m$. For example, at early stages of foregut development,  isotropic growth of the endoderm, which at that time is  characterized by being stiff and thin ($\mu_e/\mu_m=35$, $h_{e}/h_{m}=0.3$), leads to sinusoidal wrinkles (Fig.~\ref{fig11}D). At later stages, anisotropic epithelial thinning ($g_r<g_\theta$) destabilizes this wrinkled state and triggers a period-doubling instability in the foregut morphology (Fig.~\ref{fig11}D).  Therefore, the model demonstrates that the various lumen forms in the foregut may arise as spatial instabilities caused by differential growth of the tissue boundaries. The same conclusion can be obtained for the midgut and hindgut regions, where different lumen shapes emerge \cite{gill2024developmental}.

The models  reviewed here show that multicellular active processes, such as proliferation kinetics in diseased tissues, in-plane cell rearrangement in epithelial shells, and heterogeneous differential tissue growth, enable access to luminal shapes beyond those predicted by purely variational  variational theories. In addition, the model by Gill and colleagues \cite{gill2024developmental} illustrates that lumen shape formation can also be studied within a continuum theory framework.

\subsection{Lumen-tissue mechanochemical feedbacks}\label{feedbackModels}

The modeling frameworks discussed so far address how hydraulic, mechanical, electrical, and active forces sculpt lumens in diverse biological contexts. However, as reviewed in Section~\ref{mechanoprocess}, the lumen can also feed back onto the surrounding cellular envelope not only by exerting hydrostatic pressure, but also by modifying tissue contractility, gene expression, and cell fate via mechanochemical couplings. 
We review two  recent theoretical models that formalize some of these nonlinear reciprocal interactions.

\subsubsection{Role of the lumen in morphological bistability}\label{xuebistability}

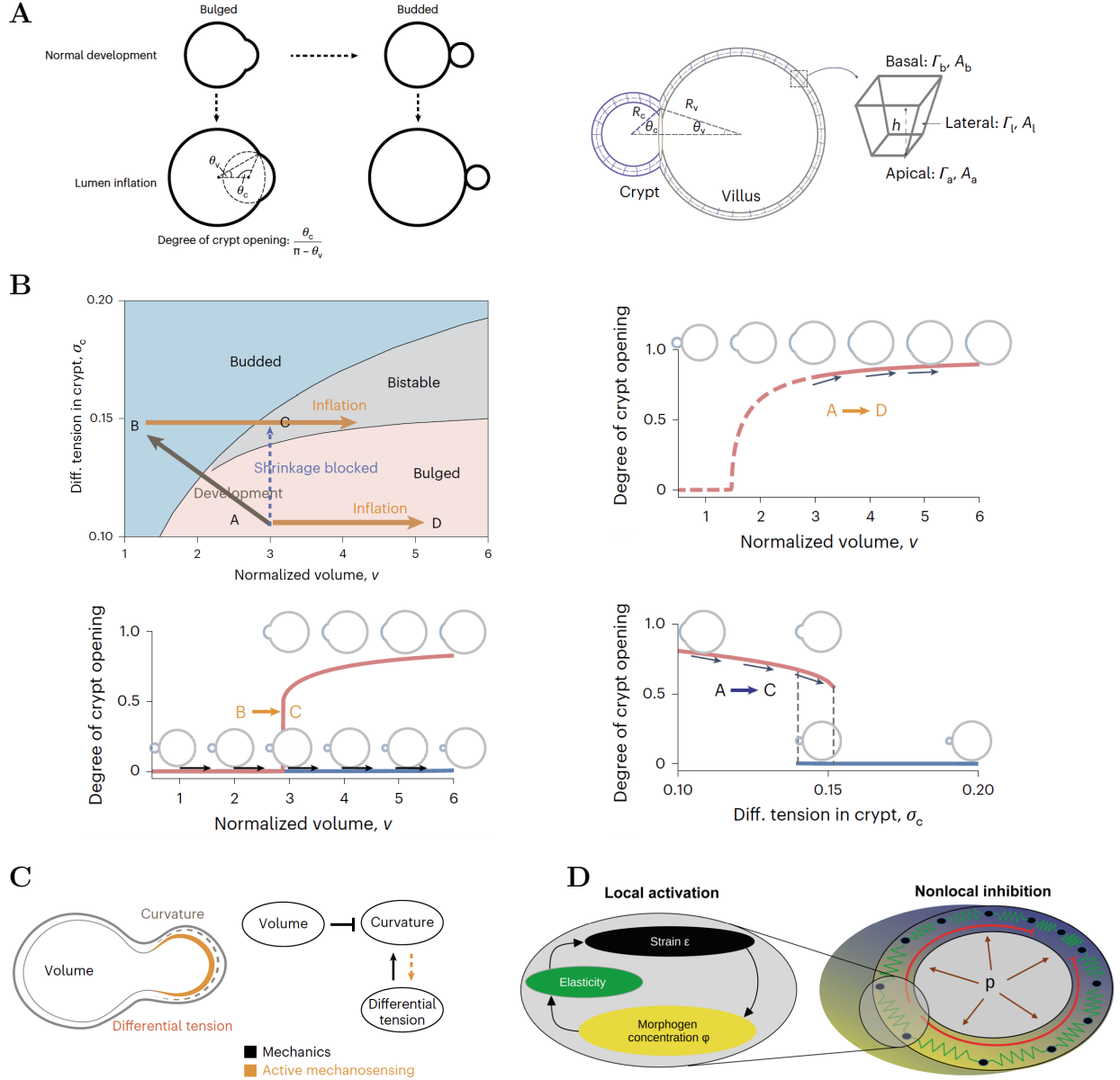
\begin{figure*}
\centering
\begin{tikzpicture}
\node[inner sep=0] (img)
{\includegraphics[width=0.96\linewidth]{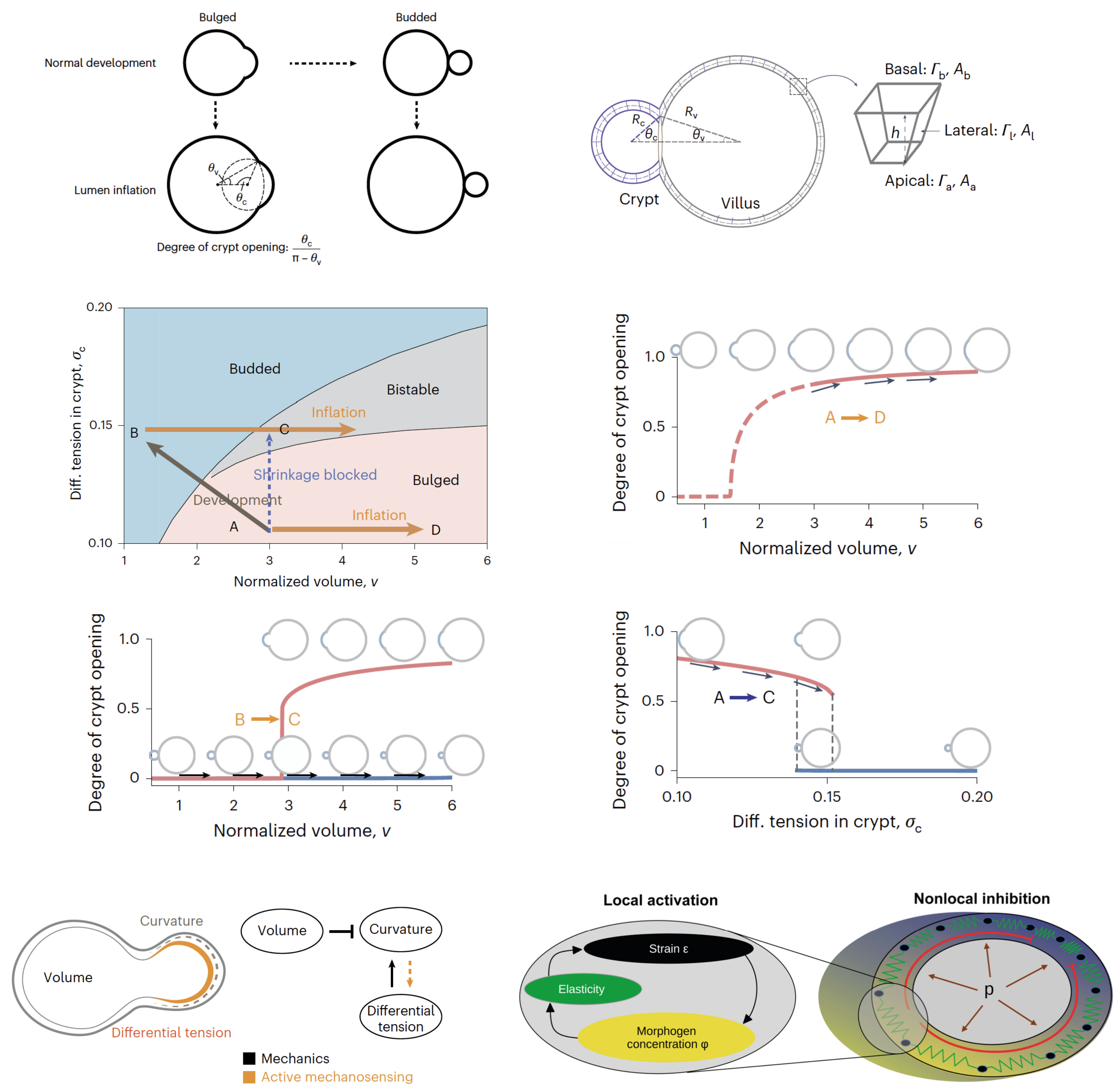}};
\node at (img.north west) [xshift=3mm, yshift=-3mm]{\large\bfseries A};
\node at (img.north west) [xshift=3mm, yshift=-43mm]{\large\bfseries B};
\node at (img.north west) [xshift=85mm, yshift=-130mm]{\large\bfseries D};
\node at (img.north west) [xshift=3mm, yshift=-130mm]{\large\bfseries C};
\end{tikzpicture}
\caption{Lumen role in mechanochemical feedback models. (A) Schematic morphologies during intestinal organoid morphogenesis and during lumen inflation experiments (left panel), and the geometrical set-up for the corresponding vertex model (right panel). (B) Phase diagram in $v-\sigma_{c}$ space illustrating the different shapes minimizing free energy, together with the morphological transitions along specific paths on the diagram. Values $v>1$ correspond to an inflated lumen. (C) Schematic of the mechanosensitive coupling involving the luminal volume, crypt geometry, and actomyosin tension. (A-C) are adapted from \cite{xue2025mechanochemical}. (D) Schematic of the local activation and nonlocal inhibition embedded in the mechanochemical model for regenerating Hydra. Adapted from \cite{weevers2025mechanochemical}.}
\label{figModelCoupling}
\end{figure*}

Xue and colleagues \cite{xue2025mechanochemical} investigated experimentally and theoretically bud formation during the development of intestinal organoids, where lumen volume arises as a key regulator of the morphological transition (see Section~\ref{mechanoprocess}). 
In this system, the transition from an initially bulged epithelial shell to a budded morphology requires lumen deflation at the appropriate time (Fig.~\ref{figModelCoupling}A): If deflation is prevented, the system fails to transition to the budded morphology; yet once budding has occurred, 
reinflating the lumen does not recover the earlier state. This history-dependent shape transition, in which an equivalent luminal volume change produces different outcomes, points to lumen-tuned bistable morphological dynamics.

The authors modeled the intestinal organoid as a hollow epithelial shell enclosing incompressible luminal fluid using a 3D vertex model \cite{yang2021cell} with two distinct regions, crypt ($c$) and villus ($v$) (Fig.~\ref{figModelCoupling}A). Similar to the vertex models reviewed above, the energy per cell of height $h$, depends on apical ($\Gamma_{a}$), basal ($\Gamma_{b}$), and lateral ($\Gamma_{l}$) surface tensions, which are region-dependent in the crypt-villus system. In the limit where the crypt size is smaller than the villus size \cite{yang2021cell,xue2025mechanochemical}, the authors derive a one-dimensional free energy that only depends on the crypt opening angle $\theta_{c}$:
\begin{equation}
\frac{\Delta F(x)}{1-\xi(x)}=\left [ 1-\sigma_{c}\left(\frac{1+x}{2\varphi}\right)^{\frac{1}{2}}\right]^{\frac{2}{3}}\left(1-\frac{1-x}{2\alpha}\right)^{\frac{1}{3}},
    \label{freeHH}
\end{equation}
where $x=\cos(\theta_{c})$ and $\xi(x)\propto v^{-1/3}$, with $v$ the normalized organoid volume relative to the initial state, which implicitly carries the luminal fluid incompressibility. $\varphi$ is the crypt fraction (number of crypt cells vs number total cells, $N_{t}$), $\alpha=(\Gamma_{a}+\Gamma_{b})_{c}/(\Gamma_{a}+\Gamma_{b})_{v}$ is the in-plane tension, and $\sigma_{c}=((\Gamma_{a}+\Gamma_{b})/\Gamma_{l})_{c}\sqrt{\pi/N_{t}}$ is the crypt differential tension, which is related to the spontaneous curvature of the crypt \cite{yang2021cell}. Depending on $\sigma_{c}$, the energy landscape accepts either one minimum, either budded or bulged morphology (Fig.~\ref{figModelCoupling}A), or both of them. 
Interestingly, as a function of the lumen volume $v$, the authors showed the morphological transitions in the different energy landscape regimes (Fig.~\ref{figModelCoupling}B), visualizing the classical path-dependent evolution of bistable dynamical systems. The positive feedback underlying this behavior is between the epithelial thickness and the out-of-plane deformations: tissue bending---due to apicobasal tension differences in the crypt region \cite{krajnc2015theory,luciano2021cell,rozman2020collective,yang2021cell}---tends to thicken the individual cells, which enhances the bending moment $M\sim(\Gamma_{a}-\Gamma_{b})h$. This effect eventually saturates as the thicker cells increase the bending stiffness of the epithelial layer. Xue \textit{et al}.~\cite{xue2025mechanochemical} confirmed this hypothesis by observing the absence of bistability in simulations where the cell thickness $h$ was kept constant. However, this mechanical bistability occurs only over a narrow range of $\sigma_{c}$ ($0.14$--$0.15$, Fig.~\ref{figModelCoupling}B), making it unlikely to be physiologically robust given the inherent variations in myosin and tension levels observed in intestinal organoids \cite{yang2021cell}.
Moreover, they discovered that the experimental tension asymmetry $\epsilon=(\Gamma_{a}-\Gamma_{b})/(\Gamma_{a}+\Gamma_{b})$ values in the crypt region, quantified through myosin intensity ratios \cite{salbreux2012actin,chugh2018actin,streichan2018global}, are below the model threshold necessary to sustain the bistable regime \cite{xue2025mechanochemical}. 

To amend the discrepancies presented above, the authors argued that luminal pressure and actomyosin tensions are not independent, as described in other model systems \cite{desprat2008tissue,fernandez2009myosin,pouille2009mechanical,okuda2018strain}. In particular, they considered a mechanotransduction mechanism in which crypt's curvature regulates tensions (Fig.~\ref{figModelCoupling}C), and introduced the geometry-dependent coupling:
\begin{equation}
\sigma_{c}=\sigma\left(\frac{R_{c}}{\bar{R}_{o}}\right)^{-n},
\label{mecs}
\end{equation}
where $R_{c}$ is the crypt radius of curvature (Fig.~\ref{figModelCoupling}A), normalized to a reference value $\bar{R}_{o}$, $\sigma$ is an intrinsic tension value dictated by stem cell fate \cite{yang2021cell}, and $n>0$ is a parameter quantifying the coupling strength. Eq.~(\ref{mecs}) encodes a positive feedback loop that reinforces the budded morphology. In the bulged state, $R_{c}$ is large and $\sigma_{c}$ is small, which makes it easier to deform the crypt upon lumen deflation. Once the system enters the budded state, however,  $R_{c}$ decreases and $\sigma_{c}$ increases, thereby stabilizing the deformed shape and locking the system into the budded geometry.
Experimentally, the authors were able to relate curvature radius and apicobasal tension assymetry---finding $n=1\pm0.5$ ($\sigma_{c}\propto\epsilon$ \cite{xue2025mechanochemical})---through the lumen inflation experiment in both bulged and budded states. The mechanosensitive model predicts a  broader  bistable region, and the experimentally measured tension asymmetries in budded crypts now lie above the (corrected) theoretical bistability threshold. Altogether, the work by Xue \textit{et al}.~\cite{xue2025mechanochemical} underscores that lumen volume can both drive and lock morphological transitions in intestinal organoid morphogenesis via mechanochemical feedback.

\subsubsection{Luminal pressure as a long-range inhibitor}

Weevers \textit{et al}.~\cite{weevers2025mechanochemical} developed a mechanochemical  model to study how the Wnt signaling organizer is positioned during Hydra regeneration (see Section~\ref{mechanoprocess}), motivated by experiments linking Wnt3  expression to tissue stretching through a positive  feedback \cite{ferenc2021mechanical,weevers2025mechanochemical}. Their model treats the regenerating spheroid as a continuum elastic thin shell that is dynamically driven by periodic luminal forcing (Section~\ref{mechanoprocess}).

The  model couples morphogen dynamics, dictated by reaction-diffusion processes, with the stress balance on the surface of the spheroid (Fig.~\ref{figModelCoupling}D). The morphogen $\phi$ evolves as
\begin{equation}
\partial_{t}\phi=f(\epsilon_{\alpha}^{\beta},\epsilon_{\beta}^{\alpha})-\xi\phi+D\nabla^{2}\phi,
    \label{eqmorpho}
\end{equation}
where its production rate $f$ increases with an invariant of the strain tensor $\epsilon_{\alpha}^{\beta}$ at a constant secretion rate. The indices $\{\alpha,\beta\}$ represent two-dimensional curvilinear coordinates. The morphogen can be degraded at rate $\xi$, and it diffuses with a coefficient $D$. 
The tissue dynamics are assumed to be fast compared with the timescales in Eq.~(\ref{eqmorpho}), so that the surface forces can be treated in quasi-static mechanical balance
\begin{equation}
0=\nabla_{\alpha}[E(\phi)\epsilon^{\alpha\gamma}\alpha_{\gamma}]+\frac{\sqrt{|a|}}{\sqrt{|A|}}p(t)n,
    \label{eqmorpho2}
\end{equation}
where the first term represents  elastic stress contributions, with a morphogen-dependent modulus $E(\phi)=E_{o}e^{-b\phi}$ ($b>0$, $E_{o}>0$), and the second term represents the time varying luminal pressure $p(t)$. Furthermore, $n$  is the outward vector normal to the spheroid surface, and $a$ ($A$) is the determinant of the metric tensor if the deformed (underformed) spherical shape. The system of equations is closed by a time-dependent equation for the lumen volume $V$, 
which only considers passive water flows with permeability $\lambda$:
\begin{equation}
\partial_{t}V=\lambda|\Delta\Pi-p(t)|A(t),
    \label{last}
\end{equation}

Numerical simulations of the model robustly produce spontaneous pattern formation in the form of a localized morphogen activation accompanied by a protrusion on the shell and to progressive stretching \cite{weevers2025mechanochemical}. Notice that from a pattern-formation equations perspective, the stabilization of this localized structure does not require local inhibition, as the luminal pressure equilibration constraint acts as a nonlocal inhibitor. Therefore, the model  offers a lumen-centered mechanistic explanation for the positioning of the signaling center in the context of Hydra regeneration. Beyond contributing to our understanding of lumenogenesis, the study  suggests that morphogen patterning need not rely exclusively on the classical local activator-inhibitor scheme involving two species. Instead, similar patterns may arise from local chemical activation combined with long-range mechanical inhibition, a possibility that is particularly appealing given that the molecular realization of the classical mechanism remains elusive \cite{mercker2016beyond,wang2023mechano}.

\section{Designer matrices to modulate lumenogenesis}\label{matrices}

As reviewed in previous sections, the ECM plays key roles in lumenogenesis: it orients apicobasal polarity (Section~\ref{s21}), provides biochemical cues for tissue sculpting (Section~\ref{sECM}), and mechanically resists lumen expansion (Section~\ref{sculpting}). Yet despite this importance, the coarse-grained models reviewed in Section~\ref{theocomp} treat the ECM only indirectly, absorbing its effects into effective parameters such as repulsive cell-ECM interactions in phase field models \cite{lu2025generic,lee2025permeability} or implicit contributions to basal tension in vertex models \cite{vasquez2021physical,guha2026control}. Because these effective terms are not directly linked to measurable ECM properties, they remain difficult to constrain experimentally. This limitation suggests a clear path forward for lumenogenesis modeling, namely to exploit recent advances in engineered, or \textit{designer}, matrices that allow precise and systematic control of ECM properties
\cite{enemchukwu2016synthetic,ritzau2023microfibrous,mitrofanova2024bioengineered}.

In this section, we review these advances by focusing on three classes of ECM control. We begin with matrix mechanics and adhesion, where synthetic environments have been used to separate the effects of stiffness from those of ligand presentation and degradability, revealing how specific biochemical and mechanical cues regulate cell polarity and lumen formation (Section~\ref{5d1}). We then discuss confinement, emphasizing how both hard and soft restrictions imposed by the environment influence epithelial organization, lumen initiation, and long-term stability (Section~\ref{5d2}). Finally, we turn to geometrical cues and spatial patterning, where technologies such as microwells, microfluidics, melt electrowriting, and bioprinting have opened new ways to guide lumen morphology (Section~\ref{5d3}). Along the way, we highlight where connections can be made to the modeling approaches discussed in Section~\ref{theocomp}.

\subsection{Mechanics and adhesion}\label{5d1}

The first step in lumen formation is the establishment of apicobasal polarity. The presence of ECM is necessary for the correct establishment of polarity and suspension cultures (no ECM) of cells yield cysts with the apical side facing outward instead of toward the lumen with inverted polarity \cite{margalef2019controlling}. There has been a growing interest in deciphering the exact ECM cues that are necessary for lumenogenesis and for recreating physiologically relevant luminal structures \textit{in vitro}. Seminal work by Paszek \textit{et al}.~\cite{paszek2005tensional} showed that mammary epithelial cells cultured in compliant (low stiffness, $<500$ Pa) reconstituted basement membrane matrices, such as matrigel, supported lumen formation, while higher stiffness activates Rho-ROCK signaling---a pathway that promotes actomyosin contractility \cite{salbreux2012actin}---and impairs lumen formation. Similarly, MDCK cells embedded in type-I collagen gels form luminal structures and are surrounded by basement membranes \cite{o2002building}. However, it is hard to decouple the effect of mechanics from biochemical cues in natural ECMs since increasing the ECM concentration to tune mechanical properties simultaneously changes the concentration of adhesion ligands. Consequently, this coupling obscures how individual properties of the matrix independently regulate lumenogenesis, including lumen formation and expansion.

Synthetic matrices that mimic key properties of the natural ECM serve as an alternative to systematically investigate the impact of various biophysical and biochemical cues on lumenogenesis. For example, matrix metalloproteinases (MMP) degradable synthetic poly (ethylene glycol) (PEG) hydrogels presented with RGD (Arg-Gly-Asp) adhesion ligands, an integrin binding motif, promoted MDCK lumen morphogenesis to the same extent as in collagen-I hydrogels \cite{enemchukwu2016synthetic}. Importantly, this degradable hydrogel system enabled control over the number of lumens formed by varying polymer density and showed RGD density to be the determinant of cyst polarity and lumen formation. One of the binding integrins of RGD is $\alpha v \beta 3$ and blocking this cell-ECM interaction using inhibitors or integrin targeting antibodies disrupted lumen formation. For lumen formation in intestinal organoids, RGD was indispensable, supporting higher cell viability with higher percentage of live cells and robust lumen formation when compared to other adhesion ligands including GFOGER, AG73, IKVAV \cite{cruz2017synthetic}. In addition, these organoids after several days maintained a central lumen and displayed epithelial budding at the interface with the hydrogel. However, in a different study, GFOGER interaction through $\alpha 2\beta 1$ was found to be necessary for creating stable lumen phenotypes \cite{hernandez2020fully}. The discrepancy in the adhesive ligand requirements were attributed to the differences in the epithelial sources (human vs. mouse crypts), and also to  GFOGER being a triple helix peptide which is structurally rigid and can have impact on mechanics and ligand presentation \cite{hernandez2020fully}.

From a modeling perspective, the effects of adhesion ligand  can enter  the theoretical descriptions of Section~\ref{cellcon} through the basal surface tension $\Gamma_{b}$ in Eq.~(\ref{energyPER}) and less directly through the stringency parameter $k_{l}$ in Eq.~(\ref{hamiltonian}). In principle, both parameters  depend on cell-ECM interactions. Therefore, designer matrices with independently tunable ligand type and density could be used to experimentally test the transition between spherical and convoluted luminal morphologies predicted by varying $\Gamma_{b}$ or $k_{l}$.

\subsection{Confinement}\label{5d2}
Confinement is a fundamental physical cue experienced by cells \textit{in vivo} and has emerged as a critical regulator of lumenogenesis. In this context, confinement refers to the mechanical restriction imposed on cells either by limiting them to adhesion patterns or through the ability of the cells to remodel their surrounding matrix over time. In synthetic hydrogels, confinement is strongly influenced by matrix degradability and viscoelasticity. Hydrogels whose polymer network is crosslinked with non-degradable molecules or cultured on adhesive patterns impose "hard" confinement as the cells are not able to relieve the mechanical restrictions. In contrast, cells cultured in stress-relaxing matrices or MMP-degradable matrices enable them to remodel the environment either mechanically or via MMP-mediated degradation resulting in a progressive decrease in confinement over time, which we refer to as "soft" confinement \cite{wolf2007multi,packard2009direct}.

To investigate the effects of hard confinement on lumen formation, Rodr\'iguez-Fraticelli \textit{et al}.~\cite{rodriguez2012cell} seeded MDCK cells on collagen micropatterns of different sizes. On large micropatterns ($\sim1600\ \mu$m$^{2}$), lumen formation was less while on smaller ones ($700\ \mu$m$^{2}$), the lumen formation efficiency was significantly increased. Mechanistically, this process was found to be mediated through Par6-atypical protein kinase C (aPKC) dependent signaling. In MMP degradable PEG gels, intestinal stem cells promoted lumen formation \cite{gjorevski2016designer}. While both hard and soft confinements supported lumen formation, cell viability was impaired in hard confinement conditions in synthetic hydrogels after $7$ days in culture, underscoring the importance of remodeling at the cell-ECM boundary in creating long term stable lumens \cite{cruz2017synthetic}.

Interestingly, different levels of soft confinement can impact the emergence of distinct lumen phenotypes in epithelial tissues, ranging from  no lumen to single lumen and multiple lumen structures \cite{enemchukwu2016synthetic}. Recent works have also demonstrated the impact of ECM stress-relaxation, which enables mechanical remodeling by cells. Fast relaxing matrices presented with high RGD ligand density allow apicobasal polarity and lumen formation in hiPSCs, independent of matrix stiffness \cite{indana2021viscoelasticity}. While low RGD density yielded thicker multicellular structures that persisted over time, high RGD density resulted in lumens that had monolayered hiPSC cysts recapitulating epiblast-like organization. Furthermore, it was determined that actomyosin contractility, actin polymerization and Rac1---a pathway involved in cell polarization \cite{rodriguez2014organization}---activity were involved in lumen formation \cite{indana2021viscoelasticity,indana2024lumen}. More broadly, in natural ECMs, confinement is inherently dynamic, as cell mediated MMP degradation can progressively alter stress-relaxation resulting in transition from initially restrictive environments to permissive states \cite{narasimhan2025matrix}. Thus, confinement should be viewed as an emergent property of cell-ECM interactions rather than a fixed material parameter.

The observation that confinement controls lumen phenotypes can be framed within the phase field approach described in Section~\ref{phasefieldS}. In this framework, the ECM is represented by a phase field with an elastic modulus $\alpha_{c}$ that penalizes deviations from its relaxed area and confines  cells through steric cell-ECM  repulsion. A stiffer or non-degradable matrix would correspond to larger values of $\alpha_{c}$, thereby restricting  cell expansion and limiting the space available for lumen coalescence. A systematic study relating $\alpha_{c}$ and matrix stiffness or degradability could improve the current ECM description within the phase field framework and test whether confinement-driven transitions between lumen phenotypes \cite{enemchukwu2016synthetic} can be quantitatively reproduced.
   
\subsection{Geometrical cues and patterning}\label{5d3}
\begin{figure*}
\centering
\begin{tikzpicture}
\node[inner sep=0] (img)
{\includegraphics[width=0.98\linewidth]{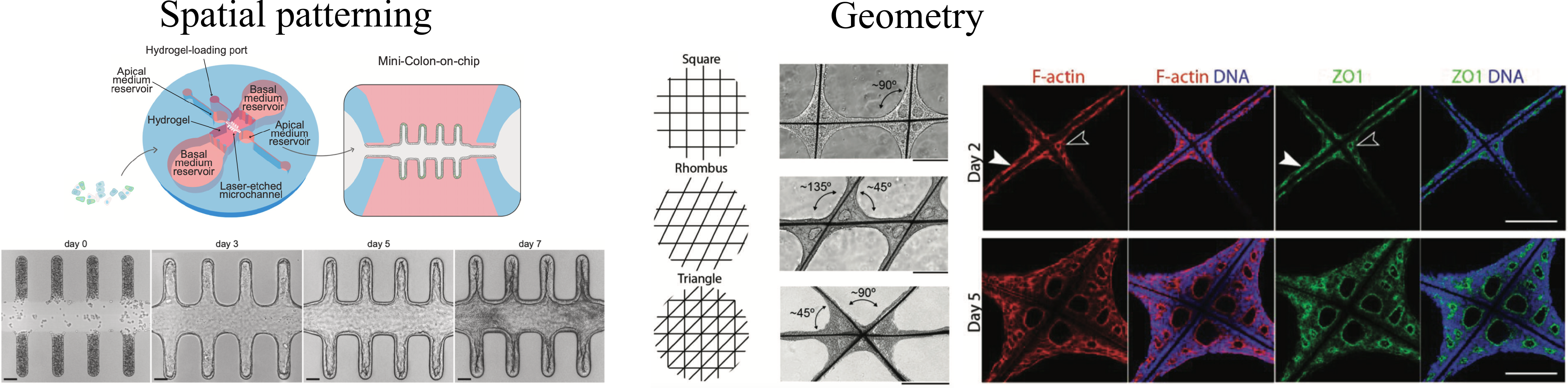}};
\node at (img.north west) [xshift=0mm, yshift=-2mm]{\large\bfseries A};
\node at (img.north west) [xshift=72mm, yshift=-2mm]{\large\bfseries B};
\node at (img.north west) [xshift=110mm, yshift=-2mm]{\large\bfseries C};
\end{tikzpicture}
\caption{Geometrical control of lumenogenesis.(A) Spatial patterning to create gut morphogenesis using microfluidics (scale bar, $100\ \mu$m). Adapted from \cite{mitrofanova2024bioengineered}. (B) Melt electrowriting to control scaffold geometry (scale bar, $1$mm). (C) Characterization of lumen formation on square grid scaffolds  at day $2$ and day $5$ by staining for F-actin, and immunostaining with apical protein marker ZO1 and nuclei (DNA). A representative example of lumen emergence at the scaffold intersections is shown on day $2$ (with white line arrowheads) and at the scaffold walls (solid white arrowheads). The bottom row shows the matured tissue comprising the lumen at day $5$ (scale bar, $200\ \mu$m). Both (B) and (C) are adapted from \cite{ritzau2023microfibrous}.}
\label{fig12}
\end{figure*}
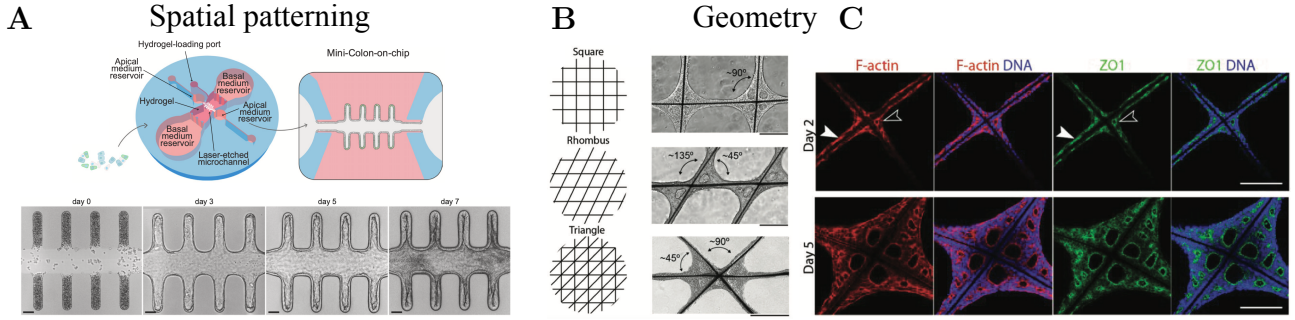

Application of bioengineering tools such as microwell fabrication \cite{karp2007controlling,gong2015generation} and bioprinting \cite{murphy20143d,mandrycky20163d} has enabled precise control over geometrical cues and spatial patterning to better replicate \textit{in vivo} conditions that drive lumenogenesis. Nikolaev \textit{et al}.~\cite{nikolaev2020homeostatic} used an open channel microfluidic device in which intestinal stem cells (ISC) along with matrigel were cultured in a microchannel that mimics the budded-geometry of native crypts (Fig.~\ref{fig12}A). This led to epithelial tube formation with a central lumen guided by the channel geometry rather than the normal self-organization of cells during development. The lumens were perfusable enabling continuous flow of ions and other nutrients, and long lived stable structures. Moreover, perfusion introduces physiologically relevant shear stress and pressure, which helps in studying transport properties and barrier function.

Additionally, Gjorevski \textit{et al}.~\cite{gjorevski2022tissue} showed that the crypt-villus formation can be spatially controlled. In this work, photosensitive PEG gels that undergo degradation and softening when exposed to $405$ nm light were used to control the spatial patterning of crypts, which emerged at places where the gels were softened. Interestingly, they also found that the initial shape of the crypt achieved through microwells dictates the spatial distribution of ISC and differentiated cells. This arises from geometry-induced differences in cell packing and spreading, which are translated into YAP signaling---a mechanosensitive pathway linking cell shape deformation to gene expression changes---and subsequent symmetry breaking mediated by Notch, a cell-cell signaling pathway.

Another promising technology for spatial patterning is the melt electrowriting (MEW) \cite{kade2021polymers,o2023decade}. In MEW, polymers that have lower melt temperatures ($60$--$90^{\circ}$C) can be molded into scaffolds through extrusion based 3D printing. This is particularly useful for creating scaffolds of tunable ECM porosity enabling modulation of spatial cues. Consistent with this, a recent work showed that by varying the scaffold geometry from square to rhombus to triangle, it was possible to control lumenogenesis as the modulation of geometry directly affects the curvature sensed by the cells \cite{ritzau2023microfibrous}. Specifically, at $45^{\circ}$ grid angles on these scaffolds, larger lumens were observed with a $26\%$ increase in cell density when compared to the $90^{\circ}$ grid angles (Figs.~\ref{fig12}B~and~\ref{fig12}C).

The dependence of lumen size on scaffold angle suggests that curvature itself can act as a mechanical cue for lumen formation. This observation may reflect a mechanosensitive coupling between tissue geometry and lumenogenesis. The coarse-grained strategy reviewed in Section~\ref{xuebistability}, linking tissue curvature and tension, offers a possibility to formalize such coupling. Alternatively, different scaffold angles impose specific basal cell curvatures that may influence the apical area available for lumenogensis, introducting an additional geometric constraint that could be explored using the models reviewed in Section~\ref{cellcon}. Testing these ideas would require measurements of cortical tension and apical area  on scaffolds with systematically varied geometry.

Although MEW has the potential for creating complex geometries, integrating cells during the printing process for 3D culture is not possible due to the high temperatures involved ($>37^{\circ}$C). A complementary approach is the use of bioprinting which makes use of cell laden bioinks, composed of synthetic or natural materials, to fabricate scaffolds with defined spatial organization. For instance, luminal structures can be generated through collective self-organization of stem cells using collagen-I as the bioink, with the final architecture controlled by cell density and print geometry \cite{brassard2021recapitulating}. However, using collagen as the bioink has several limitations including the requirement of higher matrix densities, which restricts cell infiltration, as well as the use of light based approaches to crosslink, which lacks physiological relevance. A recent work mitigated these issues by employing macromolecular crowding using high concentrations of PEG \cite{ranamukhaarachchi2019macromolecular,narasimhan2025matrix} to rapidly produce 3D printed collagen scaffolds, which has promising potential for lumen morphogenesis \cite{gong2025instant}.

A critical challenge in these bioprinting approaches is the maintenance of the structural fidelity after printing as many bioinks tend to collapse due to insufficient mechanical stability. Granular microgels that are compatible with bioprinting provide a promising alternative as these materials can rapidly recover after being deformed during printing \cite{bhattacharjee2015writing}, while allowing tunable rheological properties \cite{daly2024granular}. Such granular microgel incorporation enabled the printing of even fragile materials such as matrigel, while maintaining structural stability and reproducibility.

The designer matrices discussed in this section enable orthogonal control of ECM properties, such as stiffness, ligand density, degradability, and scaffold geometry, that are inherently coupled in natural ECMs. This experimental control may provide the missing ingredient needed both to refine the ECM description in current physical models and to constrain the next generation of lumenogenesis theories.

\section{Outlook}
In this review, we have characterized lumens as active balloons whose emergence and maintenance depend on a nonlinear and reciprocal interplay between hydraulic forces, cell mechanics, biochemical signaling, ECM properties, and cell dynamics. While significant experimental and theoretical progress has been made, several open questions remain that we believe are particularly well-suited for physics-based approaches.

A first avenue involves the ECM. Although simplified ECM descriptions in theoretical models have been useful for testing hypotheses against experiments, a more informative treatment of  lumen-ECM interaction remains particularly needed. 
As discussed throughout this review, the ECM plays a critical part in providing the  cues required for lumenogenesis, from orienting apicobasal polarity to stabilizing the cavity after growth. 
Several concrete questions remain open. How does cell-mediated ECM degradation couple to lumen dynamics? Can the interplay between different adhesion ligands---
such as the RGD versus GFOGER discrepancy discussed in Section~\ref{5d1}---be rationalized within a mathematical framework? 
Could passive mechanisms, such as differential adhesion between cell-cell and cell-ECM contacts, contribute to the initial opening of the lumen through rupture-like processes? Addressing these questions will require biophysical models that treat the ECM not as a passive background, but as a dynamic participant in lumenogenesis.

A second direction involves the coarsening dynamics reviewed in Section~\ref{coarseningModels}. The coalescence of multiple microlumens into a single lumen shares similarities with classical Ostwald ripening, but with the added richness of active transport, cellular heterogeneity, and tissue organization. 
An interesting question is whether these  processes can be understood with more general theories of active coarsening, a topic that has received considerable attention in the active matter community \cite{cates2025active}. 
Blastocyst lumenogenesis adds a further layer of specificity. Schliffka \textit{et al}.~\cite{schliffka2024inverse} showed that distinct topological sites within the hydraulic network play different roles, with inverse blebs at bicellular contacts operating as hydraulic pumps that steer luminal fluid toward multicellular sinks.
Their high-resolution imaging of the entire process, from nucleation to coarsening, also raises the possibility that microlumen nucleation at bicellular interfaces may obey a form of spatial regularity.
To our knowledge, connecting lumen nucleation to ideas from pattern-formation theory remains an open and largely unexplored direction.
    
Third, the microwell confinement technique introduced by Lu \textit{et al}.~\cite{lu2025generic} opens a route to investigate how boundary shape dictates lumen positioning. In their study, the confining shapes were spherical or ellipsoidal, but more dramatic symmetry breaking geometries---including dumbbells or domains with sharp indentations---could be used to test whether lumens can be placed away from the geometric center. This question is relevant \textit{in vivo}, where lumens emerge under highly dynamic boundary conditions. Phase field models, already used in this context (Section~\ref{phasefieldS}), should be well-suited to explore this idea.  

Fourth, and at a more fundamental level, an important open question is whether lumenogenesis can be described within a unified coarse-grained theory, or whether different tissues are fundamentally governed by distinct minimal models. Most of the models reviewed here were formulated for specific biological systems, suggesting that each context may require its own tailored description. Yet from a physics perspective, it is natural to ask whether common underlying principles---for example, involving transport, force balance, geometry, and feedback---span these different systems and could be captured within a more general framework.

Finally, and along similar lines, it is also worth asking which level of modeling is most informative and predictive for describing lumenogenesis. Cell-resolved approaches, such as Cellular Potts, vertex, and phase field models, can incorporate intracellular and extracellular details together with cell-specific morphologies. However, these descriptions quickly become computationally intractable as the number of cells increases and especially when a full three-dimensional representation is required. A continuum description is then a natural next step, but the passage from a cell-resolved to a coarse-grained model requires careful analysis and well-justified assumptions.

On the experimental side, the lack of \textit{in vitro} systems that recapitulate hierarchically organized and interconnected luminal structures, such as branching ducts in mammary tissue, currently limits our ability to investigate how lumenogenesis proceeds during branching morphogenesis. Developing such systems would be valuable not only for gaining biological insight, but also for constraining models that are in physiologically relevant settings. Along similar lines, the scaffold geometries discussed in Section~\ref{5d3} offer striking control over luminal morphology, yet both their physiological relevance and the physical mechanisms by which different scaffold angles produce distinct shapes remain poorly understood.


As a final remark, we note that most references in this review are not published in traditional physics journals, which is not uncommon in biological physics reviews \cite{plum2025dynamical,bruckner2024learning}. We hope this is seen not as a barrier, but as an invitation for physicists to engage with lumenogenesis, a topic where quantitative reasoning is becoming increasingly important.

\section{Acknowledgments}
This work was supported by NSF MCB 2426002 and NSF PHY 2310496 to W.-J.R., and by a Prebys Foundation Research Heroes grant to S.I.F. S.E.-A. thanks Magdalena Fadic Repetto for the careful reading of the text and for her help in designing Fig.~\ref{fig2}.

\vspace{2cm}

\bibliographystyle{unsrt}
\bibliography{apssampCell}

\end{document}